\documentclass[fleqn,usenatbib]{mnras}
\usepackage{newtxtext,newtxmath}

\usepackage[T1]{fontenc}
\usepackage{ulem}
\usepackage{xcolor} 
\usepackage[left]{lineno}

\usepackage{graphicx}	
\usepackage{amsmath}	
\usepackage{float}
\usepackage{multicol}        
\usepackage{multirow}        
\usepackage{bm}		
\usepackage{pdflscape}	
\usepackage{lscape}
\usepackage{array}
\usepackage[figuresright]{rotating}
\usepackage{caption}
\usepackage{subcaption}

\def\mbh{$M_{\rm BH}$\/}

\def\lledd{$L_\mathrm{bol}/L_\mathrm{Edd}$}
\def\lE{$\lambda_\mathrm{E}$}
\def\civ{{\sc{Civ}}$\lambda$1549\/}

\def\rfe{$R_{\rm FeII}$}
\def\rfeUV{$R_{\rm FeII,UV}$\/}
\def\rfeopt{$R_{\rm FeII,opt}$\/}

\def\feiiq{\rm Fe{\sc ii}$\lambda$4570\/}
\def\msol{M$_\odot$\/}

\def\chm{$c(\frac{1}{2})$\/}
\def\cqm{$c(\frac{1}{4})$\/}
\def\ltsima{$\; \buildrel < \over \sim \;$}
\def\ltsim{\lower.5ex\hbox{\ltsima}}  
\def\gtsima{$\; \buildrel > \over \sim \;$}

\def\gtsim{\lower.5ex\hbox{\gtsima}} 

\def\lya{{Ly}$\alpha$}

\def\oiiiseven{[\ion{O}{III}]$\lambda$5007\/}
\def\oiiionly{[\ion{O}{III}]\/}
\def\hb{{\sc{H}}$\beta$\/}

\def\hbbc{{\sc{H}}$\beta_{\rm BC}$\/}
\def\hbFP{{\sc{H}}$\beta_{\rm FP}$\/}

\def\hbnc{{\sc{H}}$\beta_{\rm NC}$\/}
\def\mgii{{Mg\sc{ii}}$\lambda$2800\/}

\def\mgiionly{{Mg\sc{ii}}\/}
\def\mgiionlyFP{{Mg\sc{ii}$_\mathrm{FP}$}\/}
\def\mgiionlybc{{Mg\sc{ii}$_\mathrm{BC}$}\/}

\def\oiiiopt{{\sc{[Oiii]}}$\lambda\lambda$4959,5007\/}

\def\heii{{\sc{Hei}}$\lambda$1640\/}
\def\feiiuv{{Fe\sc{ii}}$_{\rm UV}$\/}
\def\feiiopt{{Fe\sc{ii}}$_{\rm opt}$\/}
\def\feii{{Fe\sc{ii}}\/}
\def\fe{{\sc{Fe}}\/}

\def\fe76087{{\sc [Fe vii]}$\lambda$6087\/}
\def\kms{km\,s$^{-1}$}

\def\rk{$R_{\rm K}$\/}
\def\rks{$R_{\rm KS}$\/}
\def\ergss{ergs\,s$^{-1}$\/}

\def\heii{{{\sc H}e{\sc ii}}$\lambda$4686\/}

\def\o4959{{\sc{[Oiii]}}$\lambda$4959\/}

\title[Low-$z$ jetted quasars in the MS context]{Optical and near-UV spectroscopic properties of low-redshift jetted quasars in the main sequence context}

\author[S.\ Terefe et al.]{
Shimeles Terefe\,$^{1,2,3,{\thanks{Visiting researcher at the IAA-CSIC as a PhD fellow}}}$\thanks{e-mail: shimeles11@gmail.com},
Ascensi\'{o}n Del Olmo\,$^{4}$,
Paola Marziani\,$^{5}$,
Mirjana Povi\'{c}\,$^{1,4,6}$, \newauthor
Mar\'{i}a Angeles Mart\'{i}nez-Carballo$^{7}$,
Jaime Perea\,$^{4}$, 
and Isabel M\'{a}rquez\,$^{4}$
\\
$^{1}$Space Science and Geospatial Institute (SSGI),  Entoto Observatory and
Research Centre (EORC), Astronomy and Astrophysics Research \\ and Development Division,
P.O.Box 33679, Addis Ababa, Ethiopia\\
$^{2}$Addis Ababa University (AAU), P.O.Box 1176, Addis Ababa, Ethiopia\\
$^{3}$Jimma University, College of Natural Sciences, Department of Physics, P.O.Box 378, Jimma, Ethiopia\\
$^{4}$Instituto de Astrof\'{i}sica de Andaluc\'{i}a (IAA-CSIC), Glorieta de la Astronomia s/n, Granada E-18008, Spain\\
$^{5}$Istituto Nazionale di Astrofisica (INAF), Osservatorio Astronomico di Padova,vicolo dell$^{’}$ Osservatorio 5, Padova I-35122, Italy\\
$^{6}$Physics Department, Faculty of Science, Mbarara University of Science and Technology (MUST), P.O. Box 1410, Mbarara, Uganda\\
$^{7}$Department of Applied Mathematics and IUMA. Computational Dynamics group. University of Zaragoza. E-50009. Spain
}

\date{Accepted XXX. Received YYY; in original form ZZZ}

\pubyear{2022}

\begin{document}
\label{firstpage}
\pagerange{\pageref{firstpage}--\pageref{lastpage}}
\maketitle

\begin{abstract}
This paper presents new optical and near-UV spectra of 11 extremely powerful jetted quasars, with radio to optical  flux density ratio $>$ 10$^3$, that concomitantly cover the low-ionization emission of \mgii\ and \hb\ as well as the \feii\ blends in the redshift range $0.35 \lesssim z \lesssim 1$. We aim to quantify broad emission line differences between radio-loud (RL) and radio-quiet (RQ) quasars by using the 4D eigenvector 1 parameter space and its Main Sequence (MS) and to check the effect of powerful radio ejection on the low ionization broad emission lines. The \hb\ and \mgii\ emission lines were measured by using non-linear multicomponent fittings as well as by analysing their full profile. We found that broad emission lines show large redward asymmetry both in \hb\ and \mgii. 
The location of our RL sources in a UV plane looks similar to the optical one, with weak \feiiuv\ emission and broad \mgii. We supplement the 11 sources with large samples from previous work to gain some general inferences. We found that, compared to RQ, our extreme RL quasars show larger median \hb\ full width at half maximum (FWHM), weaker \feii\ emission, larger \mbh, lower \lledd, and a restricted bf space occupation in the optical and UV MS planes. The differences are more elusive when the comparison is carried out by restricting the RQ population to the region of the MS occupied by RL sources, albeit an unbiased comparison matching \mbh\ and \lledd\ suggests that the most powerful RL quasars show the highest redward asymmetries in \hb.

\end{abstract}


\section{Introduction}
\label{sec:Intro}

 Type I  Active Galactic Nuclei (AGN) exhibit distinct features that are not seen in normal galaxies, such as a power-law continuum shape in UV and optical bands with a number of broad and narrow emission lines  emitted by ionic species over a wide range of ionization potentials\citep[e.g.,][]{1986ARA&A..24..171O, 2001AJ....122..549V, 2015ARA&A..53..365N}. 
They show widely different properties among themselves \citep{1985ApJ...297..166O,1992ApJS...80..109B,2003AJ....126.1720S,2008ARA&A..46..475H, 2018A&A...613A..38S}: different line profiles, intensity ratios, and ionization levels \citep[e.g.,][]{1982LicOB.927....1G, 1997ApJS..113..245C,2007ApJ...666..757S, 2010MNRAS.409.1033M}. Since the discovery of  quasi-stellar radio sources and quasi-stellar objects (quasars and QSOs)\footnote{ In the following we will use the  quasar as an umbrella term that includes all type-1 AGN regardless of luminosity and radio power.} in the early 1960s  there is a general consensus about their nature:  they are thought to be powered by matter accreting on to a supermassive black hole (SMBH) capable of producing emission across the majority of the electromagnetic spectrum. They are among the most powerful, distant, and luminous objects in the Universe reaching luminosities $\gtrsim 10^{48}$ ergs s$^{-1}$ \citep[e.g.,][]{1984ARA&A..22..471R,1997iagn.book.....P, 2018Natur.553..473B} although much of the empirical understanding is still to be developed. The last three decades have opened promising lines of investigation  on the definition and contextualization of optical and UV properties \citep[e.g.,][]{1992ApJS...80..109B,2000ARA&A..38..521S,2014Natur.513..210S,2020MNRAS.492.3580W}, with the exploration of a spectroscopic unification for broad-line emitting AGN, like the four-dimensional eigenvector 1 (4DE1) parameter space that organizes quasar diversity  \citep{2000ApJ...536L...5S}. 

 Type 1 AGN have been classified into two distinct classes using the ratio of their radio to optical flux densities \citep{1989AJ.....98.1195K}, which are thought to correlate with the presence or absence of extended relativistic radio jets. These are known as RL and RQ, respectively \citep[e.g.,][]{1993ARA&A..31..473A,1995PASP..107..803U,2019ARA&A..57..467B}.
 
With the improvement of radio interferometry techniques, it was possible to notice that both classes are capable of producing radio jets,  
 although RQ jets are far less powerful and at least in some cases sub-relativistic  \citep[e.g.,][]{1998ApJ...496..196U,2003ApJ...591L.103B,2004A&A...417..925M,  2005ApJ...621..123U,2006AJ....132..546G, 2019MNRAS.485.3009H,2021A&A...655A..95S}.  The radio power of RQ quasars can be even 2 -- 3 orders of magnitude lower than that of their RL counterparts for the same optical power.  
 \citet{2017A&ARv..25....2P}  argues that the classification should be based on a fundamentally physical rather than just an observational difference, namely the presence (or lack) of strong relativistic jets and that we should use the term “jetted” and “non-jetted”.  From the theoretical point of view, in spite of the great advancement in the ability to collect sets of data with very long baseline interferometry (VLBI), the  formation of the relativistic radio jet in quasars is still an open question \citep[e.g. see][and references therein]{1995PASP..107..803U,
2019ARA&A..57..467B,2020ARA&A..58..407D}.
 
\begin{table*}
\centering
\caption{Summary of sample properties, and observations  with Cassegrain
TWIN spectrograph of the 3.5m telescope at the Calar Alto Observatory}
\label{tab:bspo}
\begin{tabular}{l l r c l l l c c c}   
\hline\hline  
 Object &\multicolumn{2}{c}{ Coordinates} & \multicolumn{1}{c}{z}& $ m^{a}$ & $ M^{a}_{B}$ & \multicolumn{1}{c}{ Date of} & \multicolumn{1}{c}{ Total exp.} & Airmass & S/N\\
 &RA(2000)& Dec(2000)&  &  &  & observation&  time (s) & & \\
  (1) & \multicolumn{1}{c}{(2)} & \multicolumn{1}{c}{(3)} & (4) & (5) & \multicolumn{1}{c}{(6)} & \multicolumn{1}{c}{(7)} & (8) & (9) & (10) \\
 \hline
 PHL 923 & 00 59 05.6 &  +00 06 51 &  0.7183 &  17.9 &  -24.6 & 22-10-2012 &  3600 & 1.26 &  18 \\
 B2 0110+29 & 01 13 24.2 &  +29 58 16 &  0.3625 & 17.0 & -24.2 & 23-10-2012 &  2700 &  1.08 &  12 \\
 3C 37 &  01 18 18.5 &  +02 58 06 &  0.6667 & 18.8 &  -23.7 & 23-10-2012 & 3600 & 1.29 & 20 \\
 PKS 0230-051 &  02 33 22.1 & -04 55 08 & 0.7807 & 17.0 & -25.9 & 23-10-2012 & 3600 & 1.47 & 43 \\
 3C 94 &  03 52 30.6 & -07 11 02 & 0.9648 &16.7 & -26.4 & 23-10-2012 & 2700 & 1.48 & 34\\
 PKS 0420-01 & 04 23 15.8 &  -01 20 33 & 0.9136 & 17.0 & -25.9 & 22-10-2012 & 3600 & 1.30 &38 \\
 3C 179 & 07 28 10.8 & +67 48 47 & 0.8416 & 18.4 & -24.9 & 23-10-2012 & 3600 & 1.17 & 19 \\
 3C 380 & 18 29 31.8 & +48 44 46 & 0.6919 &16.8 & -25.5 &  23-10-2012 & 2700 &1.12 & 40 \\
 S5 1856+73 & 18 54 57.4 & +73 51 19 &0.4604 & 16.8 & -25.0 & 23-10-2012 & 3600 & 1.35 & 68 \\
 PKS 2208-137 & 22 11 24.1 & -13 28 10 &  0.3912 &  17.0 & -24.4 & 23-10-2012 & 2700 & 1.62 & 18 \\
 PKS 2344+09 & 23 46 37.0 & +09 30 45 & 0.6724 & 15.9 & -26.3 & 22-10-2012 & 3600 &  1.16 & 74\\
\hline                               
\end{tabular}\\
{\raggedright \textbf{Note}: Col. 2 is in units of hours, minutes, and seconds. Col. 3 is in units of degrees, minutes, and seconds.  
\small{$^{(a)}$} From the catalogue of quasars and active galactic nuclei \citep[][$13^{\rm th}$  Ed.]{2010A&A...518A..10V}, Col. 5 corresponds to  the apparent magnitude and Col. 6 is the absolute magnitude in the B-band. 
 \par}
\end{table*}

 In this paper,  we will be concerned with debated problems associated with the possibility of a real physical dichotomy between RL and RQ quasars \citep[e.g.,][]{ 1999AJ....118.1169X, 2003MNRAS.341..993C, 2008MNRAS.387..856Z, 2017MNRAS.466..921C, 2019NatAs...3..387P, 2019ARA&A..57..467B}: the effect of radio loudness on the dynamics of the low-ionization broad line emitting regions. 

The singly-ionized iron emission is systematically fainter in RL than in RQ sources, although it is still unclear if this is an effect intrinsic to a different emitting region structure, or associated with different host galaxy properties \citep[e.g.,][and references therein]{2021Univ....7..484M}. 
 
To this aim, we utilize a parameter space that  provides spectroscopic contextualization for all classes of broad line emitting AGN, the 4DE1 concept \citep[e.g.][]{2000ApJ...536L...5S}. The four dimensions of the   space are: 
\vspace{-\topsep}
\begin{itemize}
\item FWHM of \hb\ full broad profile (\hbFP) (see Sec. \ref{sec:FPA}) related to the velocity field of the low-ionization line emitting region;
\item Intensity ratio of the optical \feiiq\AA\ (which is the sum over 4434\AA\ -- 4684\AA) and \hbFP, \rfeopt\ = I(\feiiq\AA)/I(\hbFP).   \rfe\ is affected by physical conditions such as density, ionization level, and metal content \citep{2019ApJ...882...79P}.  

\item Centroid shift at half maximum of the \civ\AA, that measures the prominence of an outflowing/wind component 
\citep[e.g.,][]{2011AJ....141..167R},  and
\item Soft X-ray photon index ($\Gamma_\mathrm{soft}$) that depends on the accretion state  \citep[see e.g.][]{2000ApJ...536L...5S,2017MNRAS.468.3663J,2019ApJ...875..133P}.
\end{itemize}
\vspace{-\topsep}

 The two first parameters define the optical plane of the 4DE1 parameter space, the quasar main sequence (hereafter MS, \citealt{2001ApJ...558..553M,2002ApJ...566L..71S,2010MNRAS.409.1033M,2014Natur.513..210S,2020MNRAS.492.3580W}). 
The main advantage of the 4DE1 parameter space formulation is its weak dependence on source luminosity \citep[e.g.,][]{2008MNRAS.387..856Z}.  

This formalism also assisted the interpretation of several observational aspects that appear puzzling if, for example, sets of spectra are indiscriminately averaged together. Spectra can be averaged, but only in a well-defined context like the 4DE1 \citep{2012ASSL..386..549S}.  In addition,  it helps to establish a connection between  an  observational set of accretion  parameters such as black hole mass (\mbh) and Eddington ratio (\lE\,=\lledd)  \citep[see e.g.,][and references therein]{2017FrASS...4....1F,  2018FrASS...5....6M}.

Exploration of the 4DE1 parameter space gave rise to the concept of two populations of quasars that present important spectroscopic differences \citep{2007ApJ...666..757S}:

$\bullet$ Population A (hereafter Pop. A), whose sources generally have \hbFP\ FWHM \,<\,4000\,\kms, show Lorentzian profiles in the broad emission lines, tend to have  \rfeopt\,>\, 0.5,  significant blue shifts in high ionization lines (HILs, {e.g. \civ}) and soft X-ray excess.

$\bullet$ Population B (hereafter Pop. B), with a very wide range of \hbFP\ FWHM, with values higher than 4000 \kms,  show Gaussian profiles (a broad and an additional very broad redshifted component), \rfeopt\, <\, 0.5. In general, at low-$z$, they do not show  significant blue shifts in HILs and no soft X-ray excess.

The physical distinction between the two populations is related to the fact that Pop. A are fast-accreting objects with a relatively small \mbh, meanwhile, Pop. B are the ones with high \mbh\ and low Eddington ratios \citep{2009A&A...495...83M}. At low $z$($\lesssim 1$), most powerful RL sources belong to Pop. B \citep{2007ApJ...658..815S,2008MNRAS.387..856Z}. 

As suggested in \citet{2008MNRAS.387..856Z}, the value of studying the RL phenomenon within the 4DE1 context is at least twofold: (i) the approach compares RL and RQ quasars in a parameter space defined by measures with no obvious dependence on the radio properties \citep{2003ApJ...597L..17S}; 
and (ii)  it allows predictions of the probability of radio loudness for any AGN population given a  specific set of spectroscopic properties.  Open questions are  whether the geometry/ kinematics of RQ and RL quasars is the same or not. A related issue is also the ability to distinguish between high and low accretors from optical and UV spectroscopy, to assess any effect  the black hole spin has, and to infer the structure of the accretion disc \citep[e.g.,][]{2007ApJ...658..815S, 2009MNRAS.395..625L, 2009ApJ...699.1789T}.

This work presents new optical and near-UV spectra and a multicomponent fitting analysis of 11 jetted Pop. B (RL) quasars  at low--$z$\  ($0.3 \lesssim z \lesssim 1$). This paper extends the study to near-UV with the coverage of the \mgii\AA\ (hereafter \mgiionly) spectral range, providing still rare simultaneous observations of  \hb\ and \mgiionly.  Both \hb\ and \mgiionly\ provide diagnostics of the low-ionization line (LIL) part of the emitting region. In addition, \mgiionly\  is known to be less affected by shifts and asymmetries than \hb\ and is expected to provide a reliable virial black hole mass estimation \citep[e.g.,][]{2012MNRAS.427.3081T}.   The sources we selected are among the most extreme radio emitters,  those with a radio to optical flux density ratio $\gtrsim 10^3$. For them, we want to test if there might be a strong effect of the relativistic jets on the low-ionization UV and optical lines emission.
Section \ref{sec:sd} describes the new AGN sample, the main comparison samples, the new observations from the 3.5\,m telescope at the Calar Alto Observatory, and the archival radio data.  

Spectral non-linear multicomponent fittings and full broad profile (FP) analysis are described in Section \ref{sec:SA}.  The results of spectral fittings and observed trends are described in Section \ref{sec:results}. 

We discuss the results in Section \ref{sec:discu} and  conclude in Section \ref{sec:con}. Throughout this paper we adopt a flat $\Lambda$CDM cosmology with $\Omega_{\Lambda}$= 0.7, $\Omega_{0}$= 0.3, and
$H_{0} = 70\,$km\,s$^{-1}$Mpc$^{-1}$. 
\vspace{-\topsep}
\section{Sample and data}
\label{sec:sd}
\subsection{Sample selection}
\label{sec:sample}

 Table \ref{tab:bspo}  presents the main properties of our sample, reporting the quasar identification (Col. 1), the equatorial coordinates at J2000 (Cols. 2 and 3), the median redshift  estimated for each quasar as explained in Sect. \ref{sec:red} (Col. 4), the apparent and absolute magnitudes taken from  \citet{2010A&A...518A..10V} (Cols. 5 and 6  respectively) from where we also took the initial redshift for the sample selection, and a summary of the observations (see Sect. \ref{sec:reduction}). 
 We selected 11 RL quasars based on their strong radio emission that satisfies a criterion of extreme radio loudness, with values of the Kellermann parameter  \citep[radio to optical flux density ratio]{1989AJ.....98.1195K}  $R_\mathrm{K}$ exceeding $10^3$ (see Sec. \ref{sec:arrd} and Table \ref{tab:radio}, Col. 6 and 7).  Such extremely powerful jetted sources permit a detailed study of the effects of powerful radio ejections on the low-ionization broad line region (BLR). The 11 sources are optically bright, with log\,$L_{Bol}$ in the range of 44.9 to 46.7 [\ergss], and were selected in a redshift range ($0.35\lesssim z\lesssim\,1$; \citealt{2010A&A...518A..10V})  that makes possible the concomitant coverage of \hb\ and \mgiionly. Such observations provide consistent information, as observed spectra at different times may suffer a significant continuum variability. 
Our high signal-to-noise spectra with spectral range 3500--10000\AA\ provide an extended view of the quasar continuum in each spectral region and allows us to make accurate measurements of emission line parameters such as the intensity of optical and UV \feii\ features. 

\begin{figure}
\centering
 \includegraphics [scale=0.35]
{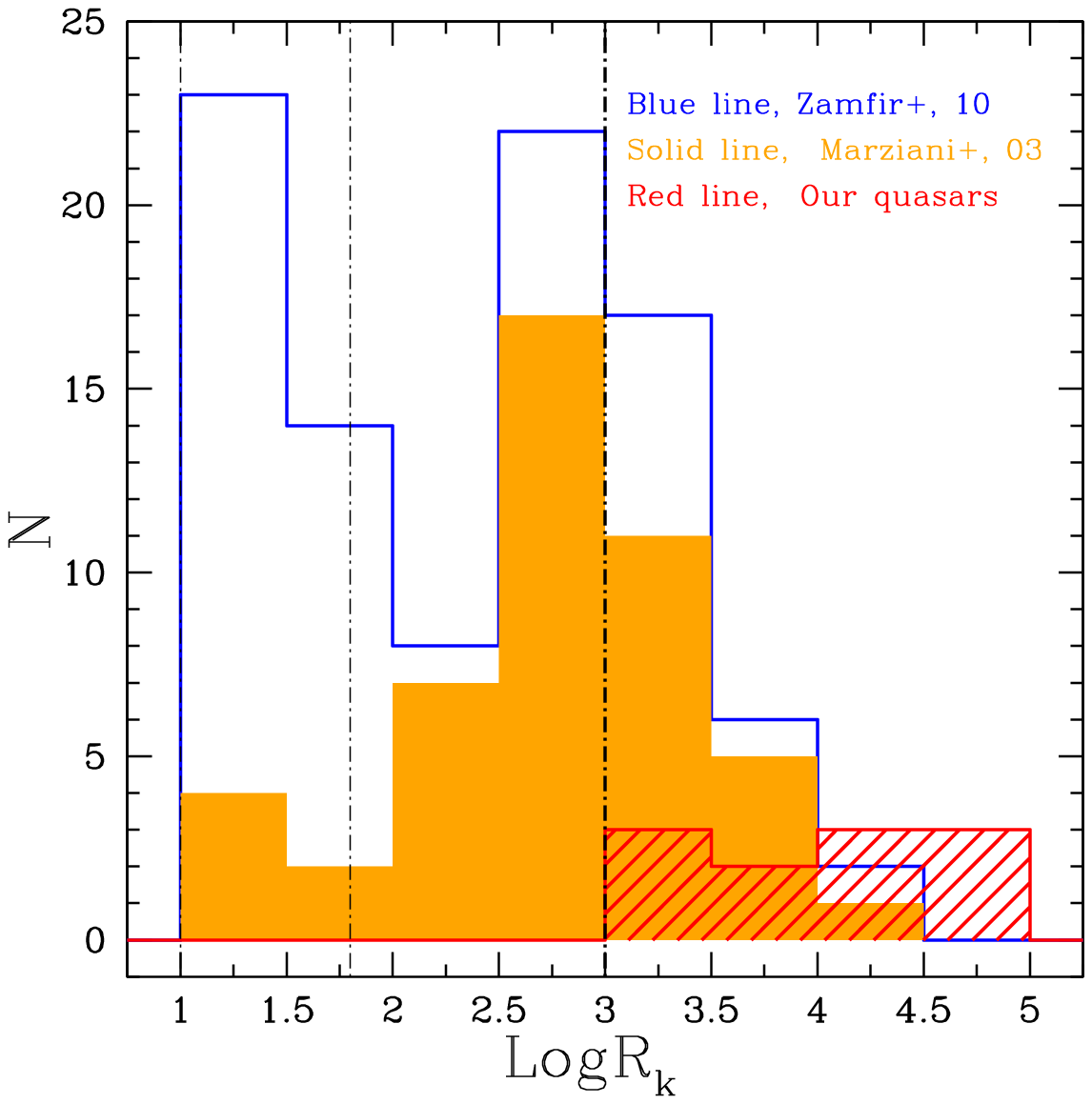} 
\caption{ Distribution of radio loudness parameter for Pop. B RLs from our sample (red striped area) 
and the main comparison samples used in this work,  \citet{2003ApJS..145..199M} (solid orange) and   
\citet{2010MNRAS.403.1759Z} (blue line). 
The vertical dot-dashed lines at 1 and 1.8 mark the nominal RQ-radio intermediate and radio intermediate-RL boundaries \citep{2019A&A...630A.110G}.  The vertical dot-dashed line at 3 marks the boundary for extreme radio loudness values.} 
\label{fig:hist}
\end{figure}

There are not so many RL AGN with radio loudness parameter, $R_\mathrm{K} \gtrsim 10^3$ (see Sect. \ref{sec:arrd}), and the uniqueness of this study rests on the selection of extremely powerful jetted sources. In our sample, all sources are not only with log$R_\mathrm{K}\gtrsim 3$ but 6 of them are with log$R_\mathrm{K}\gtrsim 4$ (see the histogram 
in Fig. \ref{fig:hist} in which we showed the distribution of the radio loudness parameter that was estimated by using the 1.4GHz radio flux density and the g-band optical flux density in all the cases).
Only  a handful of sources in the main comparison samples (see below) have $R_\mathrm{K} \gtrsim 10^4$ and a few tens have $\gtrsim 10^3$.  

In the optical plane of 4DE1 parameter space, our extreme $\log R_\mathrm{K} \gtrsim 3$\ (hereafter eRk) represent non-negligible addition to RL sub-samples of optically selected quasars that are represented in Population B spectral types 
\citep{2008MNRAS.387..856Z}. The eRk sample enhances the significance of statistical comparisons between the highest radio loudness sources and RQ  as well as other RL sub-samples.

\subsubsection{Main comparison samples}
\label{sec:compsam}
 As main comparison samples in the optical, we used the low redshift samples of \citet{2003ApJS..145..199M}  and \citet{2010MNRAS.403.1759Z}.  The redshift and luminosity ranges, the number of sources (both Pop. A and B) considered in each sample,  the number of Pop. B RL sources in different radio loudness ranges, as well as the emission lines used and profile measures available in each one are summarized in Table \ref{tab:comp}. The sample of \citet{2010MNRAS.403.1759Z} consists of 469 quasars of which 209 are classified as Pop. B. 

The sample of \citet{2003ApJS..145..199M} originally has 215 sources. Before using it as a main comparison sample, we excluded sources in common with the other comparison sample (38) and also nine sources that may create doubt in the interpretation of the spectra,  due mainly to a noisy spectrum, the presence of a strong stellar continuum or strong star formation, after a careful examination. 
The 168 remaining objects, of which 94 belong to Pop. B,    contain information for the line center (c($\frac{1}{2}$))
and the line base (c($\frac{1}{4}$)) of \hb. These measurements provide an empirical description of the full line profile, i.e., velocity shifts at different fractional intensities of the FP, e.g., at the line base, at the line center, and at peak fractional intensities (c($\frac{9}{10}$)).

We also use in particular for the \mgiionly\ region the composite spectra from \citet{2013A&A...555A..89M} built for different spectral types (STs) along the MS, i.e.  A1 to A4 for Pop. A quasars and in Pop. B for the STs B1, B2, B1$^{+}$ and B1$^{++}$ as well as within the B1 bin for those classified by the authors as RL. The composite spectra from that sample have a measurement for both UV and optical regions.

Other secondary samples used in specific points such as those of \citet{2009ApJ...707.1334W} and the {\sc QSFit} catalogue \citep{2017MNRAS.472.4051C} are described in the corresponding sections (Sec. \ref{sec:uvp} and \ref{sec:comp}).

\begin{table}
 \setlength{\tabcolsep}{1pt}
\caption{Summary of main comparison samples and  this work (eRk) sample}
\label{tab:comp}       
\scalebox{0.92}{\raggedright  
\begin{tabular}{l cccc}
\hline\hline 
{Information} & {Marziani+03} & {Zamfir+10} & {Marziani+13}$^{a}$ & { This work}$^{b}$ \\
\hline
{Redshift range} & {z < 0.8} & {z < 0.7} & {0.4 < z < 0.75} & {0.35 < z < 1} \\
{No. of sources} & {168} & {469} &{8}$^{a}$ & {11} \\
{Parameters} & {c($\frac{1}{2}$), c($\frac{1}{4}$), \rk} & {c($\frac{1}{4}$), \rk, AI} & {c($\frac{1}{2}$), c($\frac{1}{4}$)} & {c($\frac{1}{2}$), c($\frac{1}{4}$), \rk} \\
  & {FWHM,AI} & {FWHM,\rfeopt} & {FWHM,AI} & {FWHM, AI} \\
  & {} & {} & & {\rfeopt,\rfeUV}$^{b}$\\
 Lines used & {\hb} & {\hb} & {\hb\ and \mgiionly{}} & \hb{} and \mgiionly\\
 Log\,$L_{bol}$[ergs s$^{-1}$]& 43.7--47.8& 43.0--47.0&  45.7--46.9& 45.15--46.57 \\

 No. Pop. B RLs:  \\
\ \  log\rk>1.8 & 42 & 61 & --& 11\\
 
\ \ log\rk>3 & 17 & 25 & -- & 11 \\
\ \  log\rk>4 & {1} & 2 & -- & 6 \\
\hline                               
\end{tabular}
}

{\raggedright\textbf{Note}: $^{a}$This sample provides 8 composite spectra generated by using spectral binning of physically similar quasars of 680 SDSS spectra. $^{b}$All the profile parameters measured for our eRk quasars, both in the FP and the multi-component fittings, are detailed in the following sections. c($\frac{1}{2}$) and  c($\frac{1}{4}$) represent the centroid velocity shift at $\frac{1}{2}$ and $\frac{1}{4}$ fractional intensity of the line respectively. AI denotes the asymmetry index. \par}

\end{table}

\subsection{Optical and UV observations}

\label{sec:reduction}
Long slit optical observations were obtained with the Cassegrain TWIN spectrograph of the 3.5m telescope at the Calar Alto Observatory (CAHA, Almería, Spain)\footnote{\url{http://www.caha.es/}}. The TWIN spectrograph is optimized to cover a wavelength range from about 3400\AA\ to 10000 \AA, dividing the light from the slit in two channels (blue and red) by a dichroic mirror. We selected the beam splitter at 5500\AA\ and gratings T13 and T11, for the blue and the red channels, respectively, in order to obtain simultaneously the spectra corresponding to the \mgiionly\ and  \hb\ regions at the redshift of the objects of our sample. 
T13 grating provides a spectral coverage from 3500\AA\
to 5500\AA\ with a reciprocal dispersion of 2.14\AA/pixel, meanwhile, T11 grating covers the range from 5400\AA\ to 10000\AA\  with a spectral dispersion of 2.41\AA/pixel.

Table \ref{tab:bspo} with the main properties of the sample also summarizes the observations where is listed the date of observation (Col. 7), the total exposure time in seconds (Col. 8), the air mass (Col. 9), and  the signal-to-noise ratio (S/N) measured in the continuum (Col. 10). The observations were obtained with a slit width of 1.2 arcsecs and oriented at the paralactic angle in order to minimize the effects of atmospheric differential refraction in the spectra. The total exposure time required for each object was split into three exposures to reduce the number of cosmic rays and eliminate them by combining different exposures.

Data reduction was carried out in a standard way using {\tt IRAF}\footnote{{\tt IRAF} is the Image Reduction and Analysis Facility, a general purpose software system for the reduction and analysis of astronomical data, \url{iraf.net}} \citep{1986SPIE..627..733T}. Spectra were bias and overscan corrected and flat-fielded with a normalized flat-field obtained from a median combination of the flats obtained for each spectral region (blue and red). Wavelength calibration was obtained using the standard lamps (He-Ar and Fe-Ne) exposures and standard {\tt IRAF} tasks. The {\tt apall} task in {\tt IRAF} was used for object extraction and background subtraction. Instrumental response and absolute flux calibration were obtained each night through observations done with the same instrumental setup of two spectrophotometric standard stars, G191B2B and BD+28d4211. They were also used to remove telluric bands that are mainly observed in the red channel. The final calibrated rest-frame spectra of all the sources, once the blue (\mgiionly\ region) and red (\hb\ region) spectra were combined for each object, are shown in Fig. \ref{fig:hist}.

\begin{figure*}
\centering
\includegraphics [height=0.37\textheight]{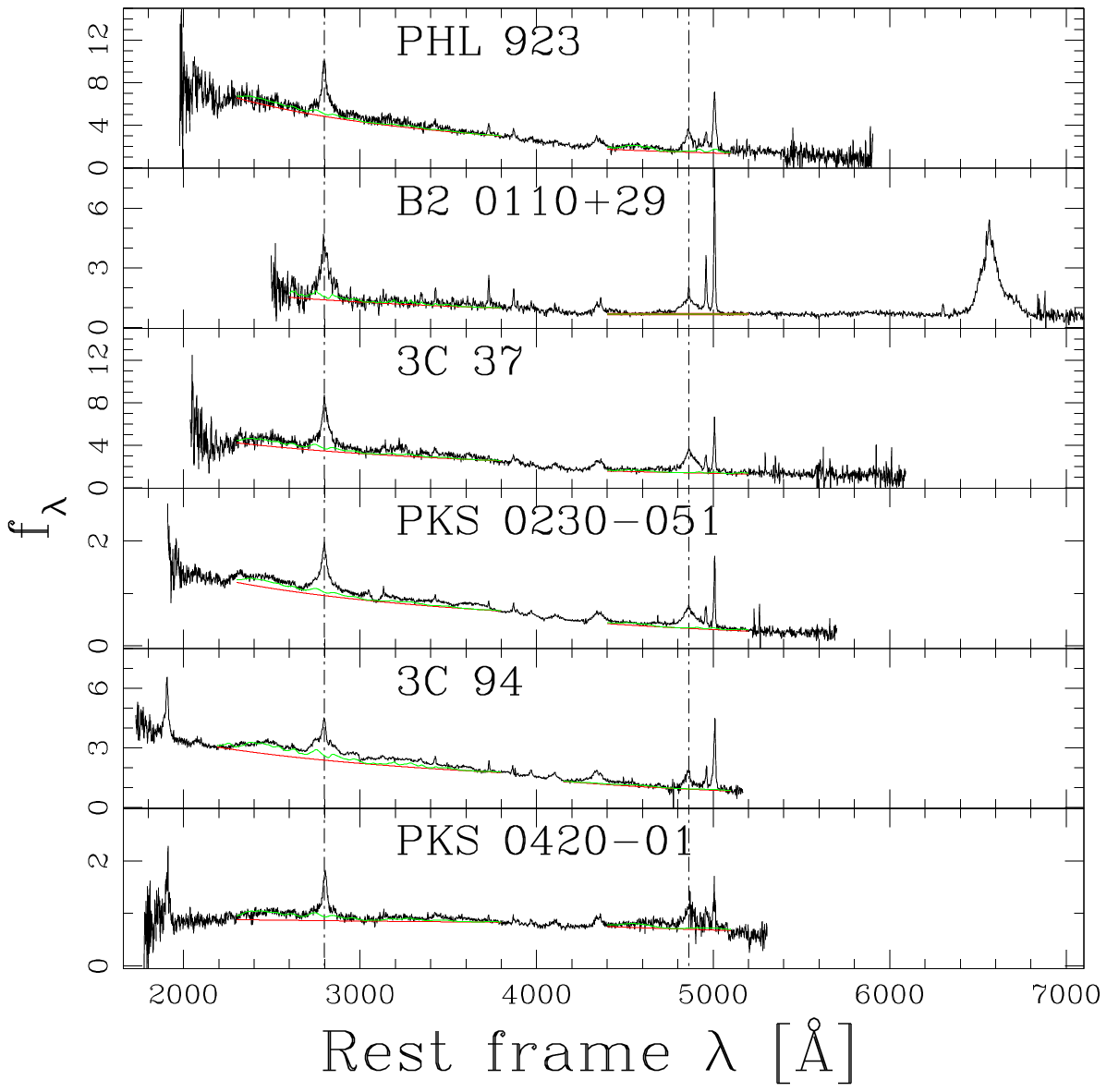}
\includegraphics [height=0.37\textheight]{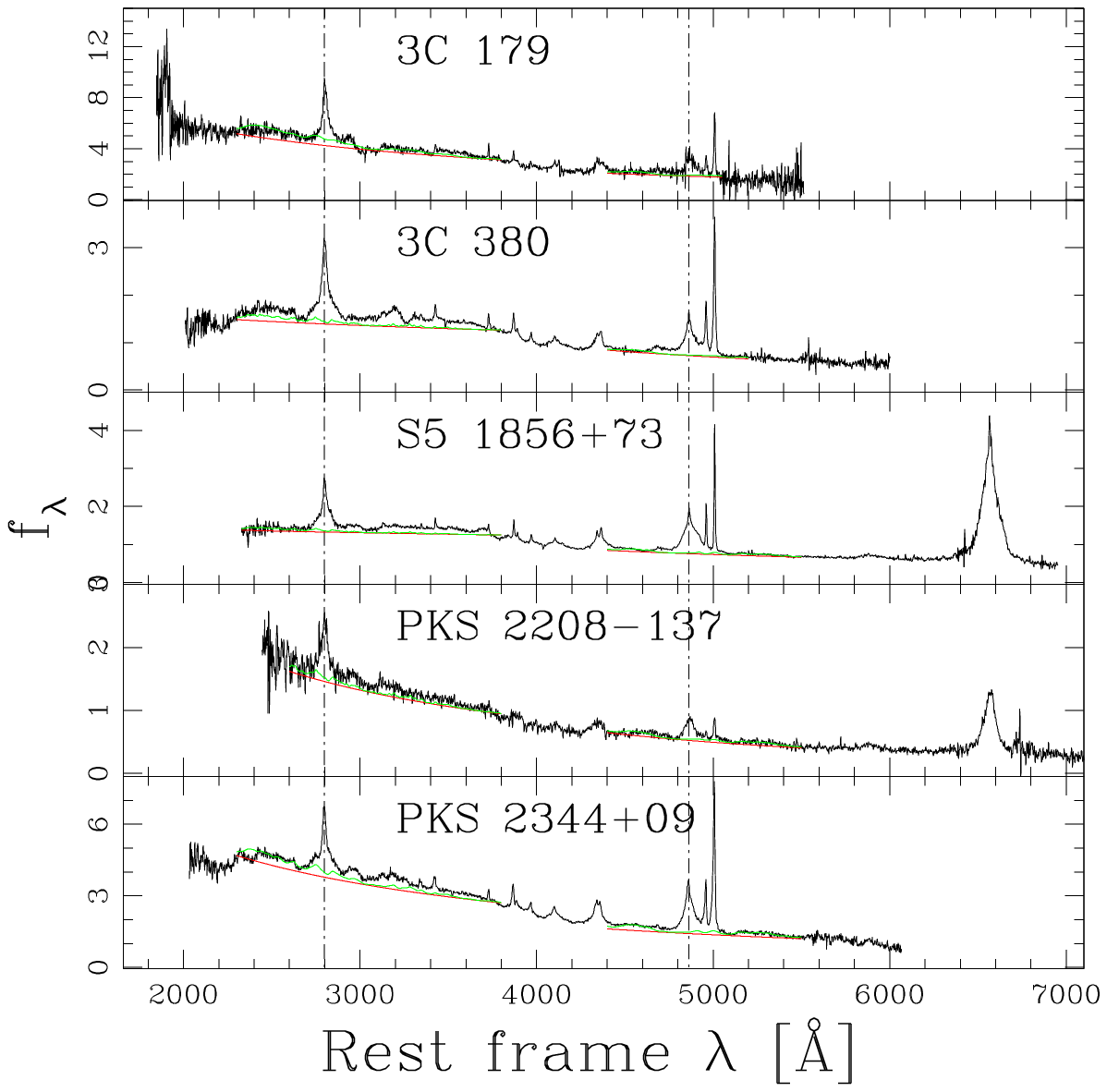}
\vspace{-0.5cm}
\caption{ Rest-frame spectra of our 11 type 1 AGN, with  \mgiionly\ and \hb\ regions after joining the two observed spectra. Abscissas are rest-frame vacuum wavelengths in \AA\ and  ordinates are specific flux in units of $10^{-15}$ergs\  s$^{-1}$ cm$^{-2}$ \AA$^{-1}$, except for PHL 923, B2 0110+29, 3C 37,  and 3C 179 that are in units of $10^{-16}$ergs\,  s$^{-1}$ cm$^{-2}$ \AA$^{-1}$. In each panel, the green line traces \feii\ emission and the red one represents the power-law continuum. Dot dashed vertical lines trace the rest-frame wavelength of \hb\ and \mgiionly.}

\label{fig:spectra}
\end{figure*} 
\subsubsection{\textbf{Redshift determination}}
\label{sec:red}
 We determine the redshift of the quasars, following \citet{2020A&A...635A.151B}, by measuring the observed vacuum wavelengths of individual narrow emission lines, in particular [OII]$\lambda$3727\AA, H$\delta$, H$\gamma$, H$\beta$, and [OIII]$\lambda\lambda$4959,5007\AA\AA\, that are available in our spectra. For that, we used the {\tt IRAF} task {\tt splot}, and by taking mainly as reference vacuum wavelengths the ones from \citet{2001AJ....122..549V}. The median value of the redshift obtained from the different narrow emission lines was adopted as the empirical redshift of each object (reported in Table \ref{tab:bspo}, Col. 4), and then used for getting the rest frame spectrum of each quasar. Typical root mean square (rms) in the estimated redshifts are between 0.0003 and 0.001. 

\subsection{Archival radio data}
\label{sec:arrd}
 Table \ref{tab:radio} presents the main radio properties of the sample. The radio fluxes were obtained at 1.4\,GHz (20\,cm) from the National Radio Astronomy Observatory (NRAO) Very Large Array (VLA) Sky Survey (NVSS)\footnote{NVSS-\url{https://www.cv.nrao.edu/nvss/NVSSPoint.shtml}}  \citep{1998AJ....115.1693C} and at 5\,GHz (6\,cm) from \citet{2010A&A...518A..10V}\footnote{\url{https://vizier.u-strasbg.fr/viz-bin/VizieR-3?-source=VII/258/vv10}} catalogue. We also list in Col. 3 the integrated flux at 1.4\,GHz compiled from the Faint Images of the Radio Sky at Twenty centimeters (FIRST)\footnote{FIRST-\url{http://sundog.stsci.edu/cgi-bin/searchfirst}} survey \citep{1995ApJ...450..559B} for the only five sources of our sample observed by FIRST.

As we mentioned in the introduction, quasars have been classified as RL and RQ by using radio and optical measurements as well. 
However a clear consensus on a boundary between RL and RQ quasars has been difficult to achieve and several radio-loudness criteria have been used in the literature \citep[see e.g.,][and references therein]{2005AJ....130..586W,2014arXiv1408.1090H,2016ApJ...831..168K,2022MNRAS.516.2824C}. One of the most commonly used criteria is the radio-loudness parameter defined by \cite{1989AJ.....98.1195K}, \rks, as the ratio of the radio flux density at 5\,GHz to the B-band optical flux density. In this work, we have used a modified version of the  radio-loudness parameter, currently commonly used, based on the ratio of the rest-frame radio flux density at 1.4 GHz to the rest-frame optical flux density in the g-band, \rk{} \citep[e.g.,][]{2008MNRAS.387..856Z, 2015MNRAS.452.3776G}. 
For completeness,  we also estimated the Kellermann parameter in rest frame units (\rks). 
 
To estimate the radioloudness parameters, the k-corrected radio flux density at 1.4\,GHz and 5\,GHz 
was found from the observed radio flux densities (reported in Table \ref{tab:radio}, Cols. 2 and 4, respectively) by using the spectral index $\alpha$ obtained from \citet{2010A&A...511A..53V} (reported in Table \ref{tab:radio}, Col. 5) and the relation from  \citet{2019A&A...630A.110G}, 
\vspace{-0.4cm}
\begin{equation} f_{\nu_{o,e}}= f_{\nu,o}[(1+z)^{\alpha-1}] \end{equation}
where the subscript \textit{"o"} refers to the observer$'$s frame, and the subscript \textit{"e"} refers to quantities in the quasar rest-frame. In all the calculations, we used the convention for the spectral index, S $\propto \nu^{-\alpha}$.   
We have FIRST radio data only for five of our sources (see Table \ref{tab:radio}, Col. 3), and hence we used the NVSS 1.4\,GHz fluxes. 

The rest-frame optical fluxes around the effective wavelengths $\lambda\,$4770\AA\, (for g band) and $\lambda$\,4450\AA\, (for B band) were obtained from our rest-frame spectra. The resulting radio loudness parameters are reported in Cols. 6 and 7 of Table \ref{tab:radio}, respectively.  Throughout this paper,  we used the modified radio loudness parameter (\rk). Also, radio power ($P_{\nu}$) has been calculated by using \citet[][their equation 1]{2008ApJ...687..859S} and is reported in Col. 8. Previous works showed a well-defined lower limit for \rk{} in which lobe dominated (LD) RL sources have a value of $\log$\rk\,>\,1.8  and a radio power log\,$P_{\nu}\,>\,31.6$ [ergs\,$\mathrm{s^{-1}\,Hz^{-1}}$] \citep[e.g.,][]{2003ApJ...597L..17S, 2008MNRAS.387..856Z}. \citet{2019A&A...630A.110G} also classified sources on the basis of \rk, as radio detected (RD, sources with log\rk\,<\,1.0), radio intermediate (RI, 1.0$\leq$log\rk\,<\,1.8) and radio-loud (RL, log\rk$\geq$\,1.8). According to these classifications, all our sources are strong radio emitters and can be labeled as strong RLs. As can be seen in Table \ref{tab:radio}, the resulting radio-power for all our quasars is also extremely high with log$P_{\nu}$ between 33.5 and 35.4\,[${\rm ergs\,s^{-1}\,Hz^{-1}}$] and with log\,\rk\,>\,3, by far meeting the requirement to classify them as RL.

\subsubsection{Radio morphology classification}
\label{sec:rmorp}
 A morphological radio classification of our sample at an angular scale of arcsecs was obtained with the NVSS images and supplemented by FIRST images for the 5 objects also observed in that survey. This is the scale related to a kpc sized scale associated with extended features such as the radio lobes and normally used in the majority of the comparison samples. We classified our sources according to the following scheme: as  core dominated (C) if the core appears unresolved; as core plus lobe/s (CL), if we have a visible bright core and at least one lobe, and as lobe-dominated (LD), if we see the lobes without core in the spatial scales of VLA or if the radio emission is dominated by lobes. This classification for our quasars is provided in Col. 9 of Table. \ref{tab:radio}. In this scale, we have 2, 6, and 3  quasars with C, CL, and LD radio morphology, respectively. Also, our quasars present inner jets at parsec and sub-parsec scales (milli-arcsecs scale, VLBA/I), as expected for strong radio emitters. \cite{2022ApJS..260....4P} provides information on the estimated distance between the VLBI core and a structure assigned to a jet/s for 8/11 of our sources.

 In this paper, we used the kpc scale, which is available for the whole sample and is the most widely used in literature and in the comparison samples. In Appendix A of the online supplemental material, we included the FIRST images and their overlay on the optical Pan-STARRS image for PHL 923, 3C 37, PKS 0230-051, 3C 94, and PKS 2344+09. In addition, for two sources (B2 0110+29 and  PKS 2208-137) that do not have the FIRST cutout image, we present NVSS radio maps, that we used for the morphological classification.

\begin{table}
\caption{Radio properties of the sample}         
\centering 
\setlength{\tabcolsep}{1.2pt}
\scalebox{0.9}{  
\begin{tabular}{l r c c c c c c c}   
\hline\hline   
 &\multicolumn{1}{c}{NVSS} & \multicolumn{1}{c}{FIRST} & \multicolumn{1}{c}{Flux at} & & & & & \\

 Object & \multicolumn{1}{c}{Int.flux} & \multicolumn{1}{c}{ Int.flux} & \multicolumn{1}{c}{ \ 5\,GHz$^{(a)}$} & $ \alpha^{(b)}$ & $ log R_{K}^{(c)}$ & $ log R_{KS}^{(d)}$ & \multicolumn{1}{c}{$ log(P_{\nu}$)} &  Morph.\\
 & \multicolumn{1}{c}{(mJy)} &\multicolumn{1}{c}{( mJy)} & \multicolumn{1}{c}{(mJy)} & & & & & \\
 (1)&\multicolumn{1}{c}{(2)}&( 3)&(4)& (5)& (6)& (7)& (8)&(9) \\
 \hline 
PHL 923    & 2508$\pm$75 & 2494 & 1410 & 0.42 & 4.63  & 4.37 & 34.63& CL \\
B2 0110+29 & 810$\pm$ 9 & -- & 311 & 0.54& 4.34 & 3.99 &33.50 & LD \\
3C 37 & 1522$\pm$46 &  1646 & 621 & 0.92 & 4.49 & 4.15 & 34.45 & CL \\
PKS 0230-051 & 210$\pm$8 & 139& 160 & 0.52 & 3.13 & 3.03 & 33.59 & CL \\
3C 94  & 3061$\pm$112 & 2798 & 790 & 1.04& 4.23 & 3.57 & 35.18 &  LD \\
PKS 0420-01 & 2726$\pm$82 & -- & 1580 &0.23 &3.94 & 3.71& 34.71 & LD \\
3C 179 & 2123$\pm$75 & -- & 1000 & 0.72& 4.63 & 4.29& 34.79 & CL \\
3C 380 & 13753$\pm$413 & -- & 5000 & 0.75& 4.73 & 4.34 &35.41 & CL\\
S5 1856+73 & 490$\pm$16 & --& 610 & 0.12& 3.03 & 3.22 & 33.42 &  C \\
PKS 2208-137 & 1330$\pm$40 & -- & 620 & 0.66& 3.71 &3.45 & 33.81 & CL \\
PKS 2344+09  & 1804$\pm$54 & 1734 & 1690 & 0.19 &3.44 &3.49 &34.37 & C \\
\hline                                   
\end{tabular}
}\\
{\small{\raggedright \textbf{Note}:  $^{(a)}$From \citep{2010A&A...518A..10V}. $^{(b)}$from \citet{2010A&A...511A..53V}. $^{(c)}$With NVSS k-corrected radio flux density at 1.4\,GHz and optical rest frame flux density in g band. $^{(d)}$With k-corrected radio flux density at 5GHz and rest frame optical flux density in B band. Units in Col 8 are ergs\,s$^{-1}$Hz$^{-1}$\par}}

\label{tab:radio}
\end{table}

\section{Spectral analysis}
\label{sec:SA}

Narrow and broad emission lines are prominent features in optical and UV spectra of type I AGN \citep{1988LNP...307....1O}.  In this work, the analysis of each spectrum was performed independently in the two observed spectral regions, corresponding to the blue and red  arms of the spectrograph, respectively.  The red arm includes both low ionization features such as \hb, \feiiq\AA, and HILs like \oiiiopt\AA\AA\  (hereafter \oiiionly) as main representatives. The low ionization \mgiionly\  doublet is located in the redshifted UV region along with prominent \feiiuv\ emission covered by the blue arm. In each region, we carried out a  multicomponent non-linear fitting and an analysis of the FP of the emission lines,  as it is described in the next subsections.

\subsection{Spectral fitting}
\subsubsection{Optical region}
\label{subsec:OR}

In order to get relevant physical parameters for the emission lines observed in our spectra, an empirical model that matches the observed spectrum was applied. Our spectral non-linear multicomponent fitting was carried out by using {\tt IRAF} {\tt specfit} routine, which can fit data spanning a large range in wavelength by using a non-linear $\chi^2$ minimization technique \citep{1994ASPC...61..437K}, and used in previous studies \citep[e.g.,][]{2010MNRAS.409.1033M, 2015ApJS..217....3M, 2017A&A...608A.122S}. This allowed us to simultaneously fit a continuum, a scalable \feii\, emission, and individual emission lines yielding FWHM, peak wavelengths, and intensities of all emission line components. To get the best fit, we first used a simplex algorithm with an iteration up to 200 and afterward, a  Levenberg-Marquardt algorithm with a lower number of iterations to ensure convergence to the global minimum $\chi^{2}$ \citep{levenberg1944method}.

For non-linear multicomponent fitting, we included the following components and conditions in the fittings: 

\begin{itemize}
   \item A power law local continuum underlying the \hb\ region to take into consideration the thermal accretion disk emission in optical \citep{1978Natur.272..706S}. We defined the continuum by using three to four regions that are free of emission lines. 
   \item A  scalable \feii\ template for modeling the \feiiopt\  emission lines that are blended with \hb, as \feii\ has a large number of transitions producing overlapping and blended lines \citep{2009A&A...495...83M}.

\item We assumed  that the Balmer line \hb\ has three components with Gaussian profiles, as appropriate for Pop. B sources \citep[e.g.,][]{2010MNRAS.409.1033M, 2010A&A...509A...6B}:
\begin{itemize}
\vspace{-0.2cm}
\item[a)] A narrow component (NC) that represents the Narrow Line Region (NLR) with low-density and more slowly moving clouds with narrow line width, which infers this region is far from the central supermassive black hole (SMBH);
\item[b)] A broad component (BC) associated with the BLR with almost unshifted component and corresponds to dense and fast-moving clouds which indicates the proximity to the central SMBH;
\item[c)] A very broad component (VBC) associated with a Very Broad Line Region (VBLR) with high ionization and large column density that corresponds to broader and redshifted components.
\end{itemize}     
\item The  \oiiionly\, emission lines, represented by two Gaussian NC set at rest-frame plus two blue-shifted semi-broad components (SBC).
\item A \heii\AA\ line when there were hints of its presence with a BC and/or VBC component.
 \end{itemize}
 
{The fitting for this region was done in a different wavelength range (see Table \ref{tab:meashb} Col. 2) for each object to account for the complex nature observed in some of the spectra, like PHL 923 and 3C 94.} In the fitting procedure, the number of free parameters was reduced by assuming constraints related to emission lines coming from the same region. All the narrow lines were assumed to have roughly the same width and shift. The two  \oiiionly\, lines were assumed to have a flux ratio  I(\oiiionly$\lambda$4959\AA)/I(\oiiionly$\lambda$5007\AA) of 1:3 \citep{2007MNRAS.374.1181D}. In all sources except PHL 923, 3C 94, and PKS 2344+09, the \heii\AA\ line contributes to the blue wing of \hb, and we constrained the shift and FWHM of the \heii\AA\ VBC to be the same as \hb\ VBC \citep{2007ApJ...669..126S}. 

 The power law that defines the continuum level and the \feii\ contribution obtained from the {\tt specfit} analysis are plotted in Fig. \ref{fig:spectra} as red and green lines respectively. 
The plot of the other components from the {\tt specfit} fitting for \hb\ and  \oiiionly\, is presented in the right plots of Fig. \ref{fig:f3}, where the black line represents the rest-frame spectrum and the dashed magenta line shows the model fit. Residuals from the fittings are shown in the bottom panels. In 3C 94, PKS 0420-01, and 3C 179, with the highest redshift in the sample, the \hb\, region is close to the edge of the observed spectrum and thus affected by more noise. This resulted in larger uncertainties in the determination of parameters related to the \hb\, line profile. 

\subsubsection{UV region}
\label{sec:uvr} 

The doublet \mgiionly\ is located in a complex spectral region of the UV known as the small blue bump \citep[e.g.,][]{2012A&AT...27..557A, 2019MNRAS.484.3180P,2022arXiv220811437G}. To determine the parameters of the broad \mgiionly\ line profiles,  we carried out, as in optical, a multicomponent fitting of the region of interest including:

\begin{itemize}
    \item A power-law continuum underlying \mgiionly, in order to approximate the thermal accretion disk emission in UV  region \citep[e.g.,][]{1983ApJ...265...92M};

   \item  An \feii\  scalable template for accounting the \feiiuv\,  (which is the integrated flux over 2200\AA\ -- 3090\AA)    emission lines blended with \mgiionly.  This template is based based on CLOUDY simulations assuming ionization parameter $\log U = -2.25$, $\log n_\mathrm{H} = 12.25$, solar chemical abundances and standard \citep[i.e.,][]{mathewsferland87} AGN continuum. For practical purposes it is analogous to the e.g.,
   \citet{2008ApJ...675...83B} ``best" template;
   
   \item A UV Balmer continuum, 
   
   found to be important at $\lambda$\,<\,3646\AA\, \citep{2014AdSpR..54.1347K, 2017FrASS...4....7K}.

\item As \mgiionly\ is a doublet, we fitted the blended line component by using multiple Gaussians:

    \item[--] Two NCs, accounting for the doublet of \mgiionly\ at $\lambda$2796.35\AA\  and $\lambda$2803.53\AA, with an assumed ratio of 1.5:1.  
    \item[--] Two BCs, where the intensity ratio between the blue and red broad component of the 
    doublet is taken to be 1.25, which is representative of the physical conditions observed in the BLR according to CLOUDY simulations \citep[see details in][]{2013A&A...555A..89M}.
    \item[--]  One VBC, instead of two due to the small doublet separation of $\sim$\,8\AA\,   which is much less than the FWHM of the VBC components.  
    This restrains us from separating two VBCs with the spectral resolution we have.
\end{itemize}

The fitting was done in a wide wavelength range, from 
2600\AA\ to 3800\AA, in order to include the narrow LIL [OII]$\lambda$3727\AA\, and to obtain measurements for \feii\  emission in UV, which is not well known for jetted sources \citep{2018FrASS...5....6M}. We also included a Gaussian profile fitting for  O{\sc{iii}}$\lambda$3133\AA\ and the two HILs of [NeV] at $\lambda\lambda$3346,3426\AA\AA\ when the lines are clearly present. For the low redshift quasars, with z < 0.4 (B2 0110+29 and PKS 2208-137), the \mgiionly\ doublet measures are affected by a higher uncertainty because it is located at the extreme blue edge of the observed spectrum, which could imply usually more noise and a worse determination of the blue continuum level.

\begin{figure*}
\centering
\includegraphics [width=3in,height=1.55in ]{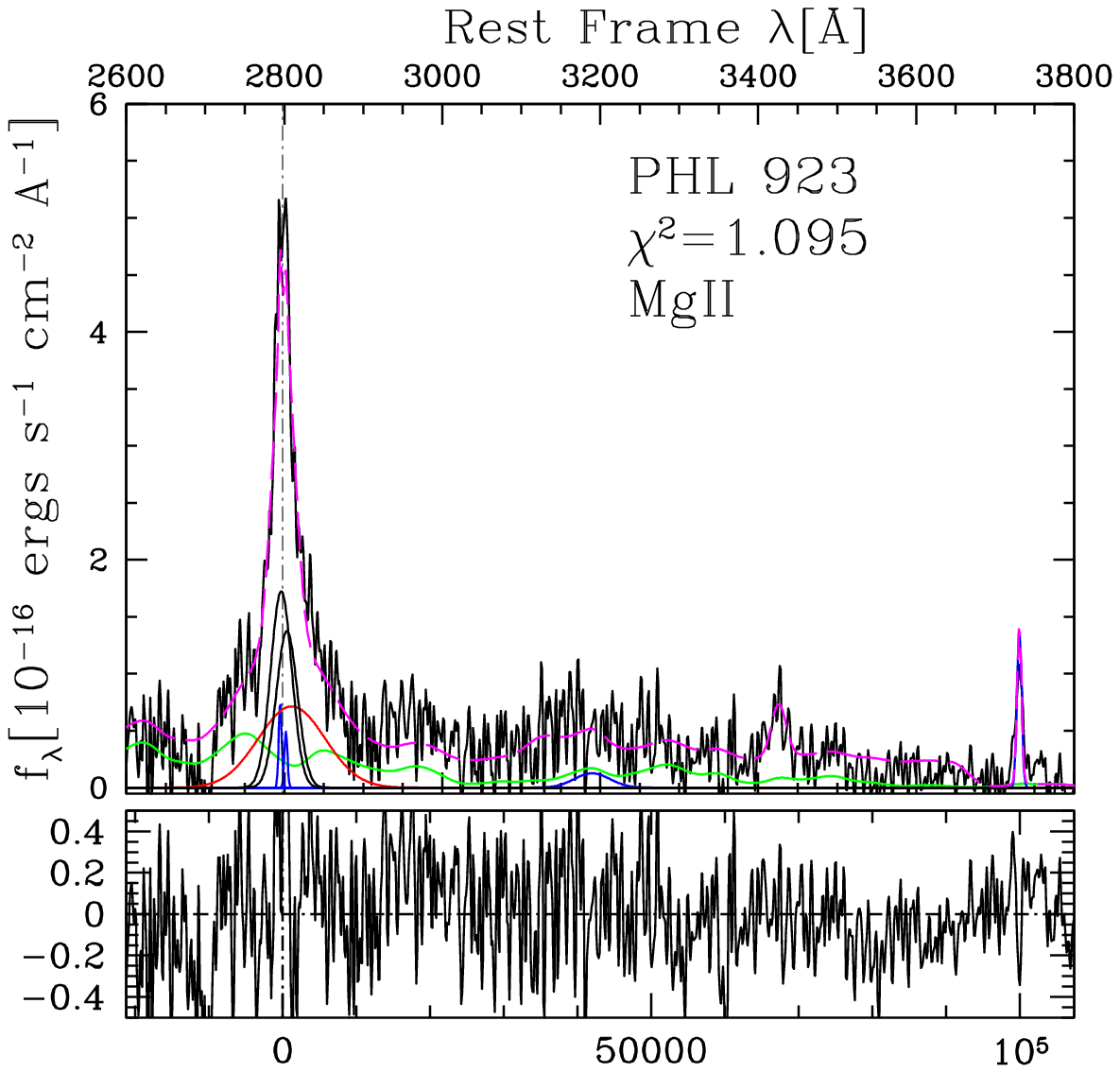}
\includegraphics [width=2in,height=1.55in ]{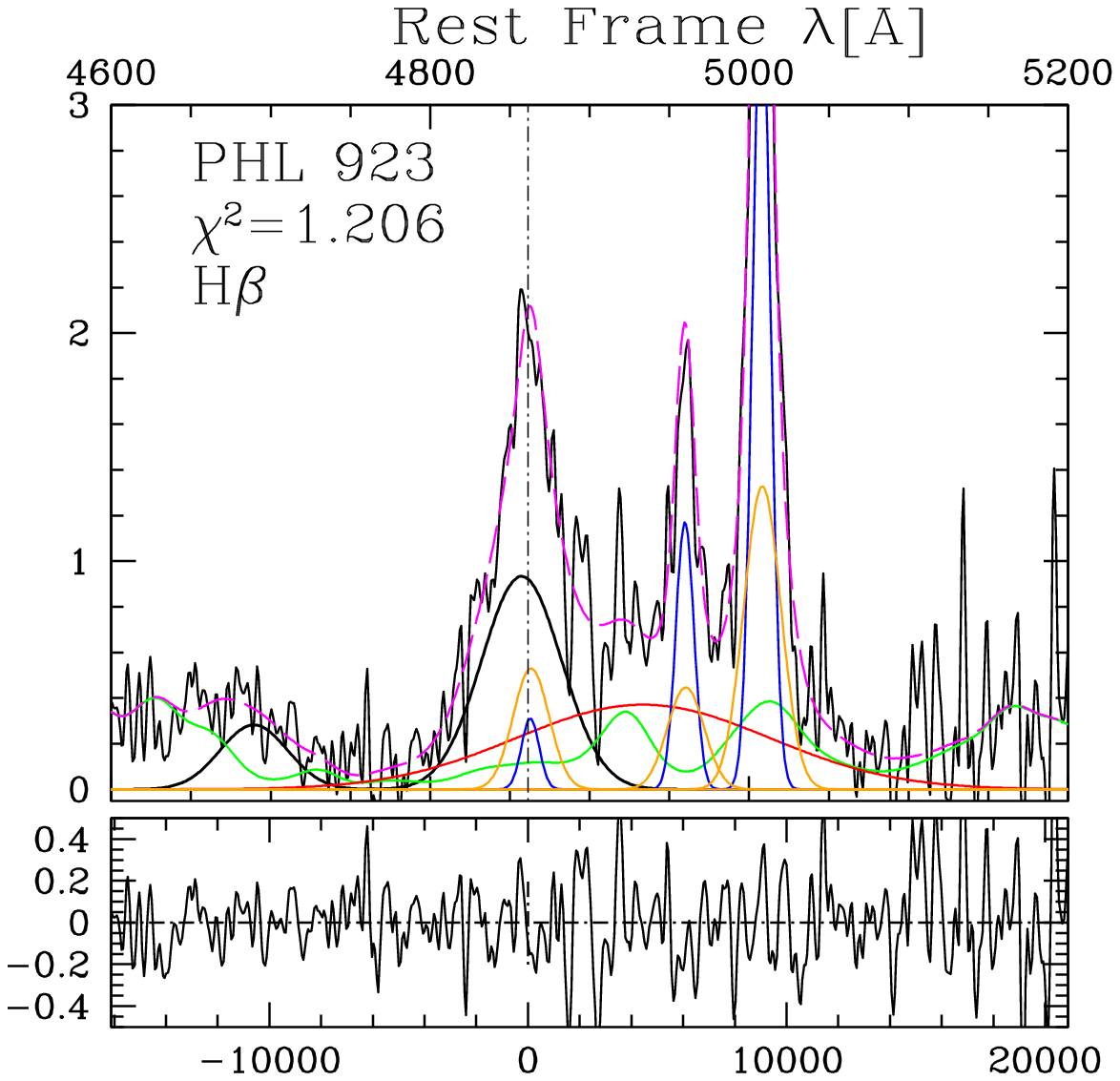}
\includegraphics [width=3in,height=1.55in ]{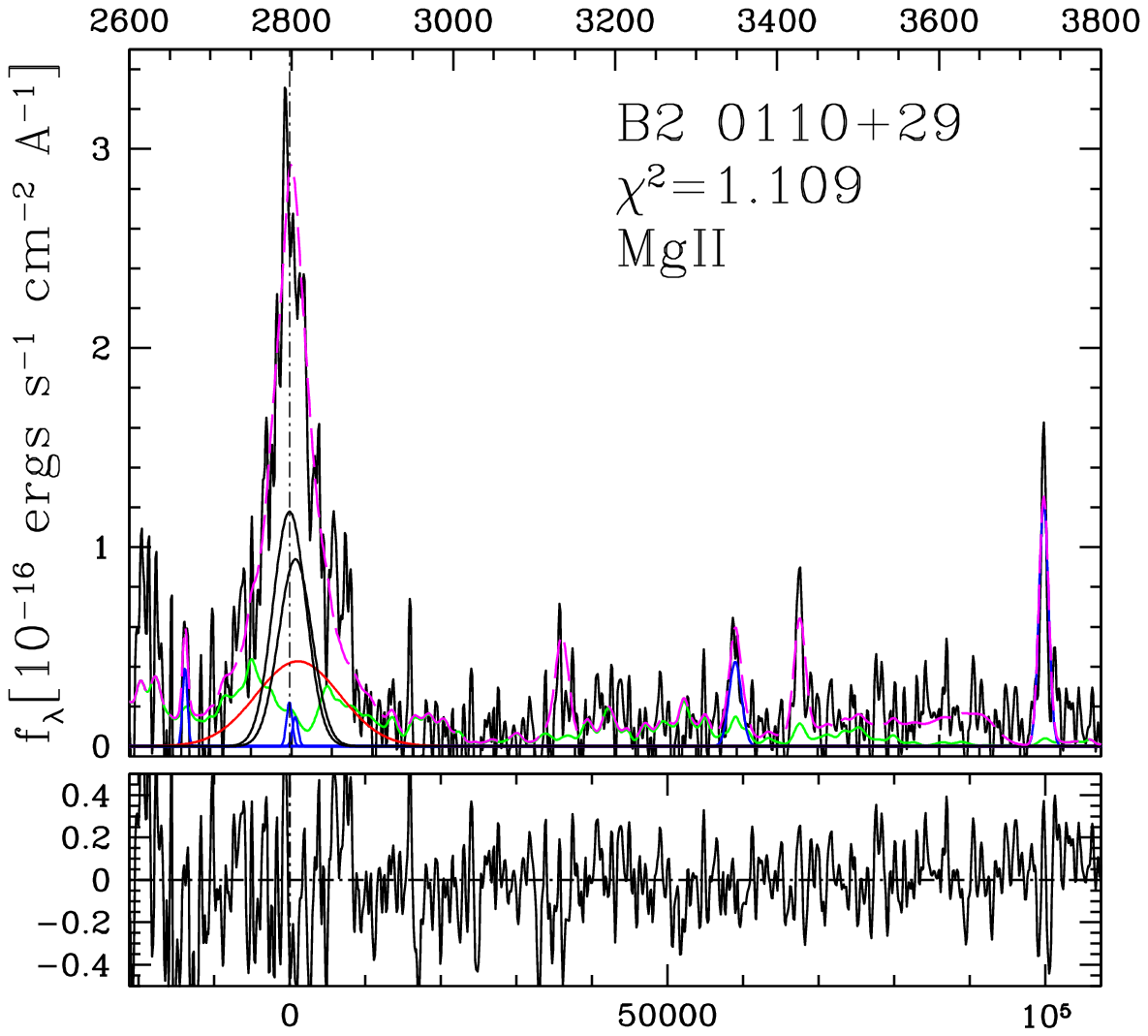}
\includegraphics [width=2in,height=1.55in ]{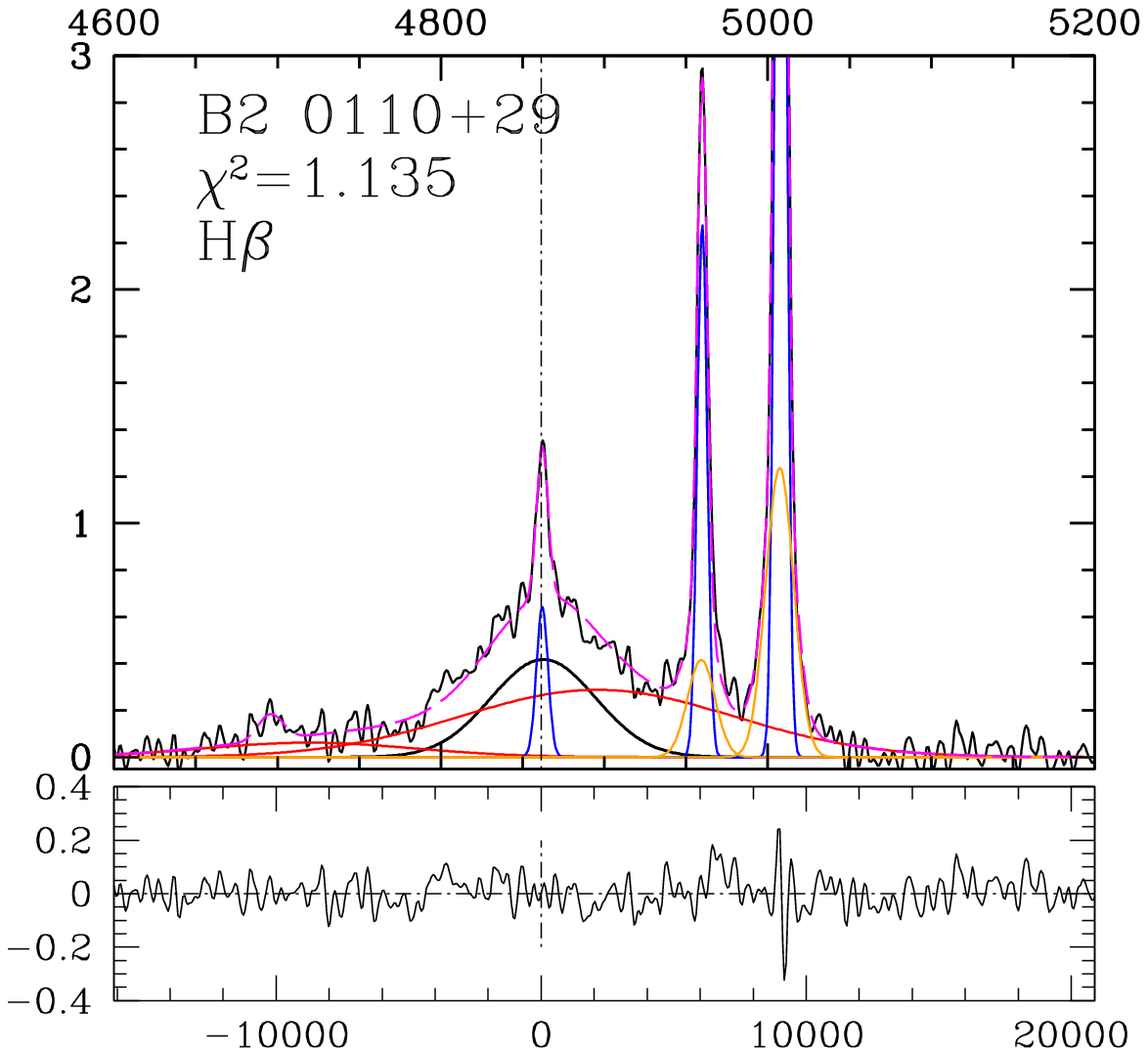}
\includegraphics [width=3in,height=1.55in ]{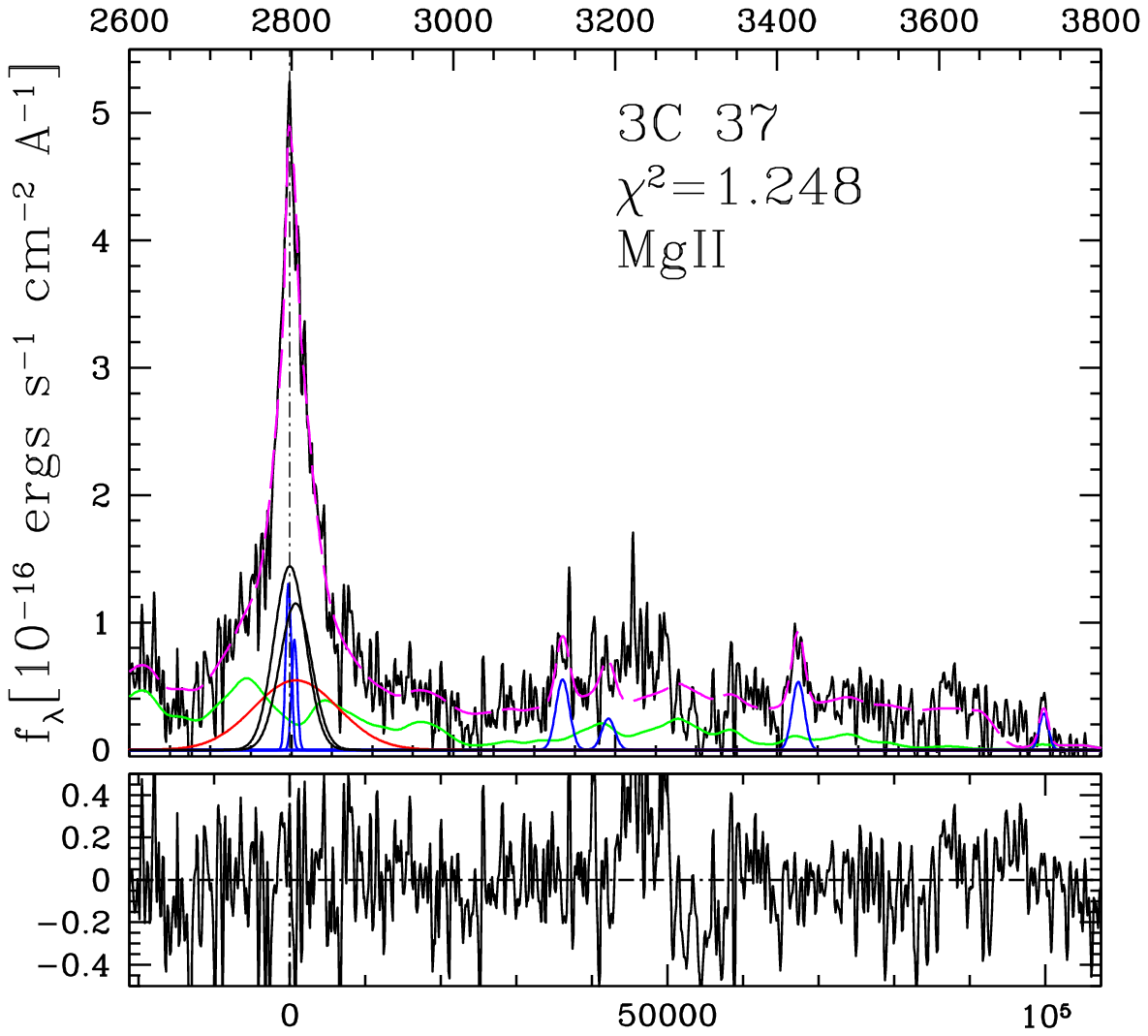}
\includegraphics [width=2in,height=1.55in ]{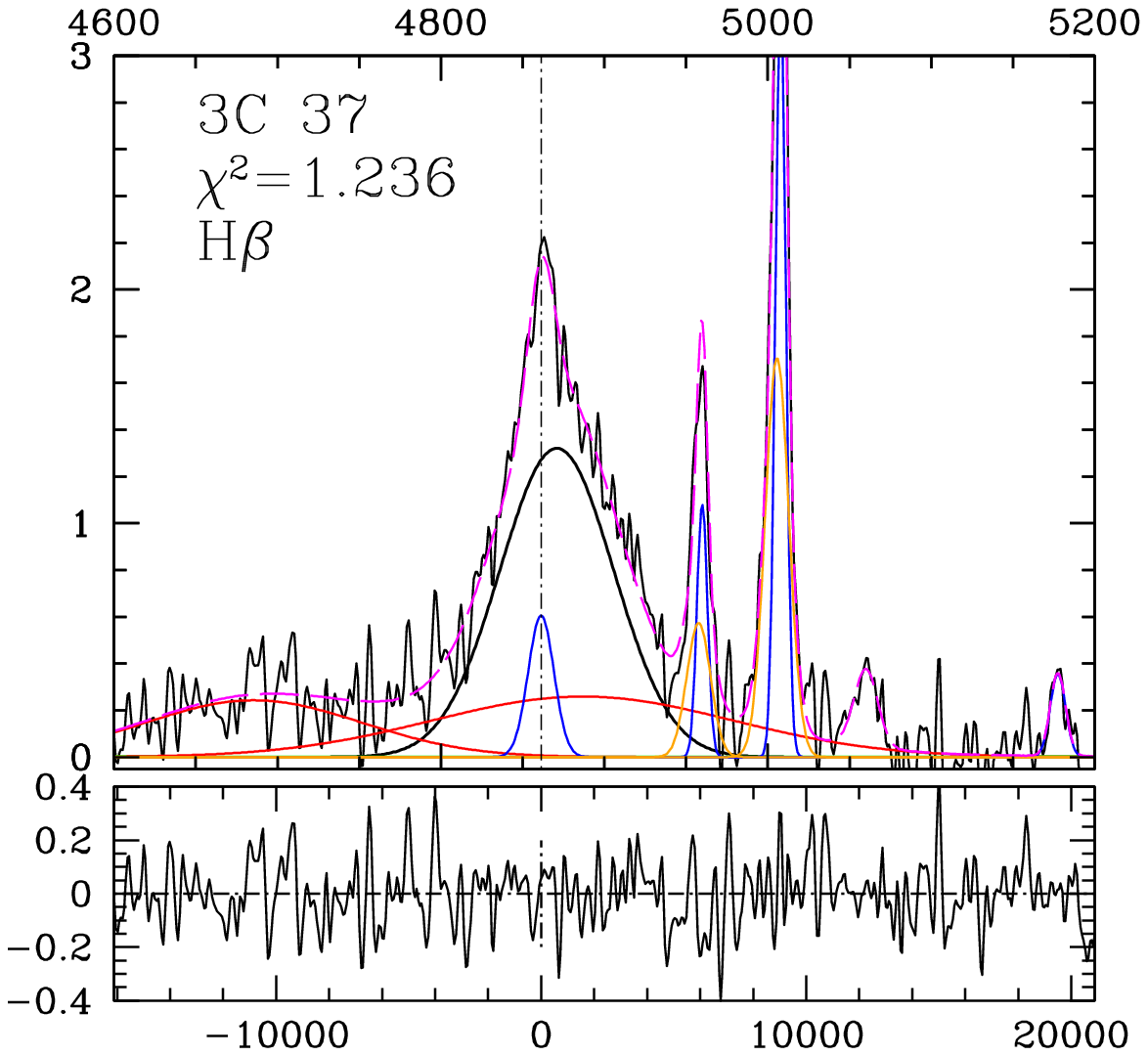}
\includegraphics [width=3in,height=1.55in ]{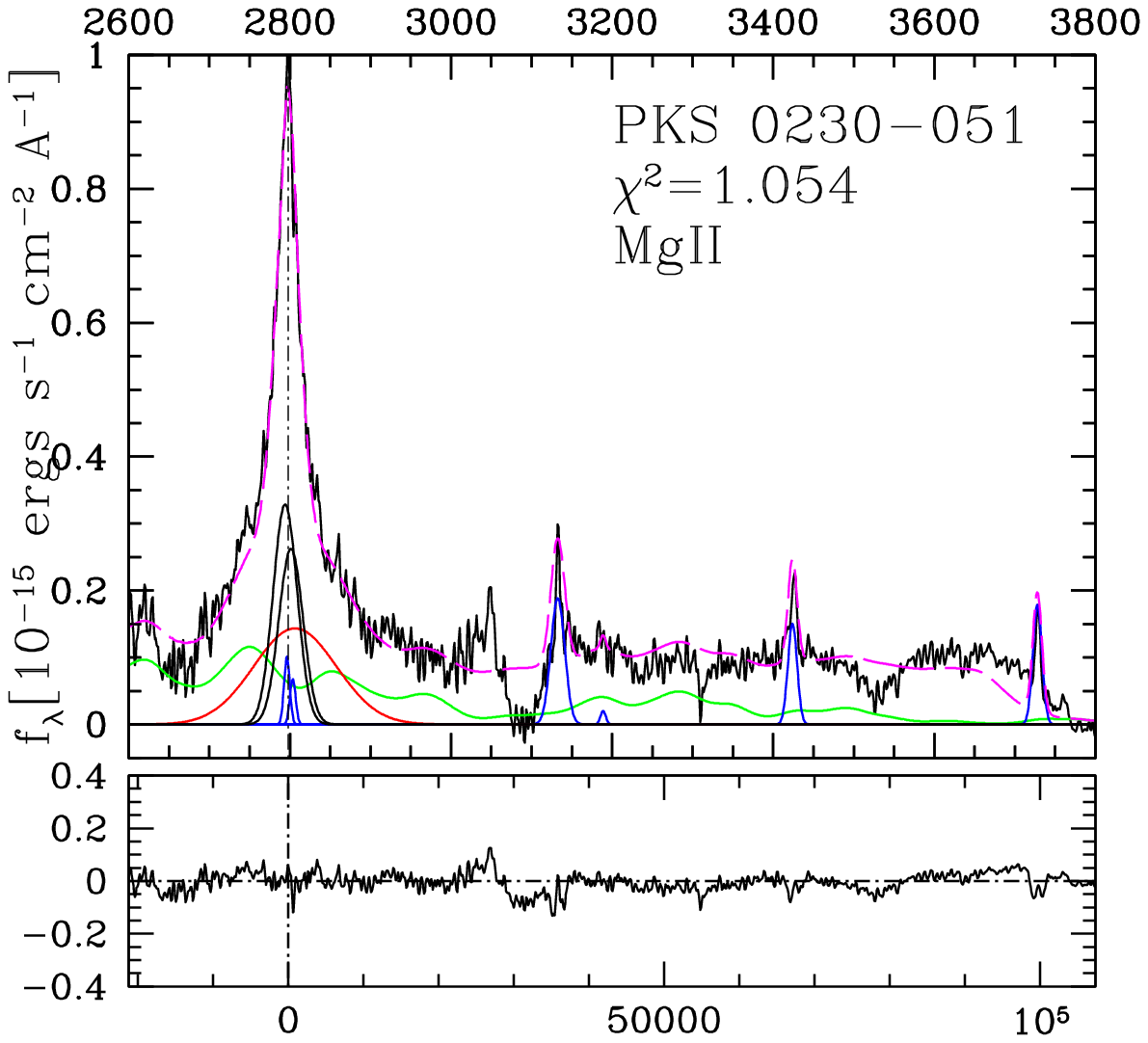}
\includegraphics [width=2in,height=1.55in ]{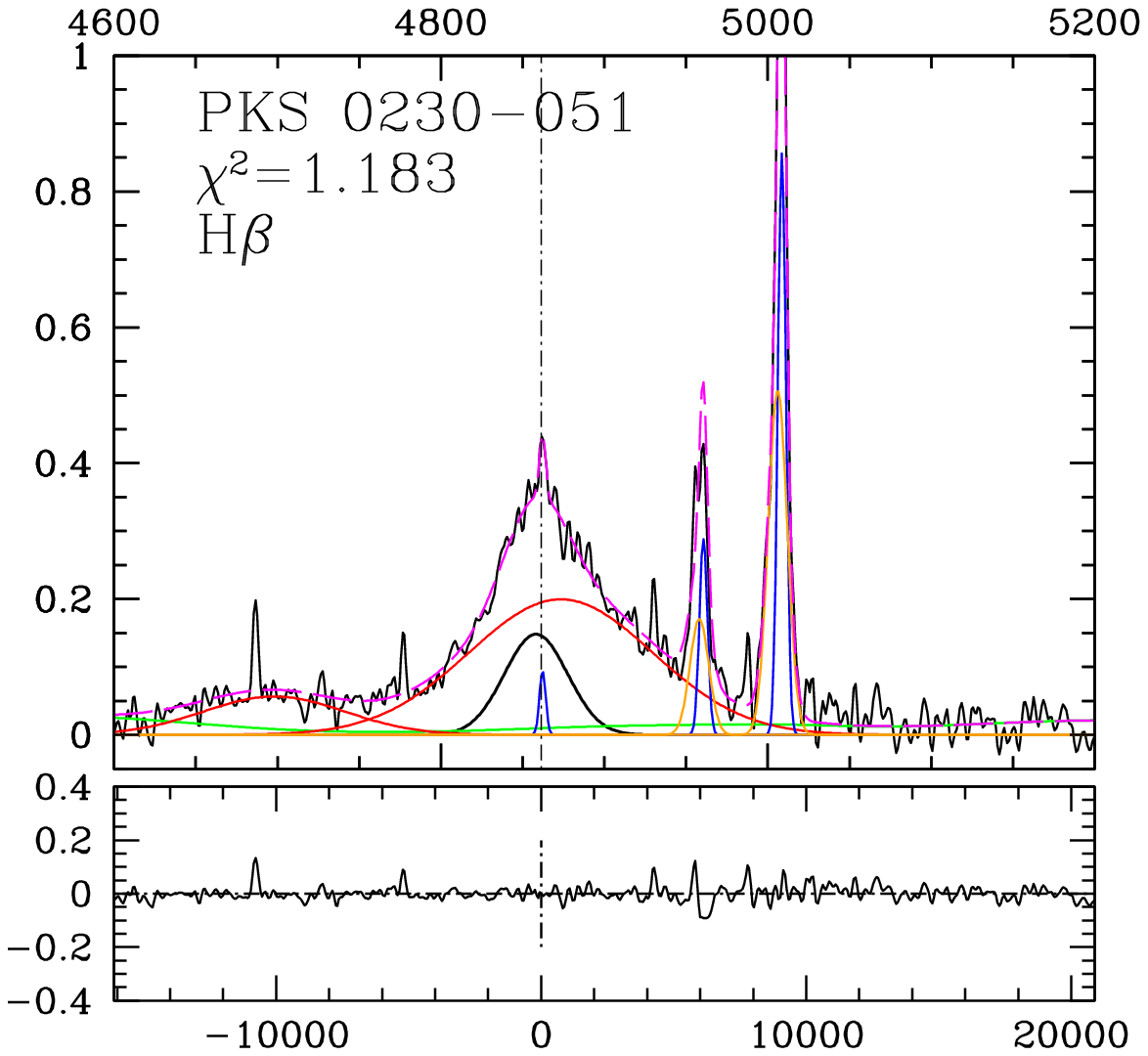}
\includegraphics [width=3in,height=1.55in ]{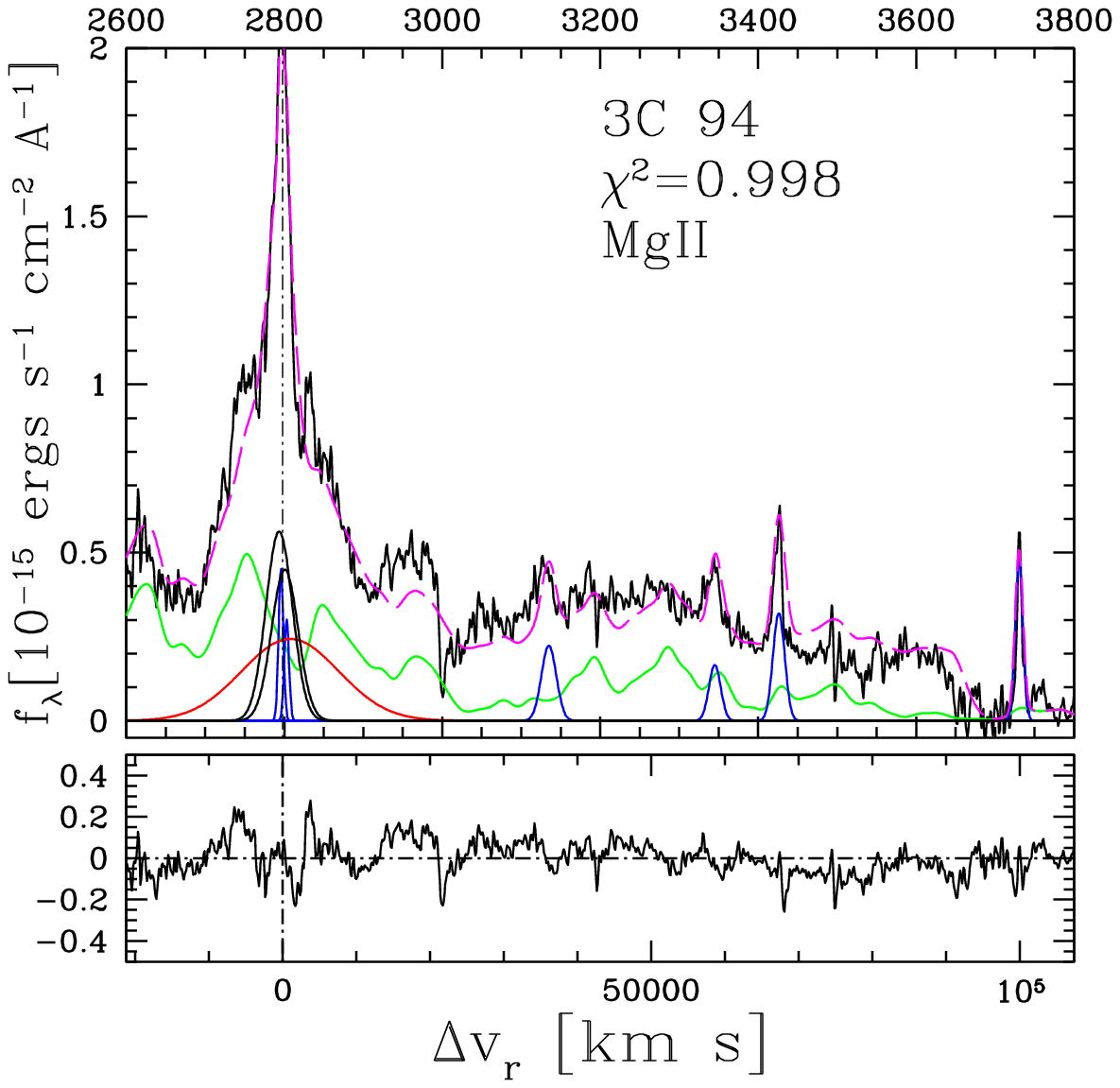}
\includegraphics [width=2in,height=1.55in ]{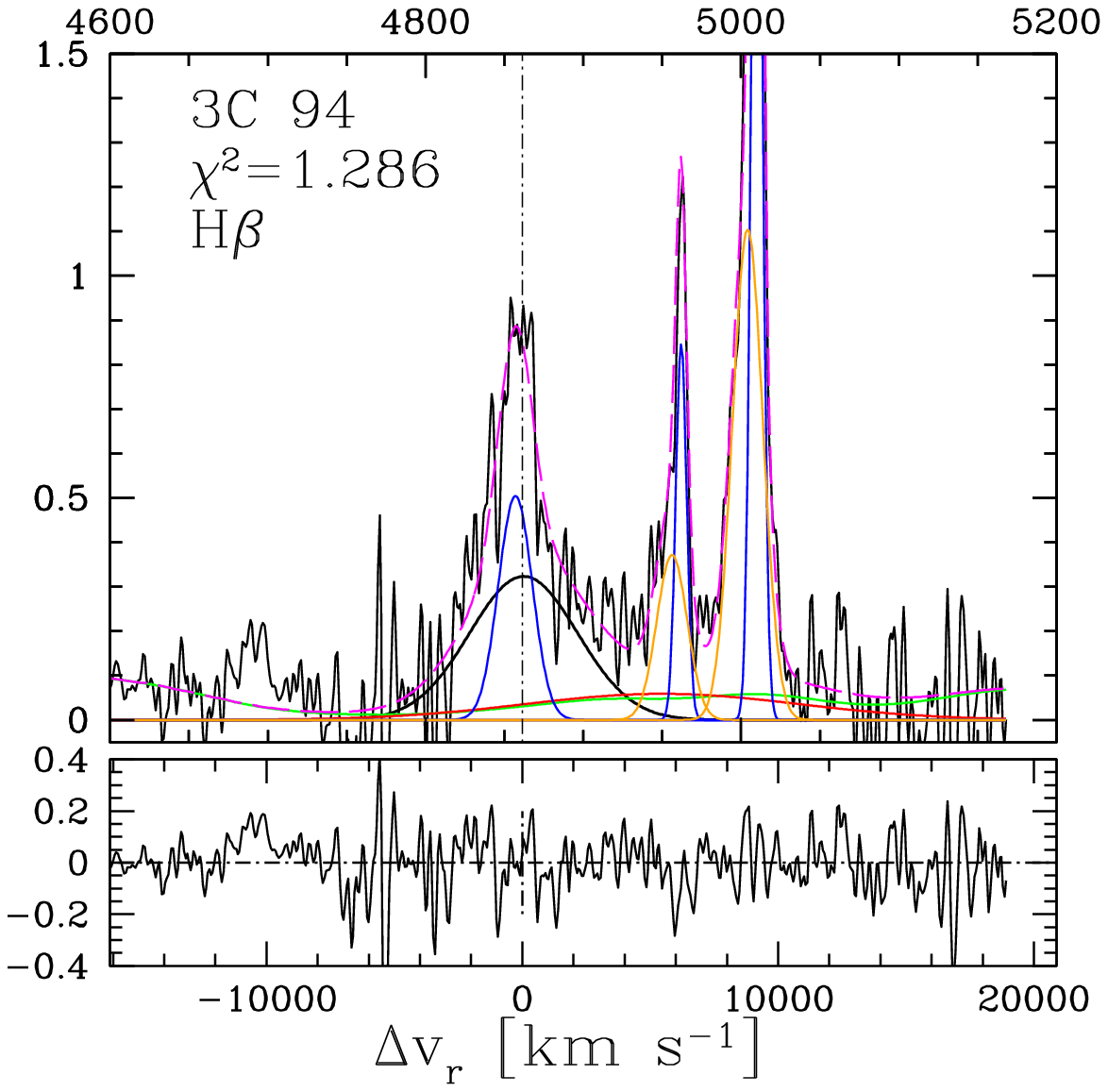}
\caption{ Multicomponent fitting results of our quasars 
in the region of \mgiionly\  and \hb\ lines (adjacent left and right panels),  represented after subtracting the continuum  obtained from the simultaneous best {\tt specfit} fit. In both cases, the upper abscissa is rest-frame wavelength in \AA\ and the lower abscissa is in radial velocity units. The vertical scales correspond to the specific flux in units of 10$^{-15}$ ergs\, s$^{-1}$\, cm$^{-2}$\, \AA$^{-1}$, and 10$^{-16}$ ergs\,s$^{-1}$\,cm$^{-2}$\,\AA$^{-1}$ (PHL 923, B2 0110+29, 3C 37,  and 3C 179) in both panels. Black continuous lines correspond to the rest-frame spectrum. The emission line components are: \feii{} (green), VBC (red), BC (black), all the NCs as blue lines, and  \oiiionly\  SBC as orange line. 
The dashed magenta line shows the model fitting. The dot-dashed vertical lines trace the rest-frame wavelength  of  \mgiionly\ and \hb. The lower panels show the residuals of the fit. The reduced $\chi^{2}$ values indicated in each panel are estimated in a  window of $\lambda$ around the main lines, \hb\ and \mgiionly.}
\label{fig:f3}
\end{figure*}

The overall fitting  result for \mgiionly\ spectral region for each object
is shown in the left plots of Fig. \ref{fig:f3} (adjacent with the \hb\ fitting, in the right panels). Residuals from the fittings are also shown in the bottom panels.

\begin{figure*}
\centering
\includegraphics [width=3in,height=1.55in ]{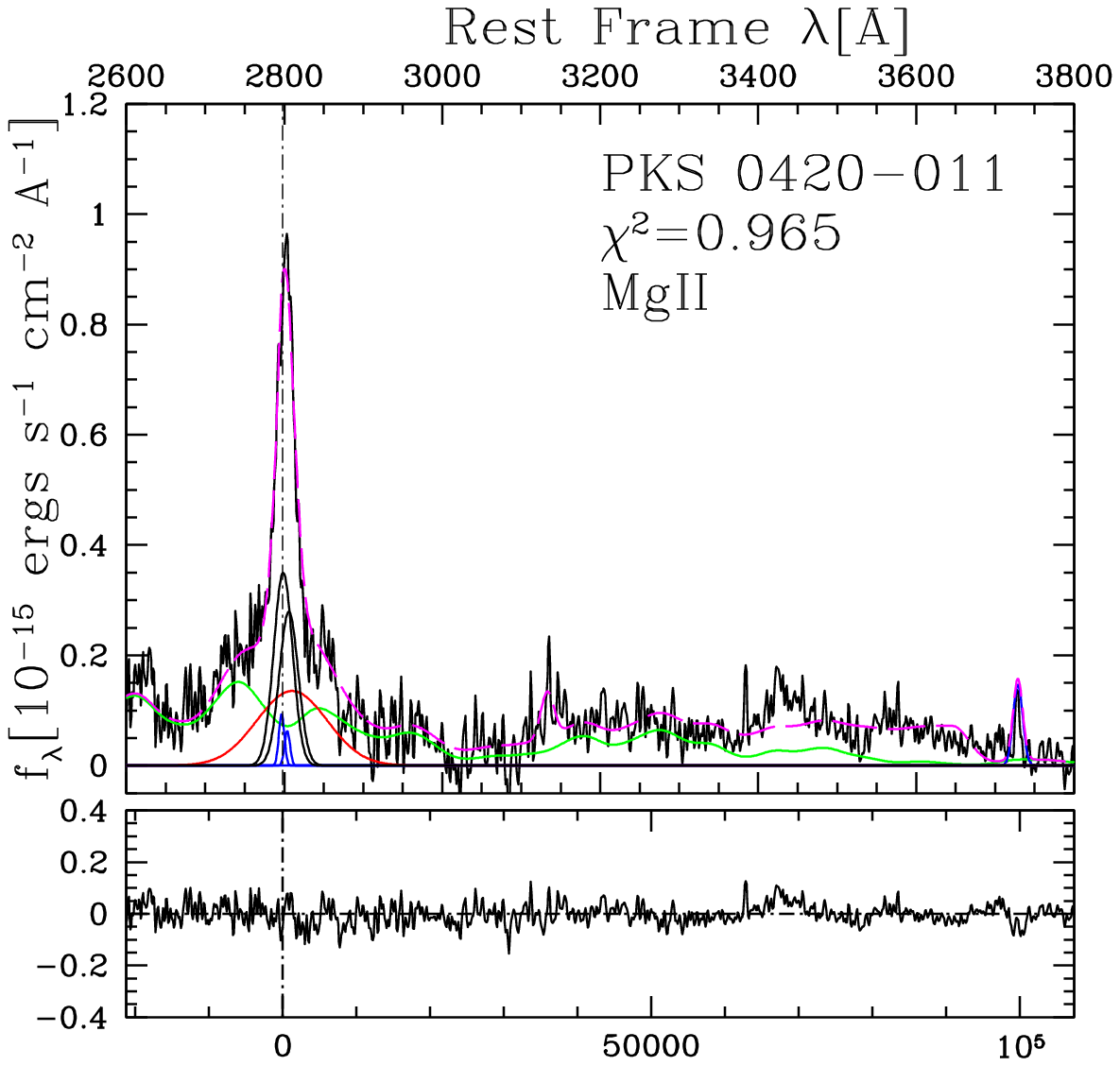}
\includegraphics [width=2in,height=1.55in ]{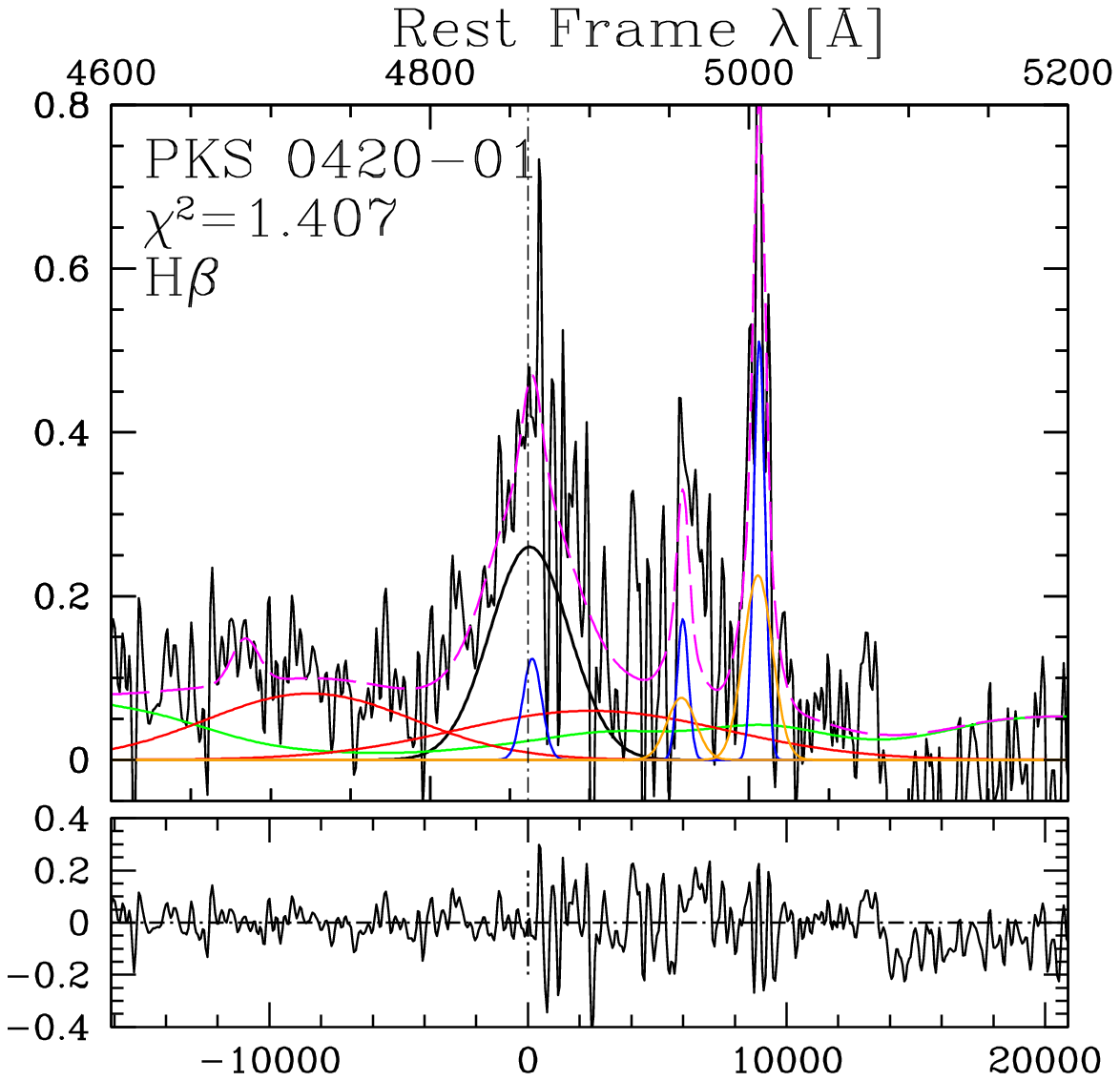}
\includegraphics [width=3in,height=1.55in ]{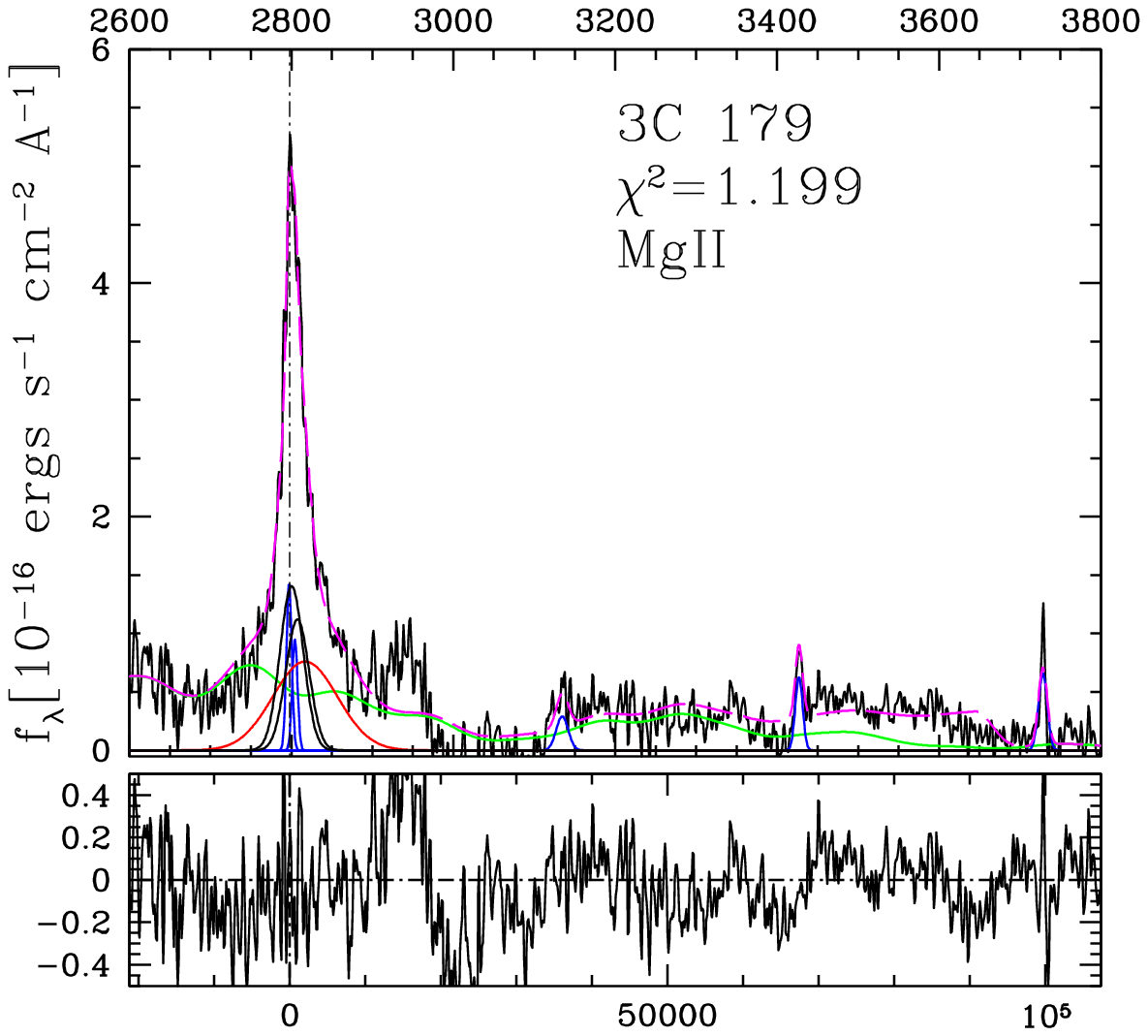}
\includegraphics [width=2in,height=1.55in ]{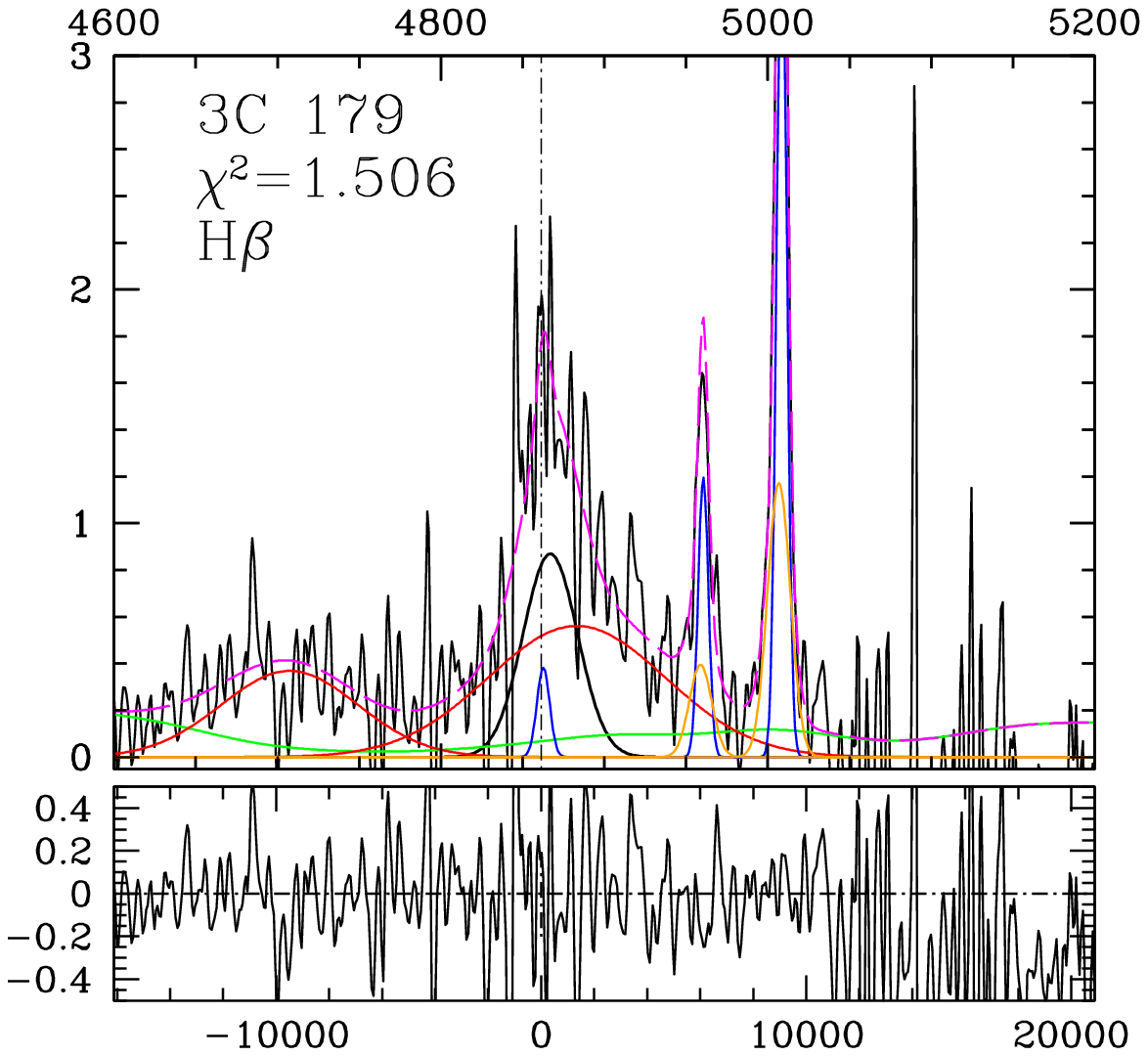}
\includegraphics [width=3in,height=1.55in ]{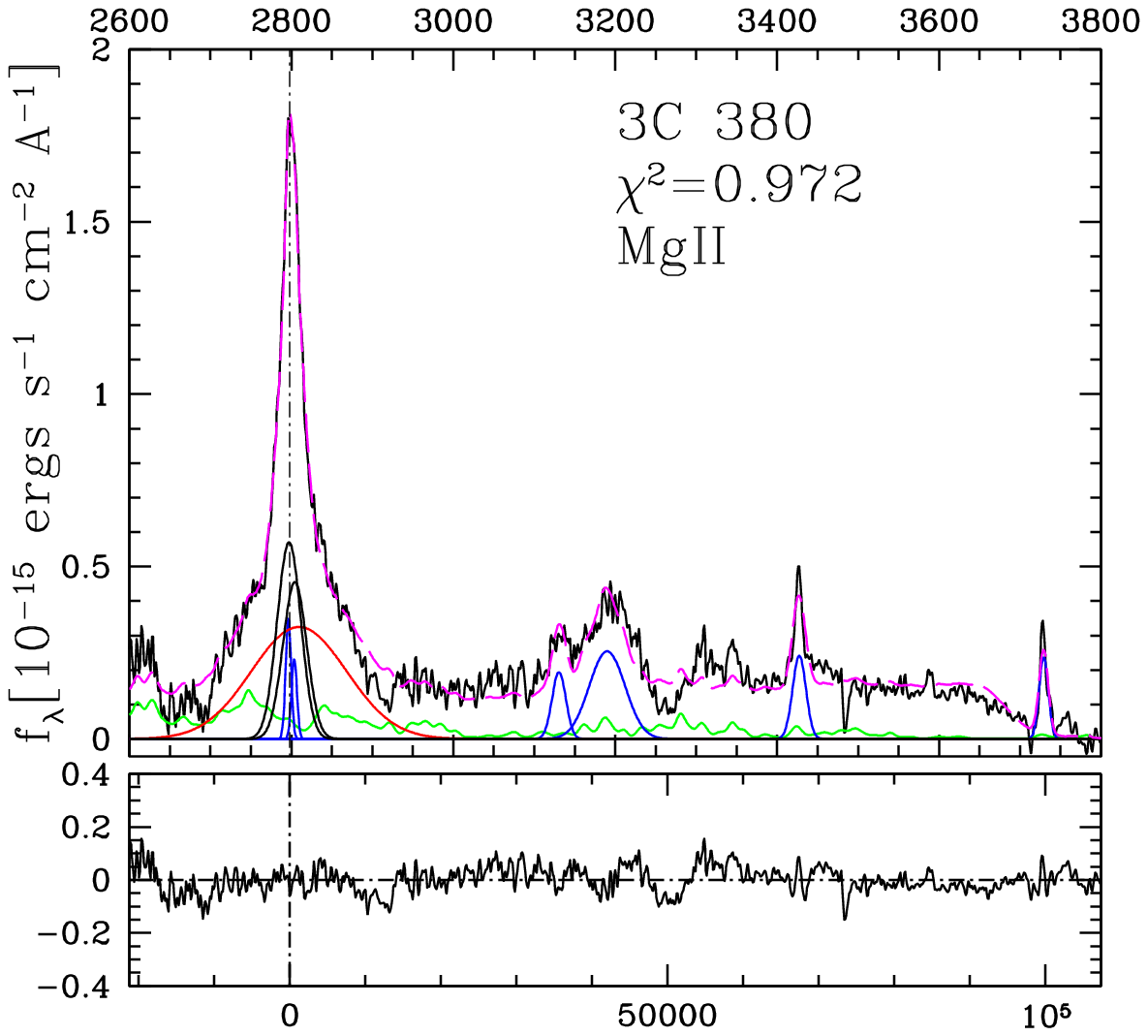}
\includegraphics [width=2in,height=1.55in ]{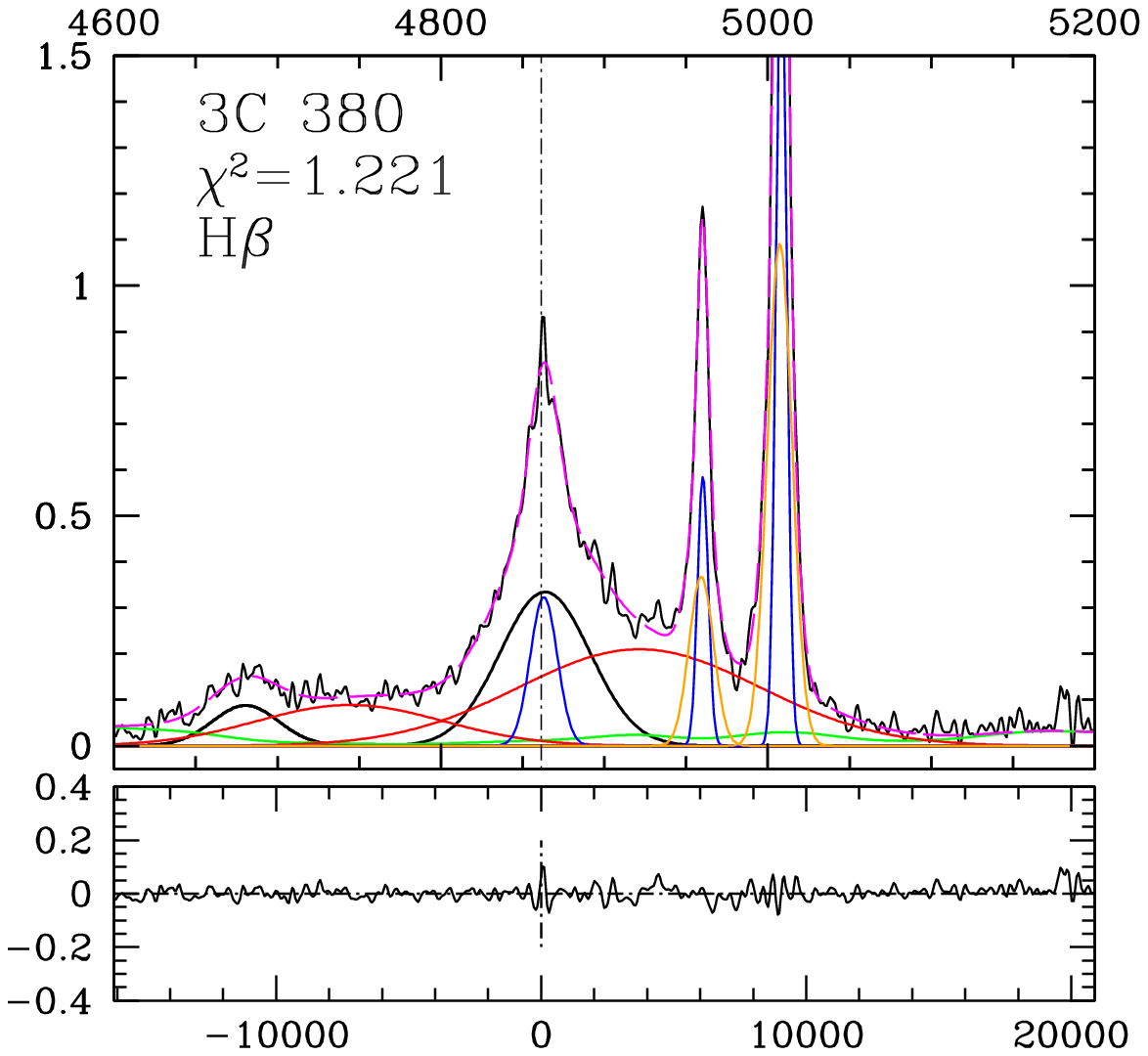}
\includegraphics [width=3in,height=1.55in ]{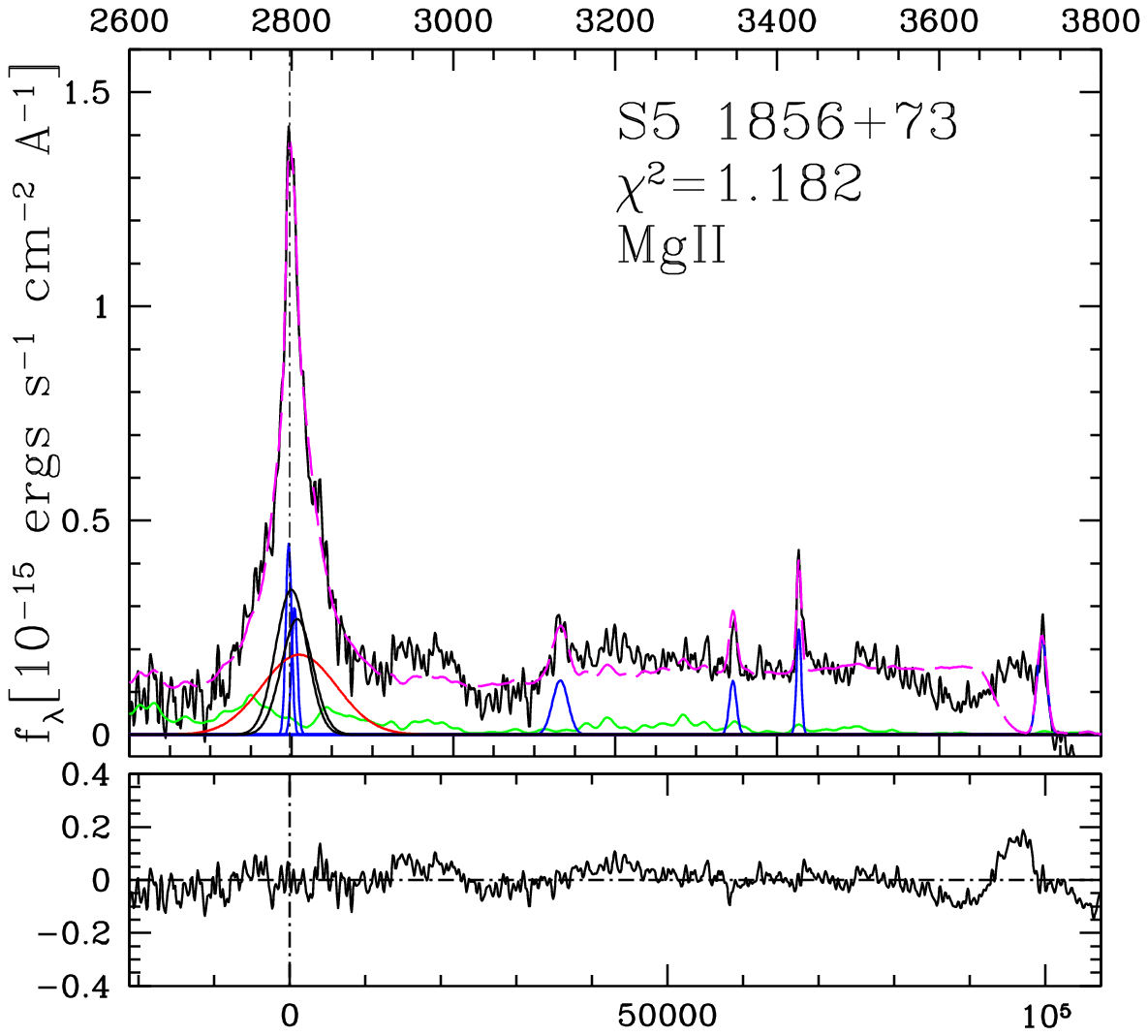}
\includegraphics [width=2in,height=1.55in ]{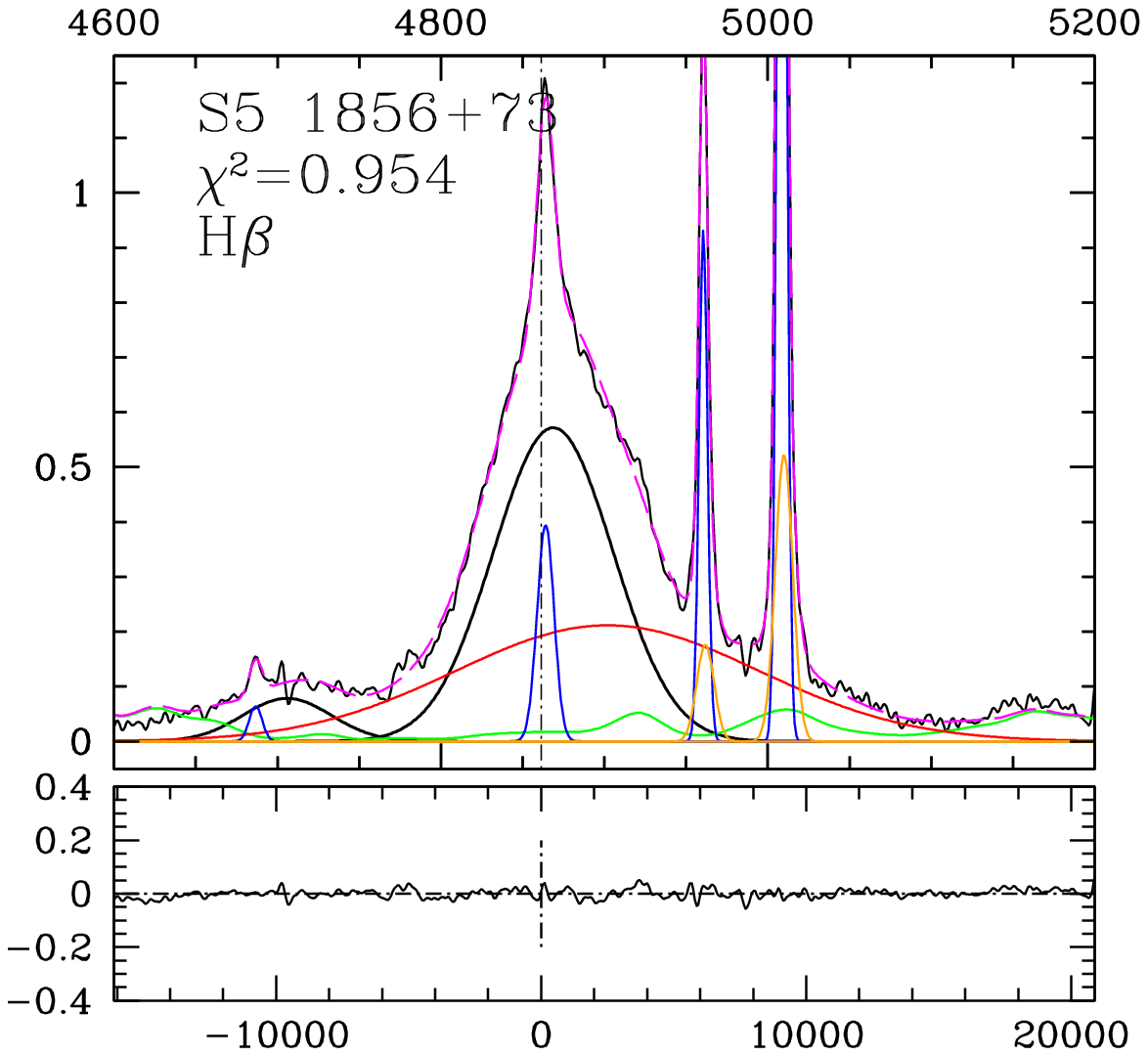}
\includegraphics [width=3in,height=1.55in ]{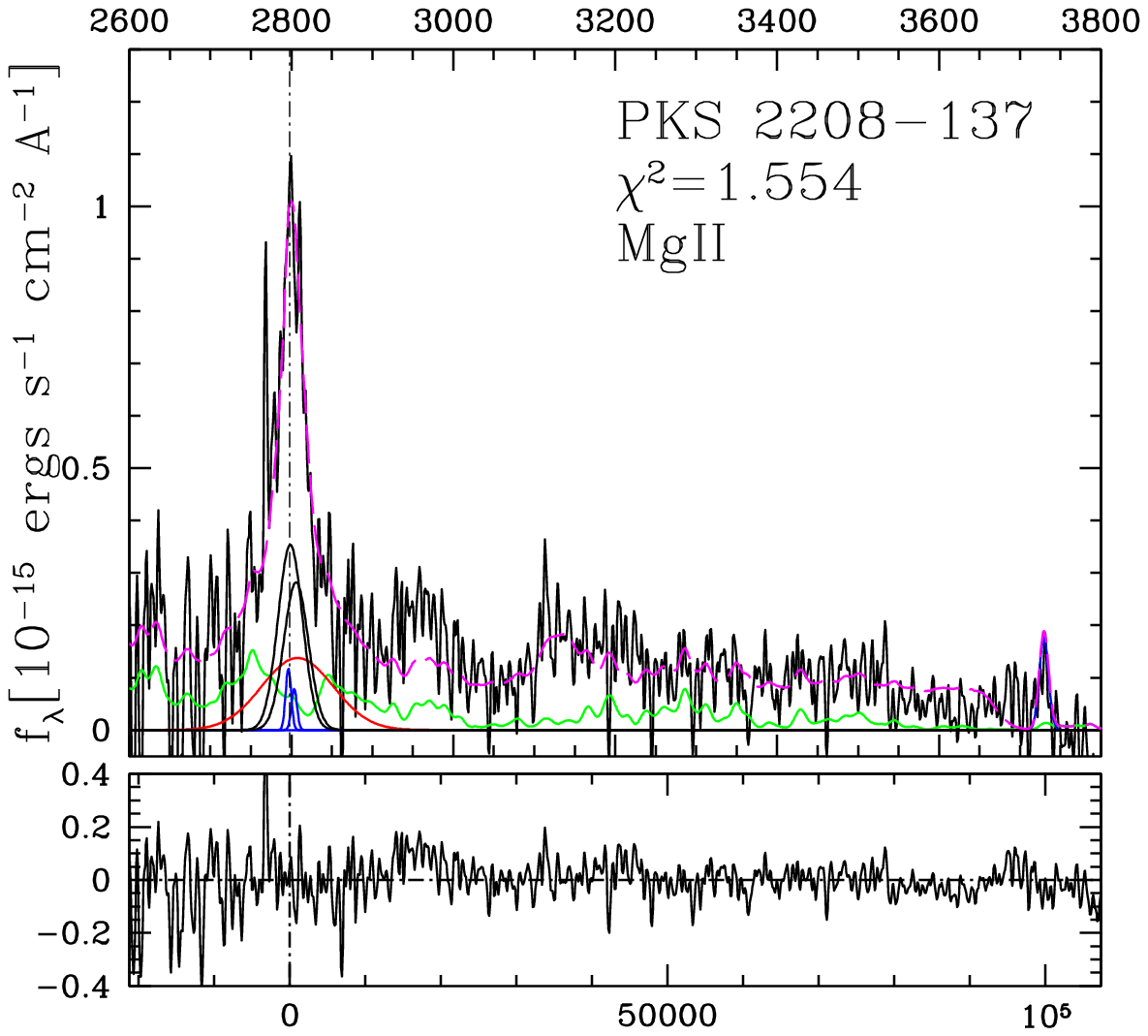}
\includegraphics [width=2in,height=1.55in ]{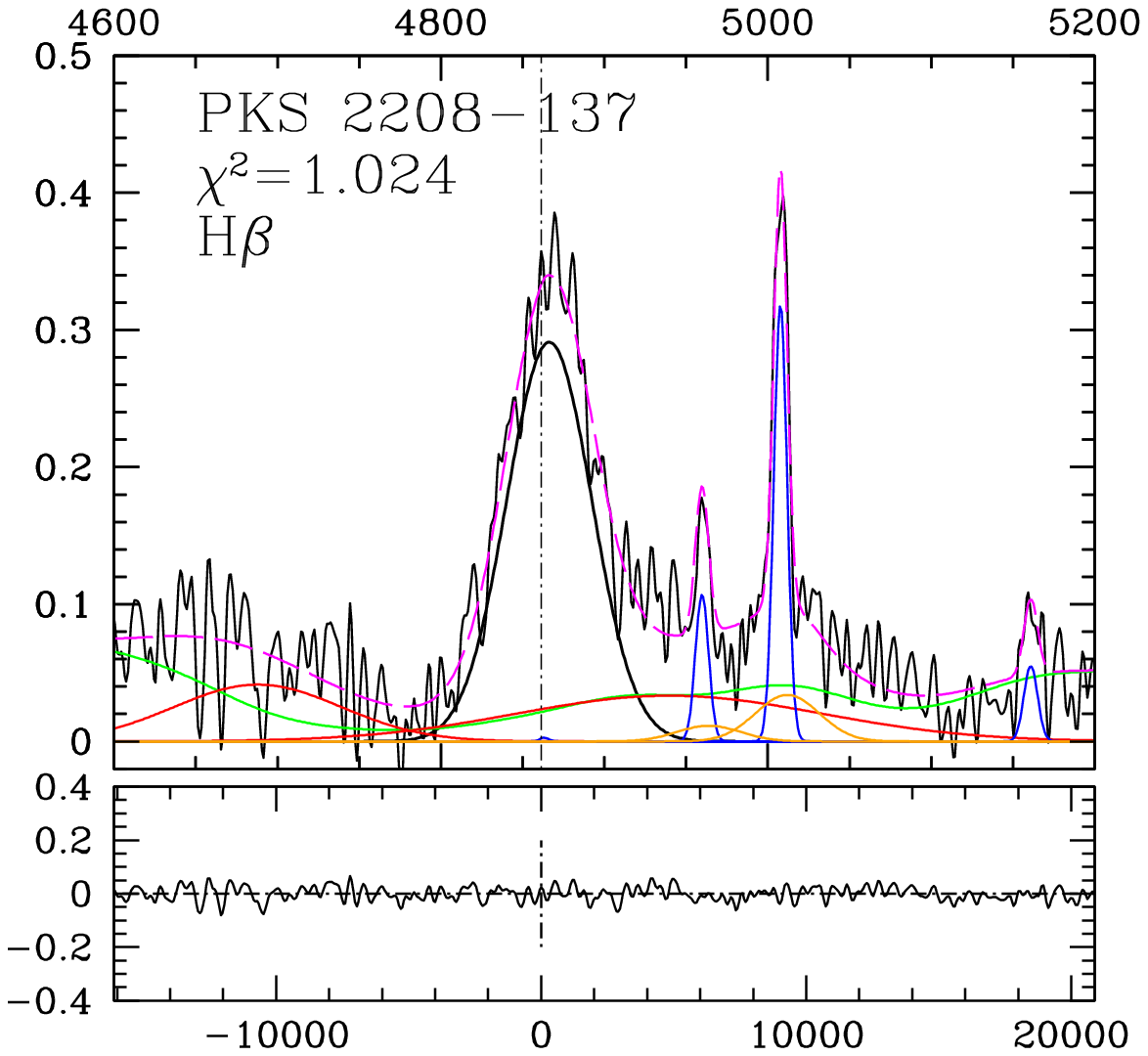}
\includegraphics [width=3in,height=1.55in ]{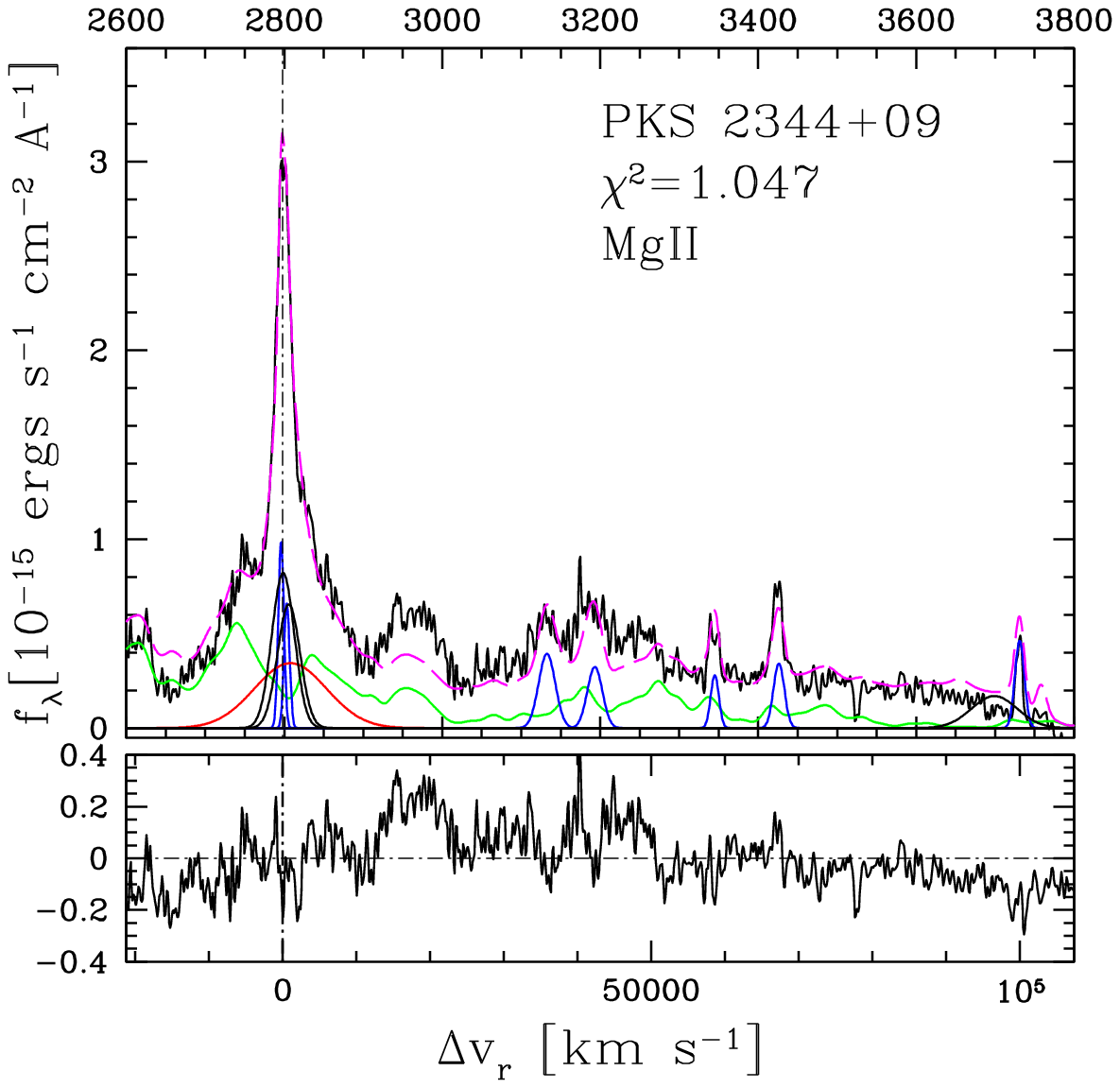}
\includegraphics [width=2in,height=1.55in ]{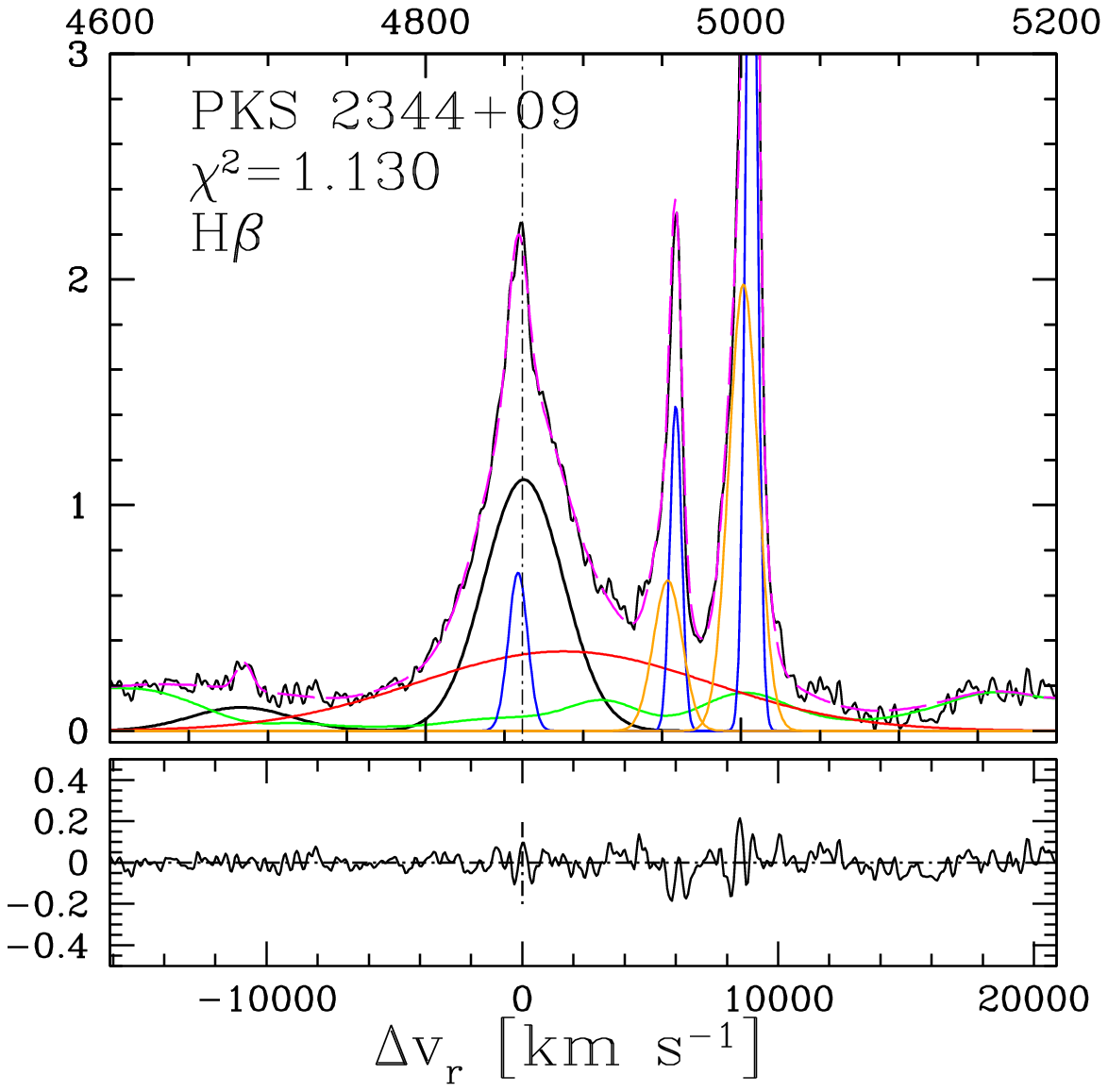}\\
\centering{\small {Figure 3. Results of the {\tt specfit} analysis (cont.)}}
\label{fig:f32}
\end{figure*}

\subsection{Full profile analysis}
\label{sec:FPA}
Besides the multicomponent spectral fitting to account for the individual components of the emission lines of \hb\ and \mgiionly\ regions, a complementary measurement that describes the FP is required.  The parametrization of the FP has been obtained by estimating the equivalent width (EW), FWHM, centroids at different fractional line intensities,  $(c(i/4)$ for i = 1, 2 and 3), the peak velocity assumed to be the centroid at 9/10 fractional intensity $(c(9/10)$), and asymmetry  (AI) and kurtosis (KI) indexes. The latter two quantities and centroids  are used as defined according to \citet{2010MNRAS.403.1759Z}.

 \subsection{Error estimation}
\label{sec:errest}
We made a coarse estimate of the fractional errors in the fluxes and relative intensities of the lines by using the interactive {\tt IRAF} task {\tt splot}. We empirically define extreme levels for line base or for continuum placements and assign them a confidence limit of $\pm$ 2$\sigma$.  The range of uncertainty and the typical errors that we derived depend on whether a feature was extended (such \feiiopt\ and \feiiuv\ emission) or sharp and prominent ( \oiiionly\, narrow lines) or faint affected by a much stronger line component (i.e., \heii\AA). Considering \feiiopt\ emission, the uncertainty ranges between 20\%\ and 50\%\ at 1$\sigma$ confidence level, the latter for quasars with extremely faint \feii\ emission like B2 0110+29.  
For \feiiuv, we have uncertainties between 15\%\ and 35\%\,, the latter for relatively faint quasars. The Balmer continuum is affected by a comparable fractional uncertainty, as it is modeled over the same spectral range of \feiiuv\ emission. Considering the individual components of \hb, the NC has uncertainties between 15\%\ and  50\%. The larger uncertainty applies to the case where the NC is weaker i.e., for quasars like 3C 179 and S5 1856+73: a small change of the BC translates into a much larger fractional change for the NC. This can also be seen from the intensity ratio between the NC and BC. For the two quasars with larger uncertainties (3C 179 and S5 1856+73), this ratio is, I$_\mathrm{NC}$/I$_\mathrm{BC}$ $\approx$ 0.125 and 0.097, respectively.  The prominent BC of \hb\ is affected by a small uncertainty between 10\%\ and  20\%. For the sharp, narrow, and well-defined  \oiiionly\, line, the uncertainty in flux is between 5\%\  and 10\%. For the UV lines, the \mgiionly\ NC, in most cases we have uncertainties between 15\%\ and 30\%. The larger uncertainty is associated with cases in which the NC merges with the BC. For the prominent \mgiionly\ BC, the uncertainty is estimated between 10\%\ and 20\%. The uncertainties for the BC can also be used for the VBC since the two components are usually of comparable strength.  

In all the cases, errors for the FP parameters have been estimated taking into account the effect of variation in the continuum on the profile parameters, by assuming a $\pm$ 5\%\ variation in the continuum. The uncertainties were derived with quadratic error propagation.

\section{Results}
\label{sec:results}

\subsection{Spectral multicomponent fitting results}

\subsubsection{\hb\, and \oiiionly  region}
\label{sec:reshb}

Spectrophotometric measurements of the optical region as well as the parameters of the FP and the individual components of \hb, represented in the right panels of Fig. \ref{fig:f3}, are presented in Table \ref{tab:meashb}. 
 Col. 2 contains the rest-frame specific continuum flux at 5100\AA\ obtained from the fitted power law; Col. 3 lists the wavelength range used for the spectral fitting in the \hb{} region; Cols. 4 to 12 report the FP measures  \hbFP\ (BC\,+\,VBC): EW (Col. 4), total intensity (Col. 5), FWHM of \hbFP\ (Col. 6), the centroids at $\frac{1}{4}$, $\frac{1}{2}$, $\frac{3}{4}$ and $\frac{9}{10}$ fractional intensity of the peak (Cols. 7 -- 10), as well as AI and KI (Cols. 11 and 12, respectively). For each line component isolated with the {\tt specfit} analysis, we report the total flux (I), the peak shift, and the FWHM in \kms: for NC (Cols. 16 -- 18), for BC (Cols. 19 -- 21), and for VBC (Cols. 22 -- 24). Additionally, we also provide in Table \ref{tab:meashb} the parameters related to \feii: the total flux of the blue blend of \feiiq\AA\ in Col. 13, and the smoothing factor applied to the \feii\ template to reproduce the observed \feii\ feature (Col. 14). In Col. 15 we report the \rfeopt\ parameter
(see Sect. \ref{sec:Intro}).
In the two most redshifted quasars in the sample at  z\,>\,0.9 (3C 94 and PKS 0420), the reddest part of the spectrum has a slightly lower S/N due to the correction of the telluric band that extends between 9250\AA\, and 9700\AA\, in the observed wavelength. This results in a kind of flat-topped and even multipeaked profile in \hb\, (see Fig. \ref{fig:f3}), and slightly larger uncertainties in the determination of the higher fractional intensity centroids of \hb, and to less extent in   \oiiionly.

Regarding \hbFP, in general the objects of  our eRk sample show FWHM(\hb)\,$\gtrsim$\,4000 \kms, with a median value of 5100 \kms, which is a typical value of  B1 spectral type (ST) quasars at z\,<\,1 (see Sect. \ref{sec:op}) with similar luminosities \citep[e.g.,][]{2002ApJ...566L..71S, 2008MNRAS.387..856Z}. In two cases (3C 179 and PKS 2208-137), the FWHM is slightly less than 4000 \kms. For PKS 2208-137, the barely lower value may be caused by the uncertainty in identifying the \hb{} NC, which may have been included in the BC  producing a slightly narrower and sharper BC profile. In the case of 3C 179, with the lowest FWHM ($\sim$ 3600 \kms) in   the eRk sample, that value may be influenced by a slightly high continuum location on the red side due to the correction of the telluric band in that region, preventing an accurate continuum determination on the redward  \oiiionly.
 
In our quasars, the \hbFP\ presents an asymmetry towards the red, with a median AI of 0.2, which is related to the presence of a VBC. 
This fact is clearly reflected in the centroid velocities at different fractional intensities (see Table \ref{tab:meashb}), which are significantly shifted to the red wing of the broad profile in the majority of quasars, being more pronounced towards the base of the line, with a median value of +1160\,\kms. This is also verified in the centroid velocity at half peak intensity, c($\frac{1}{2}$), with a median value of 230\,\kms.

\begin{sidewaystable}[tp]

\setlength{\tabcolsep}{1.2pt}
\caption{\ \  Results of the full broad profile and the {\tt specfit} analysis for \hb.}
\label{tab:meashb}
\scalebox{0.83}{
\begin{tabular}{@{\extracolsep{3pt}}l c c c c c c c c c c c c c c c c c c c c c c c@{}} 

\hline\hline 

&&&\multicolumn{9}{c}{Full broad profile (FP) (BC + VBC)}&&&&\multicolumn{3}{c}{NC} & \multicolumn{3}{c}{BC}& \multicolumn{3}{c}{VBC} \\

\cline{4-12}\cline{16-18}\cline{19-21}\cline{22-24}

Object & $f_{\lambda,5100}$ &\multicolumn{1}{c}{Fitting range} & \multicolumn{1}{c}{EW} & \multicolumn{1}{c}{I} & \multicolumn{1}{c}{FWHM} & \multicolumn{1}{c}{C($\frac{1}{4}$)} & \multicolumn{1}{c}{C($\frac{1}{2}$)} & \multicolumn{1}{c}{C($\frac{3}{4}$)} & \multicolumn{1}{c}{C($\frac{9}{10}$)} & \multicolumn{1}{c}{AI} & \multicolumn{1}{c}{KI} & \multicolumn{1}{c}{\feiiq} & \multicolumn{1}{c}{SF} & \multicolumn{1}{c}{\rfeopt} &  \multicolumn{1}{c}{I} & \multicolumn{1}{c}{Shift} &\multicolumn{1}{c}{FWHM}&  \multicolumn{1}{c}{ I} &\multicolumn{1}{c}{Shift} & \multicolumn{1}{c}{FWHM} &  \multicolumn{1}{c}{I} & \multicolumn{1}{c}{Shift} & \multicolumn{1}{c}{FWHM} \\

\multicolumn{1}{l}{(1)}& \multicolumn{1}{c}{(2)} & \multicolumn{1}{c}{(3)} & \multicolumn{1}{c}{(4)} & \multicolumn{1}{c}{(5)} & \multicolumn{1}{c}{(6)} &\multicolumn{1}{c}{(7)} & \multicolumn{1}{c}{(8)} & \multicolumn{1}{c}{(9)} & \multicolumn{1}{c}{(10)} & \multicolumn{1}{c}{(11)} &\multicolumn{1}{c}{(12)} & \multicolumn{1}{c}{(13)} &\multicolumn{1}{c}{(14)} &\multicolumn{1}{c}{(15)} &\multicolumn{1}{c}({16)} & \multicolumn{1}{c}{(17)} & \multicolumn{1}{c}{(18)} & \multicolumn{1}{c}({19)} &\multicolumn{1}{c}{(20)} & \multicolumn{1}{c}{(21)} &\multicolumn{1}{c}({22)} &\multicolumn{1}{c}({23)} &\multicolumn{1}{c}{(24)}\\
\hline
PHL 923   & 0.13 & 4400 - 5100 & 92 & 13.2 & 4420$\pm$80 & 2280$\pm$880 & -60$\pm$140 & -190$\pm$120 & -230$\pm$160 & 0.47$\pm$0.11 & 0.25$\pm$0.05 & 8.40 & 2700 & 0.63 & 0.4 & 10 & 810 & 5.8 & -330 & 3640 & 7.4 & 4370 & 11400 \\ 
B2 0110+29  & 0.07 & 4400 - 5200 & 138 & 9.4 & 6370$\pm$740 & 1690$\pm$540 & 450$\pm$250 & 130$\pm$170 & 60$\pm$210 & 0.28$\pm$0.08 & 0.31$\pm$0.04 & 0.01 & 4000 & 0.001 & 0.6 & -40 & 510 & 3.4 & -1 & 4650 & 6.0 & 2030 & 12000 \\ 
3C 37   & 0.13 & 4400 - 5200 &126 &17.7 & 5540$\pm$450& 550$\pm$300 & 490$\pm$160 & 470$\pm$150 & 470$\pm$200 & 0.02$\pm$0.13 & 0.41$\pm$0.05 & 2.8 & 2800 & 0.16 & 1.2 & -80 & 1140 & 10.2 &-30 & 4770 & 7.5 & 890 & 11800 \\ 
PKS 0230-051 & 0.30 & 4400 - 5200 & 126 & 41.5 & 5100$\pm$490 & 1370$\pm$740& 190$\pm$170& 70$\pm$140 & 40$\pm$180 & 0.26$\pm$0.12 & 0.30$\pm$0.05 & 7.8 & 3200 & 0.19 & 0.6 & -30 & 340 & 15.7 &-80 & 3880 & 25.8 & 2140 & 12900 \\ 
3C 94$^{(1)}$  & 0.82 & 4150 - 5100 & 48 & 44.6 & 5480$\pm$290& 6 30$\pm$350& 200$\pm$150 & 120$\pm$150 & 90$\pm$200 & 0.14$\pm$0.08 & 0.40$\pm$0.05 & 14.7 & 5200 & 0.34 & 13.7 &  -330:: & 1570 & 28.9 &-30: & 4960 & 15.7 &5 400 & 12500 \\ 
PKS 0420-01$^{(1)}$ & 0.70 & 4400 - 5100 & 42 & 28.9 & 4110$\pm$330 & 250$\pm$250 & 60$\pm$120 & 30:$\pm$110 & 10:$\pm$150 & 0.07$\pm$0.06 & 0.4$\pm$0.1 & 14.4 & 5000 & 0.50 & 1.9 & 80: & 860 & 16.2 &-20: & 3600 & 12.7 & 2360 & 12200 \\ 
3C 179  & 0.17 & 4400 - 5050 & 62 & 11.4 & 3570$\pm$350 & 1160$\pm$440 & 500$\pm$120 & 410$\pm$100 & 390$\pm$120 &0.22$\pm$0.16 & 0.31$\pm$0.05 & 4.3 & 5000 & 0.37 & 0.5 & -10 & 620 & 4.0 & 310 & 2620 & 7.4 & 1460 &  8300 \\ 
3C 380  & 0.68 & 4400 - 5200 & 86 & 61.9 & 5740$\pm$750 & 2580$\pm$470 & 780$\pm$250 & 400$\pm$150 & 320$\pm$190 & 0.4$\pm$0.08 & 0.28$\pm$0.04 & 8.5 & 4800 & 0.14 & 6.7 &  20& 1200 & 23.1 & 60 & 4000 & 38.8 & 3630 & 10600 \\ 
S5 1856+73 & 0.72 & 4400 - 5500 & 135 & 102.4 & 5730$\pm$560 & 1350$\pm$540 & 530$\pm$190 & 380$\pm$150 & 340$\pm$200 & 0.2$\pm$0.12 & 0.33$\pm$0.05 & 12.3 & 2600 & 0.12 & 5.1 &  80 & 750 & 52.4 &370 & 5300 & 50.0 & 2420 & 13300 \\ 
PKS 2208-137  & 0.47 & 4400 - 5500 & 54 & 27.8 & 3850$\pm$300 & 360$\pm$180 & 230$\pm$100 & 190$\pm$100 & 180$\pm$140 & 0.07$\pm$0.09 & 0.43$\pm$0.05 & 12.6 & 4900 & 0.46 & 0.1:: & 70::  & 490:: & 18.1 & 160 & 3590 & 9.7 & 5380 & 12100 \\ 
PKS 2344+09  & 1.34 & 4400 - 5500 & 108 & 151.9 & 4150$\pm$380 & 350$\pm$370 & 40$\pm$130 & -20$\pm$110 & -30$\pm$150 & 0.11$\pm$0.11 & 0.35$\pm$0.05 & 42.9 & 3500 & 0.28 & 10.6 & -250: & 880 & 70.0 & -30 & 3640 & 81.9 & 1530 & 13400 \\  
\hline
\end{tabular}
}
\\ \\
{\raggedright \textbf{Note}: 
{Col. 2 is in units of  $10^{-15}$ ergs\, s$^{-1}$\,  cm$^{-2}$\, \AA$^{-1}$. Cols. 3 and 4 are in units of \AA. Cols. refereed to intensities as Col. 5, 13, 16, 19, and 22 are in units of $10^{-15}$ ergs\, s$^{-1}$\,  cm$^{-2}$. Cols. refereed to centroid velocities at different fractional intensities, FWHM, and shift as Cols. 6 - 10, 14, 17 - 18, 20 - 21, 23 - 24 are in units of \kms. $^{(1)}$ Centroid and peak shift determinations might be affected by the presence of a telluric band from 9250 to 9700 \AA\ at observed wavelengths. Values ending with a colon (:)  means that the values are highly uncertain  and "::" that the feature is poorly defined. \par}}

\end{sidewaystable}
\begin{sidewaystable}
\setcounter{table}{6}
\caption{\ \ Measurements of the \mgiionly\ region.}
\setlength{\tabcolsep}{1.2pt}
\scalebox{0.88}{
\begin{tabular}{@{\extracolsep{3pt}}l c c c c c c c c c c c c c c c c c c c c c c c@{}} 
\hline\hline  
 & & & \multicolumn{9}{c}{Full broad profile parameters (FP) (2BC + VBC)} & \multicolumn{2}{c}{\feiiuv}& { } &\multicolumn{3}{c}{NC(red)} & \multicolumn{3}{c}{BC(red)} & \multicolumn{3}{c}{VBC(red)}\\ \cline{4-12}\cline{13-14}\cline{16-18}\cline{19-21}\cline{22-24}
 
Object&$f_{\lambda,3000\AA}$&\multicolumn{1}{c}{$f_{\lambda,3646\AA}$}&\multicolumn{1}{c}{EW}&\multicolumn{1}{c}{I}&\multicolumn{1}{c}{FWHM}&\multicolumn{1}{c}{C($\frac{1}{4}$)}&\multicolumn{1}{c}{C($\frac{1}{2}$)}&\multicolumn{1}{c}{C($\frac{3}{4}$)}&\multicolumn{1}{c}{C($\frac{9}{10}$)}&\multicolumn{1}{c}{AI}&\multicolumn{1}{c}{KI}&\multicolumn{1}{c}{I}&\multicolumn{1}{c}{SF} &\multicolumn{1}{c}{\rfeUV}&\multicolumn{1}{c}{I} &\multicolumn{1}{c}{Shift}&\multicolumn{1}{c}{FWHM}&\multicolumn{1}{c}{I} &\multicolumn{1}{c}{Shift}&\multicolumn{1}{c}{FWHM}&\multicolumn{1}{c}{I}&\multicolumn{1}{c}{Shift}&\multicolumn{1}{c}{FWHM}\\

\multicolumn{1}{l}{(1)}& \multicolumn{1}{c}{(2)} & \multicolumn{1}{c}{(3)} & \multicolumn{1}{c}{(4)} & \multicolumn{1}{c}{(5)} & \multicolumn{1}{c}{(6)} &\multicolumn{1}{c}{(7)} & \multicolumn{1}{c}{(8)} & \multicolumn{1}{c}{(9)} & \multicolumn{1}{c}{(10)} & \multicolumn{1}{c}{(11)} &\multicolumn{1}{c}{(12)} & \multicolumn{1}{c}{(13)} &\multicolumn{1}{c}{(14)} &\multicolumn{1}{c}{(15)} &\multicolumn{1}{c}({16)} & \multicolumn{1}{c}{(17)} & \multicolumn{1}{c}{(18)} & \multicolumn{1}{c}({19)} &\multicolumn{1}{c}{(20)} & \multicolumn{1}{c}{(21)} &\multicolumn{1}{c}({22)} &\multicolumn{1}{c}({23)} &\multicolumn{1}{c}{(24)}\\

 \hline
 PHL 923  & 0.43 & 0.02 & 37 & 17.9 & 4000$\pm$320 & 120$\pm$300 & 20$\pm$180 & -10$\pm$180 & -10$\pm$150 & 0.05$\pm$0.12 & 0.41$\pm$0.05 & 25.7 & 4660 & 1.44 & 0.2 & -130 & 480 & 4.6 & 10 & 3370 & 7.6 & 1070 & 10610 \\
B2 0110+29 & 0.13 & 0.02 & 118 & 16.3 & 5810$\pm$450 & 250$\pm$280 & 190$\pm$160 & 170$\pm$150 & 160$\pm$210 & 0.02$\pm$0.09 & 0.42$\pm$0.05 & 10.9 & 2490 & 0.67 & 0.1 & 150 & 890 & 4.8 & 180 & 5130 & 5.5 & 940 & 13010 \\
3C 37  & 0.32 & 0.03 & 57 & 19.8 & 5600$\pm$440 & 250$\pm$280 & 210$\pm$160 & 200$\pm$150 & 200$\pm$210 & 0.02$\pm$0.09 & 0.42$\pm$0.05 & 29.1  & 4170 & 1.47 & 0.7 & 50 & 850 & 5.6 & 220 & 4910 & 7.2 & 650 & 13140 \\
PKS 0230-051 & 0.88 & 0.07 & 42 & 40.5 & 4470$\pm$360 & -40$\pm$280 & -130$\pm$130 & -150$\pm$120 & -160$\pm$160 & 0.03$\pm$0.13 & 0.40$\pm$0.05 & 65.7 & 5720 & 1.62 & 0.6 & 30  & 890 & 9.8 & -220 & 3750 & 18.5 & 750 & 12910 \\
3C 94  &  2.22 &  0.19 & 34 & 79.9 &  5000$\pm$410 & -60$\pm$320 & -160$\pm$140 & -180$\pm$130 & -190$\pm$180 & 0.03$\pm$0.09 & 0.40$\pm$0.05 & 247:: & 3560 & 3.09 & 2.2 & -30 &  750 & 18.8 & -270 & 4230 & 37.4 & 1050 &15460 \\
PKS 0420-01  & 0.86 & 0.07 & 40 & 33.9 & 3600$\pm$190 & 330$\pm$210 & 260$\pm$100 & 240$\pm$100 & 240$\pm$130 & 0.02$\pm$0.28 & 0.41$\pm$0.05 & 85.8 & 5720 & 2.53 & 0.5 & 50 & 870 & 8.4 & 260 &3010 & 15.1 & 1120 & 11140 \\
3C 179 & 0.40 & 0.03 & 40 & 16.7 & 4410$\pm$360 & 830$\pm$280 & 600$\pm$130 & 550$\pm$120 & 540$\pm$160 & 0.10$\pm$0.13 & 0.40$\pm$0.05 & 46.9 & 6840 & 2.81 & 0.8 & 120 & 830 & 4.6 & 460 & 3630 & 7.6 &1880 & 10000 \\
3C 380  & 1.36 & 0.16 & 61 & 85.3 & 4610$\pm$400 & 320$\pm$420 & 140$\pm$140 & 120$\pm$120 & 110$\pm$170 & 0.07$\pm$0.09 & 0.37$\pm$0.05 & 59 & 2490 & 0.69 & 1.9 & 20 & 850 & 16.7 & 110 & 3680 & 47.8 & 1070 & 14750 \\
S5 1856+73  & 1.31 & 0.16 & 39 & 51.2 & 5770$\pm$460 & 530$\pm$310 & 430$\pm$160 & 400$\pm$150 & 390$\pm$210 & 0.03$\pm$0.06 & 0.41$\pm$0.04 & 35.9 & 2490 & 0.70 & 2.8 & 60 & 900 & 13.1 & 410 & 4870 & 21.7 & 1100 & 11630 \\
PKS 2208-137 & 1.32 & 0.07 & 27 & 39.4 & 4440$\pm$350 & 300$\pm$240 & 240$\pm$120 & 230$\pm$120 & 220$\pm$160 & 0.04$\pm$0.07 & 0.42$\pm$0.04 & 45.3 & 2490 & 1.15 & 0.7 & 20 & 860 & 10.7 & 270 & 3810 & 15.2 & 870 & 11100 \\
PKS 2344+09 & 3.46 & 0.21 & 25 & 92.1 &4240$\pm$340 & 400$\pm$240 & 340$\pm$120 & 320$\pm$110 & 320$\pm$160 & 0.04$\pm$0.25 & 0.41$\pm$0.05 & 271 & 3020 & 2.94 & 5.8 & 0 & 890 & 29.2 & 270 & 3580 & 39.5 & 910& 11530 \\             
\hline
\end{tabular}
}
\\ \\
 {\raggedright \textbf{Note}: 
 {Cols. 2 and 3 are in units of  10$^{-15}$ ergs\,s$^{-1}$\,cm$^{-2}$\,\AA$^{-1}$. Col. 4 is in units of \AA. Cols. 5, 13, 16, 19, and 22 are in units of  10$^{-15}$ ergs\,s$^{-1}$\,cm$^{-2}$. Cols. 6 - 10, 14, 17 - 18, 20 - 21, and 23 - 24 are in units of \kms. Cols. 16\,-\,24 show measurements of the red component of \mgiionly\ doublet. Measurements of the blue component can be obtained by using the line ratios given in Sect. \ref{sec:uvr}.}} 

\label{tab:measmg}
\end{sidewaystable}

The multicomponent analysis corroborates the result obtained in the FP. In all cases, a VBC needs to be included (apart from the BC) to account for the observed \hb\, profile. 
The contribution of the VBC represents in 4/11 quasars more than 60\% of the total intensity of \hbFP{} (for the whole sample the median value for VBC contribution is 57\%) and that reaches FWHM(VBC)\,$\gtsim$\,10000\,\kms. As a consequence of the presence of the VBC,
FWHM(\hbbc) is always smaller than the one corresponding to the  FP. The ratio between both $\xi$\,=\, FWHM(\hbbc)/FWHM(\hbFP) has a mean value of 0.83\,$\pm$\,0.09 for our quasars, in agreement with the values reported by \citet{2013A&A...555A..89M} for an SDSS composite spectrum of sources belonging to Pop. B, B1 ST quasars. Also, a narrow component is detected in all quasars, although in general, its intensity relative to the broad profile is small, with a median value of 6\%. Considering the individual components of \hb, both the NC and BC are almost always unshifted and centered around the rest-frame of each object. In 3C 94 and PKS 2344+09 the NC appears slightly blueshifted. Two factors may be contributing to the shift: the first one and more probable is the possible presence of an unresolved SBC included in the NC of \hb, in correspondence with the observed SBC of  \oiiionly{} (see below), which would also explain the relatively high value obtained for the FWHM \hbnc\, $\sim$ 1600 \kms{} for 3C\,94. A second factor to take into account in the case of 3C 94, as we mentioned previously, 
is that both the \hb\ profile and the \feii\ blue blend are affected by 
a telluric band extending between $\approx$ 9250\AA\ and 9700\AA\ in observed wavelengths (that correspond to $\sim$ 4700\AA\, - \,4900\AA\ in the rest-frame), whose correction can also introduce greater uncertainty in the determination of the position of the peak and the centroids at the highest fractional intensities.

A similar analysis was carried out for  \oiiionly, as described in Sect. \ref{subsec:OR}. The results for the FP and the individual components for \oiiiseven\AA\ are reported in Table \ref{tab:valueso3}. We present the EW (Col. 2), total intensity (Col. 3), FWHM (Col. 4), centroid velocities at different fractional intensities (Cols. 5 -- 10), AI (Col. 9), and KI (Col. 10). Intensity, shift, and FWHM from the {\tt specfit} analysis are also presented for the  NC (Cols. 11 -- 13), and the SBC (Cols. 14 -- 16) of each quasar. The resulting measurements for  [OIII]$\lambda$4959\AA\ can be found by using appropriate line ratios detailed in Sect. \ref{subsec:OR}.

\begin{table*}
\setcounter{table}{4}
\centering
\caption{Results of FP and the {\tt specfit} analysis for [OIII]$\lambda$5007\AA.} 
\scalebox{0.98}{
\setlength{\tabcolsep}{3.5pt}
\begin{tabular}{l c c c r r r r c c c c r r c r r r} 
\hline\hline 

\multicolumn{1}{c}{Object} & \multicolumn{9}{c}{Full broad profile (FP)(NC + SBC)} && \multicolumn{3}{c}{NC} && \multicolumn{3}{c}{SBC} \\ 
\cline{2-10} \cline{12-14} \cline{16-18}
& \multicolumn{1}{c}{EW} & I & \multicolumn{1}{c}{FWHM} & \multicolumn{1}{c}{C($\frac{1}{4}$)} & \multicolumn{1}{c}{C($\frac{1}{2}$)} & \multicolumn{1}{c}{C($\frac{3}{4}$)} & \multicolumn{1}{c}{C($\frac{9}{10}$)} & AI & KI && I & \multicolumn{1}{c}{Shift} & \multicolumn{1}{c}{FWHM} && \multicolumn{1}{c}{I} & \multicolumn{1}{c}{Shift} & \multicolumn{1}{c}{FWHM} \\

\multicolumn{1}{c}{(1)} & \multicolumn{1}{c}{(2)} & \multicolumn{1}{c}{(3)} & \multicolumn{1}{c}{(4)} & \multicolumn{1}{c}{(5)} & \multicolumn{1}{c}{(6)} & \multicolumn{1}{c}{(7)} & \multicolumn{1}{c}{(8)} & \multicolumn{1}{c}{(9)} & \multicolumn{1}{c}{(10)} && \multicolumn{1}{c}{(11)} & \multicolumn{1}{c}{(12)} & \multicolumn{1}{c}{(13)} && \multicolumn{1}{c}{(14)} & \multicolumn{1}{c}{(15)} & \multicolumn{1}{c}{(16)} \\
 \hline
PHL 923  & 64 & 8.9 & 860$\pm$80 & -40$\pm$50 & -50$\pm$30 & -50$\pm$30 & -50$\pm$40& 0.04$\pm$0.12 & 0.42$\pm$0.04 && 4.99 & -30 & 806 && 3.86 & 10 & 1640 \\
B2 0110+29 & 117 & 7.9 & 480$\pm$40 & -40$\pm$20 & -40$\pm$10 & -40$\pm$10 & -40$\pm$20& -0.04$\pm$0.09 & 0.42$\pm$0.05 && 5.24 & -20 & 430 && 2.7 & -60 & 1200 \\ 
3C 37  & 41 & 5.6  & 580$\pm$50 & -100$\pm$40 & -60$\pm$20 & -50$\pm$20 & -40$\pm$20 & -0.17$\pm$0.09 & 0.38$\pm$0.04 && 2.41 & -20 & 440 && 3.23 & -150 & 980  \\ 
PKS 0230-051  & 37 & 11.4 & 450$\pm$40 & -40$\pm$30 & 20$\pm$20 & 30$\pm$10 & 40$\pm$10 & -0.21$\pm$0.13 & 0.37$\pm$0.04 && 4.94 & 40 & 340 && 6.5 & -130 & 690  \\
3C 94 & 57 & 49.1 & 620$\pm$60 & -70$\pm$80 & 70$\pm$20 & 90$\pm$20 & 90$\pm$20 & -0.29$\pm$0.20 & 0.32$\pm$0.05 && 22.0 & 110 & 490 && 27.1 & -240 & 1380   \\ 
PKS 0420-01  & 14 & 9.7 & 620$\pm$50 & -140$\pm$50 & -130$\pm$20 & -130$\pm$20 & -120$\pm$30 & -0.02$\pm$0.28 & 0.39$\pm$0.05 && 4.56 & -110 & 500 && 5.1 & -160 & 1280  \\ 
3C 179  & 27 & 4.8 & 520$\pm$40 & 10$\pm$30 & 20$\pm$10 & 30$\pm$10 & 30$\pm$20 & -0.09$\pm$0.13  & 0.41$\pm$0.05 && 2.74 & 20 & 440 && 2.1 & -80 & 940  \\ 
3C 380 & 50 & 36.1 & 610$\pm$50 & -30$\pm$40 & -10$\pm$20 & -10$\pm$20 & -10$\pm$20 & -0.04$\pm$0.09 & 0.38$\pm$0.05 && 14.4 & 10 & 470  && 21.7 & -60 & 1070 \\ 
S5 1856+73  & 34 & 25.3 & 410$\pm$30 & 10$\pm$20 &  1$\pm$10 & 1$\pm$10 & 10$\pm$20 & 0.05$\pm$0.06& 0.44$\pm$0.05 && 18.3 & 30 & 370 && 7.0 & 120 & 760    \\ 
PKS 2208-137  & 7 & 3.4 & 600$\pm$40 & -70$\pm$20 &  -60$\pm$20 & -50$\pm$20 & -50$\pm$20& -0.12$\pm$0.07 & 0.45$\pm$0.04 && 2.43 & 20 &  530 && 1.0 & -210 &  620 \\ 
PKS 2344+09  & 63 & 85.3 & 660$\pm$60 & -270$\pm$80 & -160$\pm$20 & -140$\pm$20 & -130$\pm$30 & -0.23$\pm$0.25 & 0.34$\pm$0.05 && 39.6 & -100 & 520 && 45.7 & -400 & 1300 \\  
\hline
\end{tabular}
}
\vspace{0.05truecm}

{\raggedright \textbf{Note}:
{ Col. 2 is in unit of \AA. Cols. 3, 11, and 14 are in units of $10^{-15}$ erg\ s$^{-1}$cm$^{-2}$. Cols. 4 -- 8, 12 -- 13, and 15 -- 16 are in units of \kms. \par}}
\label{tab:valueso3}
\end{table*}

Our sources show strong  \oiiionly, lines clearly separated from \hb, presenting a spectrum characteristic of Pop. B quasars as listed in Table \ref{tab:valueso3}. The FP of  \oiiionly{} shows a slight blue asymmetry for most of the objects, with a negative AI and a median value of -0.1. This is due to the presence of a weaker SBC that is blueshifted (for 9/11 of our sources). As can be seen in Table \ref{tab:valueso3} where the {\tt specfit} measures are presented, while the NC is, within the uncertainties, unshifted in the rest-frame of the objects, the SBC appears blueshifted with a median shift of -130 \kms, and a slightly broader FWHM of around 1100 \kms. The presence of blue-shifts in HILs like  \oiiionly{} is considered as one of the main detectors of outflowing gas \citep[e.g.,][]{2002ApJ...576L...9Z,2008ApJ...680..926K,2011ApJ...737...71Z,2016Ap&SS.361....3M,2020A&A...644A.175V,2022AN....34310084D, 2022A&A...659A.130K}. Our quasars present moderate   \oiiionly\ blueshifts, indicating perhaps the presence of outflows in the inner NLR. In no case do we have the named "blue-outlier", which is defined by a blue-shift in  \oiiionly\ larger than -250 \kms\ \citep{2002ApJ...576L...9Z} and are preferentially observed at low-z in the Pop. A quasars of the MS. 

In some objects \heii\AA\ is also detected in the blue side of \hb\ as a residual emission in the fit. We have fitted it with a Gaussian profile. In the majority of the cases, it seems to correspond to a VBC, but in some cases, it is not clear whether this extra emission is actually \heii\AA\ or corresponds to the \feii\ blend. Only in the case of 3C 380 two components (BC and VBC) can be identified, and for PKS 0230-51 a NC is clearly seen overlaid on the VBC (see Fig. \ref{fig:f3}). The results of the {\tt specfit} analysis for \heii\AA\ are reported in Table \ref{tab:tabheii}. The values given in the table are only to be considered, for most objects, as an indication of the \heii\AA\ detection.

\begin{table}
\setcounter{table}{5}
\setlength{\tabcolsep}{1.pt}
\caption{{\tt Specfit} result of HeII$\lambda$4686\AA.}       
\centering     
\scalebox{0.98}{                   
\begin{tabular}{@{\extracolsep{3pt}}lcccccc@{}}    
\hline\hline   

 Object&\multicolumn{3}{c}{BC}&\multicolumn{3}{c}{VBC}  \\\cline{2-4}\cline{5-7}
 & I & Shift &FWHM& I&Shift&FWHM \\

  (1)&(2)&(3)&(4)&(5)&(6)&(7) \\
\hline
PHL 923 & 1.5 & 160 & 3190 & -- & -- & -- \\
B2 0110+29 & -- & -- & -- & 1.0 & 2070 & 9630 \\
3C 37  & -- & -- & -- & 4.0 & 990 & 9600 \\
PKS 0230-051 & -- & -- & -- & 8.7 & 2180 & 10310 \\
3C 94 & -- & -- & -- & -- & -- & -- \\
PKS 0420-01 & -- & -- & -- & 13.2: & 2410: & 9730: \\
3C 179  & -- & -- & -- & 3.9 & 1510 & 6660: \\
3C 380  & 4.5 & -420 & 3070 & 12.7 & 3670 & 8470 \\
S5 1856+73 & 5.3 & 1200 & 4060 & -- & -- & -- \\
PKS 2208-137 & -- & -- & -- & 4.6 & 540 & 7110 \\
PKS 2344+09 & 8.2 & -260 & 4760 & -- & -- & -- \\
\hline                                   
\end{tabular}
}\\
{\raggedright \textbf{Note}: Cols. 2 and 5 are in units of 10$^{-15}$ ergs\ s$^{-1}$cm$^{-2}$. Cols. 3, 4, 6, and 7 are in \kms.  For (:)  measurements see the notes in Table \ref{tab:meashb}. \par}
\label{tab:tabheii}
\end{table}

\subsubsection{\mgiionly\ region}
\label{sec:resmg}

The results of the spectrophotometric measurements for \mgiionly\ region and measured parameters including both \mgiionly\ FP and the individual components from the {\tt specfit} analysis (plotted in left panels of Fig. \ref{fig:f3}), following the approach described in Sect. \ref{sec:uvr}, are presented in Table \ref{tab:measmg}. Cols. 2 and 3 contain respectively the rest-frame specific continuum flux at 3000\AA, obtained from the fitted power law, and the Balmer continuum whose intensity was estimated at the Balmer edge at 3646\AA. From Cols. 4 -- 12 are reported the FP measurements of \mgiionly, including 2BC and a VBC: EW (Col. 4), total intensity (Col. 5), FWHM of \mgiionlyFP\ (Col. 6), the centroids at $\frac{1}{4}$, $\frac{1}{2}$, $\frac{3}{4}$ and $\frac{9}{10}$ fractional intensity of the peak (Cols. 7 -- 10), as well as the AI and KI in Cols. 11 and 12, respectively. The {\tt specfit} fitting parameters of the individual components of the reddest line of the doublet are reported from Cols. 16 -- 24. The corresponding measures of the blue component of \mgiionly, can easily be found by taking into account the appropriate line ratio (see  Sect.\ref{sec:uvr}). For each red line component analyzed 
we report the total flux (I), the peak shift, and the FWHM of the NC  (Cols. 16 -- 18), BC (Cols. 19 -- 21), and VBC (Cols. 22 -- 24). In addition, we also provided the parameters related to \feiiuv: the total flux of the blue blend (Col. 13), and the smoothing factor applied to \feiiuv\, template to reproduce the observed \feii\, feature (Col. 14). In Col. 15 we report the \rfeUV\, parameter, i.e. the ratio between the intensities of \feiiuv\ (in the 2200\AA\, to 3090\AA\, range) and \mgiionly\  (\rfeUV\ = I(\feiiuv)/I(\mgiionlyFP)).

As in the case of \hb, the FWHM of \mgiionly{} FP is higher than FWHM of \mgiionly{} BC  obtained from the {\tt specfit} analysis. The median value for the ratio $\xi_{\mathrm{\mgiionly}}$ = FWHM(\mgiionlybc)/FWHM(\mgiionlyFP)  is 0.84$\pm$0.02 for our quasars, in good agreement with the value found by \citet{2013A&A...555A..89M} for the ST B1 by using composite spectra. This is due to the presence also of a VBC in \mgiionly, with a FWHM $\gtrsim$\,10000\,\kms, though \mgiionly\ shows a weaker VBC, with a median flux representing about the 42\% of the total broad intensity.  This manifests itself in a redward asymmetry, although showing a more symmetric profile than \hb, with <AI> = 0.04, and a shift towards the red of the centroid velocities of \mgiionly, more pronounced towards the base of the line, with <c($\frac{1}{4}$)> $\approx$ 300\,\kms.           

In addition, Table \ref{tab:tabuv} presents the parameters obtained from the fit to a single Gaussian profile of other lines detected in the UV region, such as the OIII$\lambda$3133\AA, HIL [NeV]$\lambda$3426\AA\, and [OII]$\lambda$3728\AA\, doublet.  For each of these lines the total flux, the peak shift, and FWHM are reported when the corresponding line is detected in the spectrum. 

\begin{figure}
\centering
\includegraphics [scale=0.4]{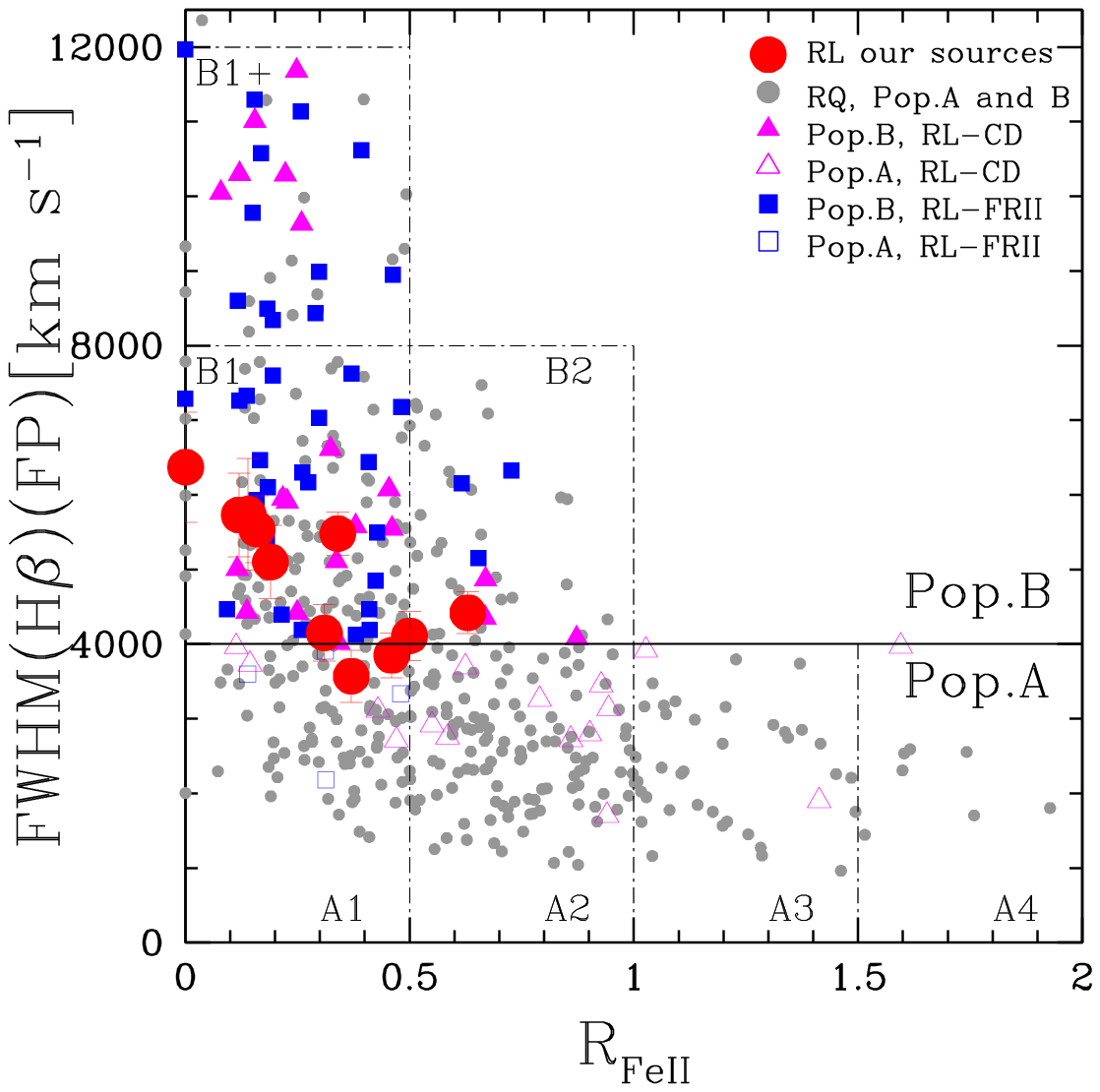}
\vspace{-0.2cm}
\caption{Location of our  eRk quasars (large red solid circles) in the optical plane of the 4DE1 space, traced by the measures from the \citet{2010MNRAS.403.1759Z} sample,  where grey dots represent the RQ (Pop. A \& B) and blue and magenta symbols corresponds to RLs. The horizontal line at 4000\,\kms\ marks the nominal Pop. A\,-\,B boundary. For the explanation of different STs see Sect. \ref{sec:op}. The vertical axis is truncated at 12500\,\kms\, for clarity.}

\label{fig:optplane}
\end{figure}

\subsection{Main Sequence optical plane}
\label{sec:op}

The optical plane of the 4DE1 parameter space is defined by the \hbFP\ FWHM  and  \rfeopt. The quasar MS allows contextualization of the observed empirical spectroscopic properties of type I AGN and their connection with the physical conditions of the BLR \citep[see e.g.,][]{2000ARA&A..38..521S,2018FrASS...5....6M,2019ApJ...882...79P,2020MNRAS.492.3580W}.
After performing the analysis of the \hb\ region (see section \ref{sec:reshb}), that provides us with both the FWHM(\hbFP) and \rfe\ (Table \ref{tab:meashb} Cols. 6 and 15, respectively), we can locate our quasars in the MS. It is possible to subdivide the optical plane into a grid of bins (or STs) formed from FWHM(\hbFP) and \rfeopt\ that shows different spectral line profile properties \citep{2002ApJ...566L..71S}: bins A1 to A4 are defined in terms of increasing \rfeopt\ with bin size $\Delta$\rfeopt\ = 0.5 (from A1 with \rfe\,<\,0.5 to A4 with 1.5\,<\,\rfe\,$\leq$2), while bins B1, B1$^{+}$ and B1$^{++}$ are defined in terms of increasing FWHM with $\Delta$FWHM=\,4000\kms, from B1 with 4000\,$<$\,FWHM$\leq$\,8000\,\kms\ to B1$^{++}$ with FWHM\,>\,12000\,\kms.

Fig. \ref{fig:optplane} shows the locus of our quasars on the MS as well as the Pop. A/B separation and the identification of some of the most populated STs, where we have also included as a comparison the low-z SDSS sample by \citet{2010MNRAS.403.1759Z}. 
 In Fig. \ref{fig:optplane},  grey symbols represent the Pop. A and B RQs, meanwhile coloured symbols represent 
  the RLs, splited into core-dominated (CD; magenta triangles) and lobe-dominated (LD) FRII (blue squares).  As clearly seen in this figure, the RQ sources show a broad domain in the MS, covering the entire range of observed values in both FWHM and \rfe, and in both A and B populations. On the converse, the RL (“jetted”) sources are mainly Pop. B and populate particular bins of the optical plane, mostly in B1 and B$^{+}$  \citep{2008MNRAS.387..856Z, 2021ApJ...922...52K}. 

Our quasars are well located in the MS domain of the CD RL sources.  
The majority lie in the B ST (more than 80\%), mostly in B1. For 3C 179 and PKS 2208, 
FWHM(\hbFP) value is only marginally below 4000\,\kms\, placing them on the upper edge of Pop. A, at the ST A1. Since in both quasars there is no doubt about its Pop. B classification with a clearly recognized VBC in the \hb{} profile, its position may be due to an inclination effect (as we mentioned in section \ref{sec:reshb}) since  the broadening of \hb\ is orientation-dependent \citep{1986ApJ...302...56W,2003ApJ...597L..17S,2008MNRAS.387..856Z}. 
Our quasars also show weak \feiiopt\ intensities. All except one case (PHL 923 with \rfeopt\ = 0.63 that locates the object in B2 bin) present low \rfeopt\ values, with a mean of 0.29 and a $\sigma$ = 0.19. This is in agreement with the results obtained by \citet{2021Univ....7..484M} who using composite spectra found that RL sources, both CD and FRII, present weaker \feiiopt\ emission ($\sim$ a factor 2 lower) compared to the composite spectrum of RQ quasars sharing the same B1 ST.

\begin{table}
\setlength{\tabcolsep}{1.0pt}
\setcounter{table}{7}
\caption{{\tt Specfit} analysis results for the other UV lines} %
\scalebox{0.96}{
\begin{tabular}{@{\extracolsep{2pt}}l c c c c c c c c c} 
\hline\hline 
\multirow{1}{*}{Object} &\multicolumn{3}{c}{OIII$\lambda3133$\AA\,  }&\multicolumn{3}{c}{[NeV]$\lambda3426$\AA\,}&\multicolumn{3}{c}{[OII]$\lambda3728$\AA} \\ \cline{2-4}\cline{5-7} \cline{8-10}
& I & \multicolumn{1}{c}{Shift} & \multicolumn{1}{c}{FWHM} & \multicolumn{1}{c}{I} & \multicolumn{1}{c}{Shift} & \multicolumn{1}{c}{FWHM} & \multicolumn{1}{c}{I} & \multicolumn{1}{c}{Shift} & \multicolumn{1}{c}{FWHM} \\ 
(1)&\multicolumn{1}{c}{(2)} & \multicolumn{1}{c}{(3)} & \multicolumn{1}{c}{(4)} & \multicolumn{1}{c}{(5)} & \multicolumn{1}{c}{(6)} & \multicolumn{1}{c}{(7)} & \multicolumn{1}{c}{(8)} & \multicolumn{1}{c}{(9)} &\multicolumn{1}{c}{(10)} \\
 \hline
PHL923 & --  & -- & -- & -- & -- & -- & 1.0 & 170: & 550: \\
B2 0110+29 & -- & -- & -- & -- & -- & -- & 1.8 & 70 & 1130 \\
3C37 & 1.2 & 130: & 1880: & 1.0 & -70  & 1550 & 0.4: & 70 & 1090:: \\
PKS 0230-051 & 4.1 & -110 & 1940 & 2.1: & -260: & 1170: & 2.6 & -40 &  1080 \\
3C94 & 4.4 & 110 & 1950 & 6.1 & 60 & 1560 & 6.1 & 190 & 1030 \\
PKS 0420-01 & -- & -- & -- & -- & -- & -- & 2.2:& 30: & 1150: \\
3C179 & 0.1:: & 80:: & 1680:: & 0.1: & 20: & 940: & 0.1 & -20 & 1030 \\
3C380 & 4.1:: & -310:: & 1910:: & 5.0  & 40 & 1700 & 3.4 & 60 & 1070 \\
S5 1856+73 & 3.0:: & -150:: & 2150:: & 1.8 & -1 & 600 & 3.4: & -110: & 1130:  \\
PKS 2208-137 & -- & -- & -- & -- & -- & -- & 2.6 & 80 & 1090  \\
PKS 2344+09 & 10.4:: & -110::& 2380:: & 5.9 & -70 & 1420 & 7.0: & 200:&1140: \\

\hline
\end{tabular}
}

 {\raggedright \textbf{Note}: 
 {Cols. 2, 5 and 8 are in units of 10$^{-15}$ ergs s$^{-1}$ cm$^{-2}$. Other columns are in \kms. For (:) and (::) measurements see the notes in Table \ref{tab:meashb}. }}
 \label{tab:tabuv}
\end{table}

\begin{figure}
\centering
\includegraphics [height=7cm]{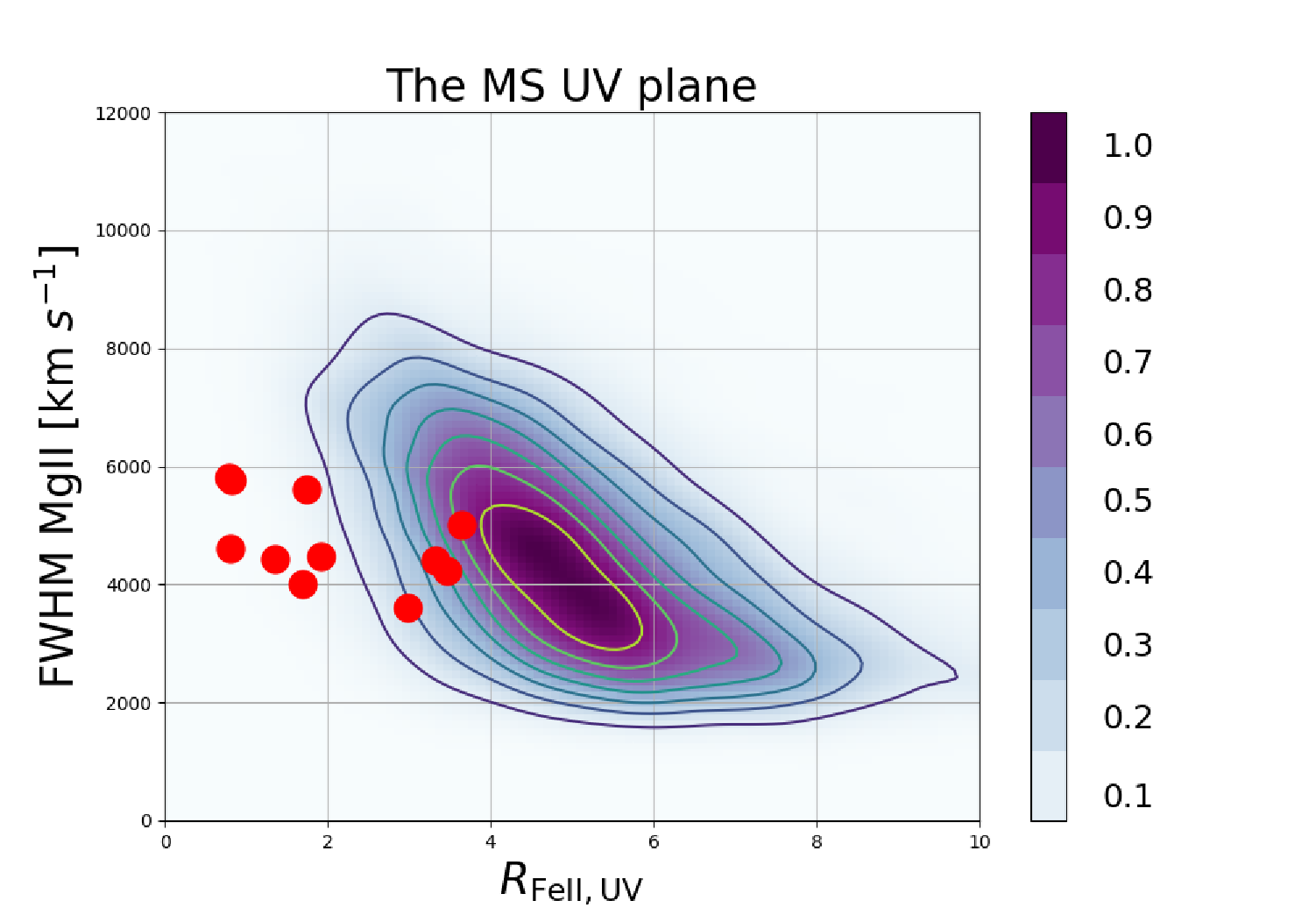}
\vspace{-0.5cm}
\caption{Location of our eRK quasars (red solid circles) in the UV plane defined by \mgiionly{} FWHM and \rfeUV{}. A comparison sample was taken from \citet{2017MNRAS.472.4051C} catalogue  by considering sources that are found around our redshift range ($0.4 \lesssim z \lesssim 1$) and is shown after kernel smoothing to account for a large number of sources (10,344). The color scale is normalized to the peak density in the UV parameter plane. 
} 
\label{fig:denscald}
\end{figure}

\subsection{Main sequence UV plane}
\label{sec:uvp}

In a close analogy to the optical plane, it is also possible to analyse the UV plane by using \mgiionly,  and \feiiuv\ line parameters. This is possible mainly due to the fact that both \hb\ and \mgiionly\ lines belong to LILs and are supposed to be emitted from a similar region \citep{1988MNRAS.232..539C}. The UV plane is also formed by using \mgiionlyFP\ FWHM and the strength of the \feiiuv.  Figure \ref{fig:denscald} shows the location of our eRK quasars in the UV plane where we used as a comparison sample the recently available parameters determined by the software  {\sc{QSFit}} from \citet{2017MNRAS.472.4051C} using SDSS-DR10 spectra, in which they incorporated spectral information in the UV, and in particular for \mgiionly\, and \feiiuv\, to produce a publicly available catalogue of AGN spectral properties. See also a discussion about this catalogue and the MS UV plane in \citet{2020ApJ...900...64S}. For the {\sc{QSFit}} catalogue we selected a sample consisting of the sources in the redshift range from 0.4 to 1 and with good quality measures in \mgiionly{} and \feiiuv\, according this catalogue.  The available comparison sample is large in number (10,344 spectra) and we used therefore a kernel smoothing for better visualization. As compared to the optical plane, the UV plane shows a larger range in \rfeUV. This is mainly due to the fact that the range  of the integrated flux in \feiiuv{} is broader as compared to the optical, \feiiopt.  For the sole purpose of the representation of our sample in the UV plane together with the data obtained from the {\sc{QSFit}} catalogue, we have estimated a \feiiuv\ modified total flux for our RL sources by  extrapolating the \feiiuv\  model fitted to our spectra to the wavelength range $\lambda$1250\AA\ to $\lambda$3090\AA\ used by  \citet{2017MNRAS.472.4051C}. Our quasars are placed in the FWHM(\mgiionlyFP) $\gtrsim$ 4000\,\kms\  and in the lowest \feiiuv{} emission, as in the case of the optical plane. 

\begin{figure}
\centering
\includegraphics[width=0.87\columnwidth]{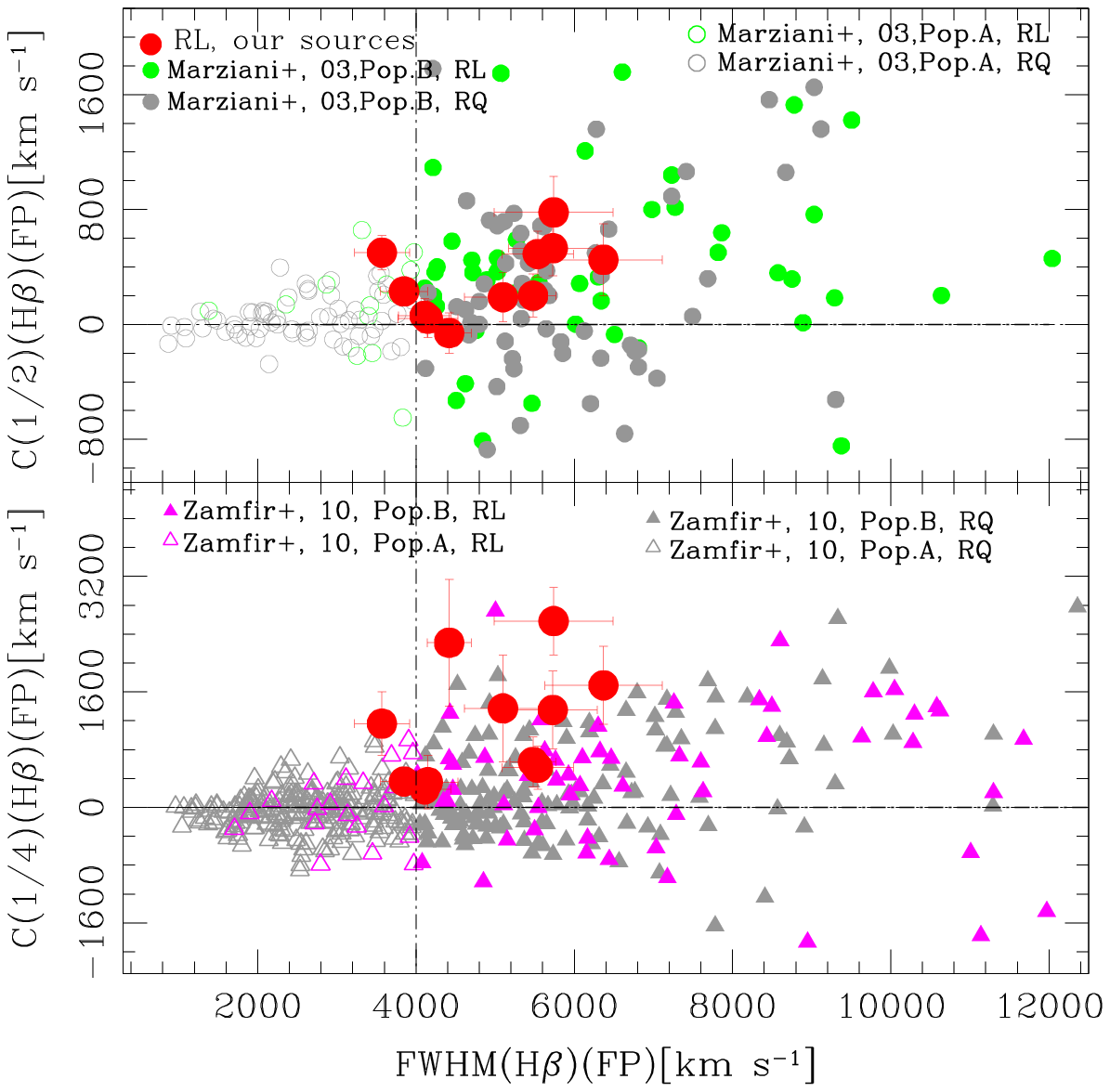}

\caption{ Centroids at \chm\,(upper plot) and \cqm\,(bottom plot) peak intensity of \hbFP{} versus the FWHM of \hbFP{}. The large red solid circles represent our RL objects. Comparison samples from \citet{2003ApJS..145..199M} and \citet{2010MNRAS.403.1759Z} are also represented.  
The legends identify  different populations and radio classes. 
The vertical dot-dashed line at 4000 \kms\ marks the nominal population A/B boundary. The horizontal dot-dashed line traces the symmetric line in  \chm{}  and \cqm.
}
\label{fig:chcqfwhmhb}
\end{figure} 

\begin{figure}
\centering
\includegraphics[width=0.85\columnwidth]{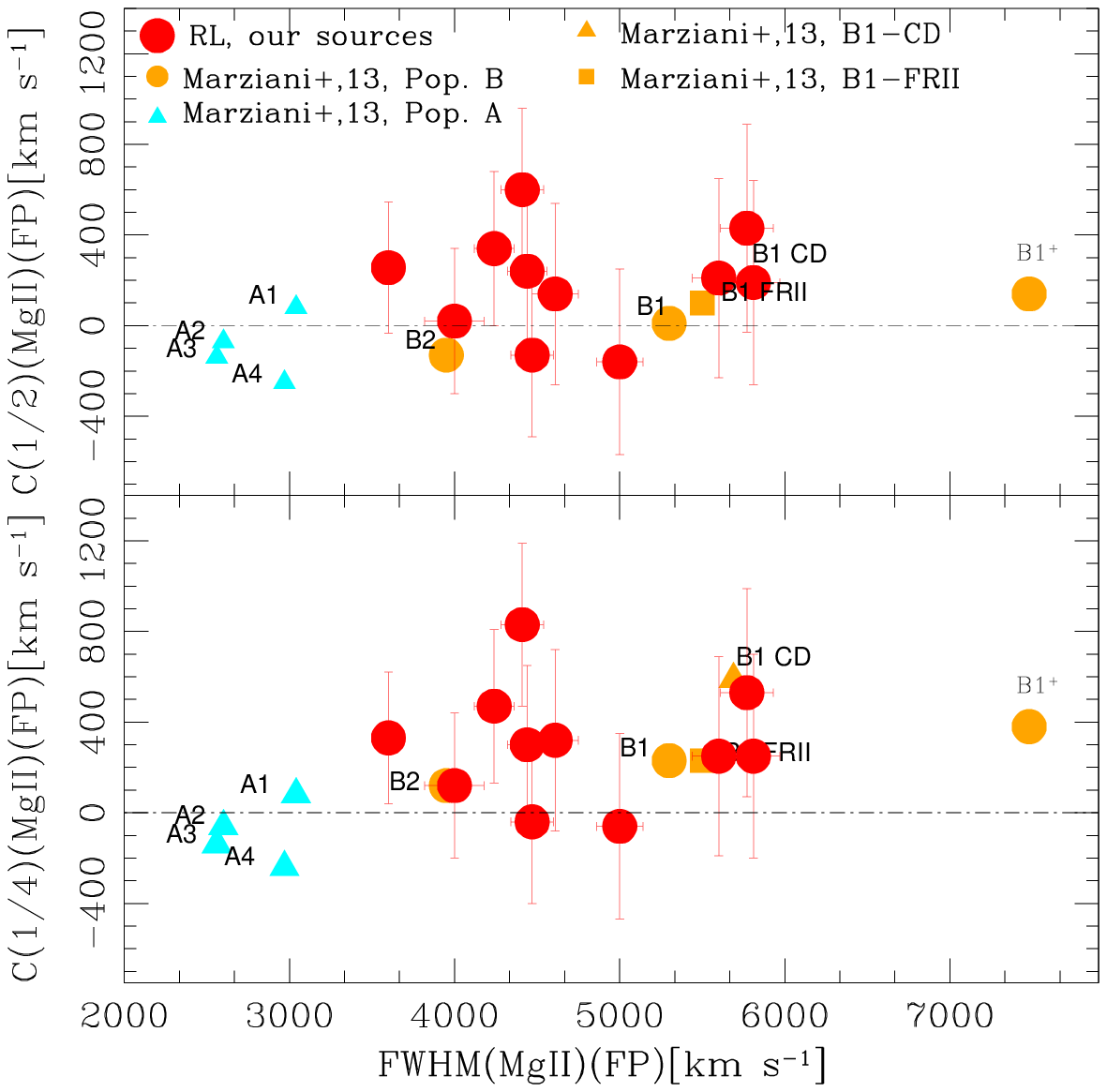}
\caption{Relation between \chm\,(upper) and \cqm\, (lower) versus FWHM of \mgiionly. Red solid circles from our RL spectra. The  additional comparison sample shown in the plot was taken from \citet{2013A&A...555A..89M}. 
 Orange symbols represent Pop. B and cyan solid triangles for Pop. A. Orange solid triangle and square represent Pop. B CD and FRII respectively.
}
\label{fig:chcqfwmg}
\end{figure} 

\subsection{Comparison between the \hb{} and \mgiionly\ spectral ranges}
\label{sec:comp}

\subsubsection{ \mgiionly{} and \hb\ line profiles}

Based on the measurements reported in Tables \ref{tab:meashb} and \ref{tab:measmg}, we analyse and compare here the line profile parameters of \hb\ and \mgiionly. The upper and bottom panels of Fig. \ref{fig:chcqfwhmhb} show the location of our eRk quasars in the c($\frac{1}{2}$) and c($\frac{1}{4}$) versus FWHM(\hbFP) plots, respectively. The comparison sample containing the c($\frac{1}{2}$) information for the RL and RQ was taken from \citet{2003ApJS..145..199M} and the c($\frac{1}{4}$) from \citet{2010MNRAS.403.1759Z}.  
Our  eRk quasars are located in the upper right part of both diagrams, mainly occupied by Pop. B quasars, but showing more extreme shift values in the c($\frac{1}{4}$) when compared to \citet{2010MNRAS.403.1759Z} sample. Fig. \ref{fig:chcqfwhmhb} also shows that our quasars follow the trend observed in the other two samples in the sense that larger velocity centroids, particularly in c($\frac{1}{4}$), correspond to wider FWHM(\hbFP). Similarly, Fig. \ref{fig:chcqfwmg} shows the centroid at c($\frac{1}{2}$) (upper plot) and c($\frac{1}{4}$) (lower plot) peak intensity of the \mgiionly{} profile versus the FWHM of \mgiionlyFP\ for our quasars and for the composite spectra by \citet{2013A&A...555A..89M} corresponding to the different STs of the MS, with cyan solid triangles denoting Pop. A  and orange-filled circles for Pop. B. Also in MgII our eRk objects are located in the Pop. B region and showing a larger shift towards the red in the base of the line, although it is less pronounced than in \hb.

One of the first differences between the two lines is observed in the broadening estimator, FWHM of the FP, in which the FWHM(\mgiionlyFP) is narrower than FWHM(\hbFP): the median value of \mgiionlyFP\  FWHM is 4470\,\kms, about 10\%\ less than the FWHM of \hbFP, 5100\,\kms{}. Since for the FWHM measurements of \mgiionlyFP, a single unresolved line is assumed for the doublet, those values can be converted to the FWHM of a single component by subtracting 300\,\kms{} \citep{2012MNRAS.427.3081T}. Taking into account this correction, we obtained a median value for the FWHM of \mgiionly\ single component of $\approx$  4170 \kms, that corresponds to  $\approx$ 0.84$\pm$0.12 the median \hbFP\ FWHM, of our eRK quasars, in agreement with previous results obtained by \citet{2009ApJ...707.1334W},  claiming that FWHM of \hb\ is larger than FWHM of \mgiionly, and later clearly identified by \citet{2013A&A...555A..89M} for the Pop. B quasars.  As explained in \citet{2013A&A...555A..89M},  \mgiionly\ might be emitted predominantly farther out from the central continuum source than \hb. This means only part of the gas emitting \hb\ is emitting \mgiionly. For instance, the innermost regions moving with the largest velocities close to the \hb\ line base could be too highly ionized to emit \mgiionly.  This is quite true as much of the broad line emission's kinematics can be dominated by Keplerian motion  \citep{1999ApJ...521L..95P, 2000ApJ...540L..13P}. 

\begin{figure}
\centering 
\includegraphics[width=0.95\columnwidth, height=0.95\columnwidth] {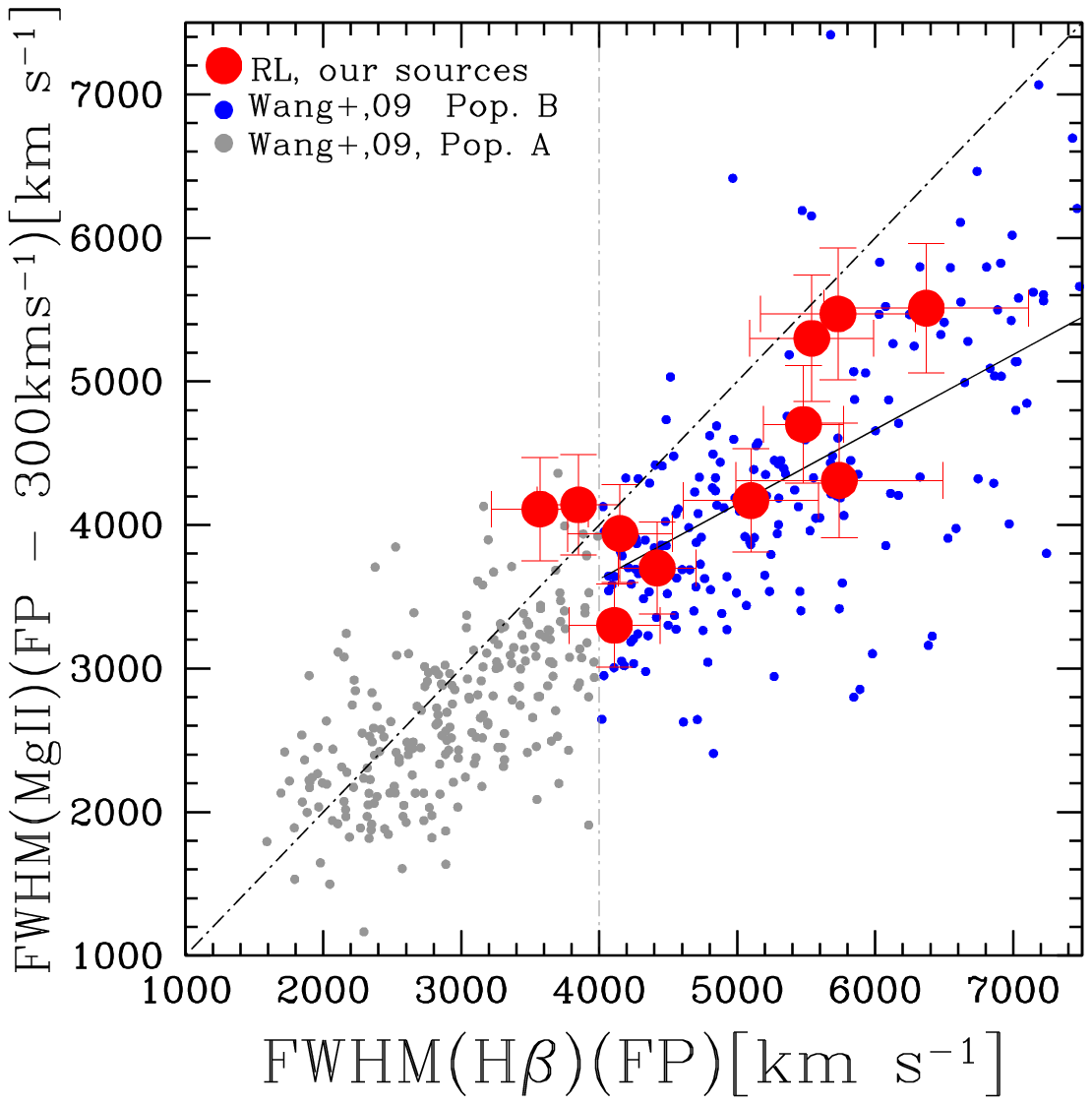}
\caption{Comparison between \mgiionly{} and \hb{} FWHM. The ordinate is the FWHM of \mgiionlyFP{} subtracted by 300\,\kms. 
A comparison sample was taken from \citet{2009ApJ...707.1334W} that we subdivided into Pop. A (gray dots) and Pop. B (blue dots) by using the 4000 \kms{} as a separation limit. The solid line represents the correlation of both FWHM including only Pop. B sources from \citet{2009ApJ...707.1334W} and our quasars. The two  values are highly correlated with a Pearson's correlation factor $r \approx 0.78$  and p-value $\approx$\,0.005.} 
\label{fig:fwhmfwhm}
\end{figure}

Figure \ref{fig:fwhmfwhm} shows the relation between FWHM of one single component of \mgiionly\ and \hbFP, along with a  comparison sample built by \citet{2009ApJ...707.1334W}. This figure shows that the correlation deviates from the one-to-one line as the profile widths are affected by the  VBC present in both \hb\ and \mgiionly, significantly more pronounced in \hb, (see Sect. \ref{sec:reshb} and \ref{sec:resmg}). This deviation is larger for higher FWHM(\hb) and in particular for FWHM\,>\,4000\,\kms\ corresponding to Pop. B sources where VBC is detected.

\begin{figure*}
\centering 
\includegraphics [width=0.32\textwidth]{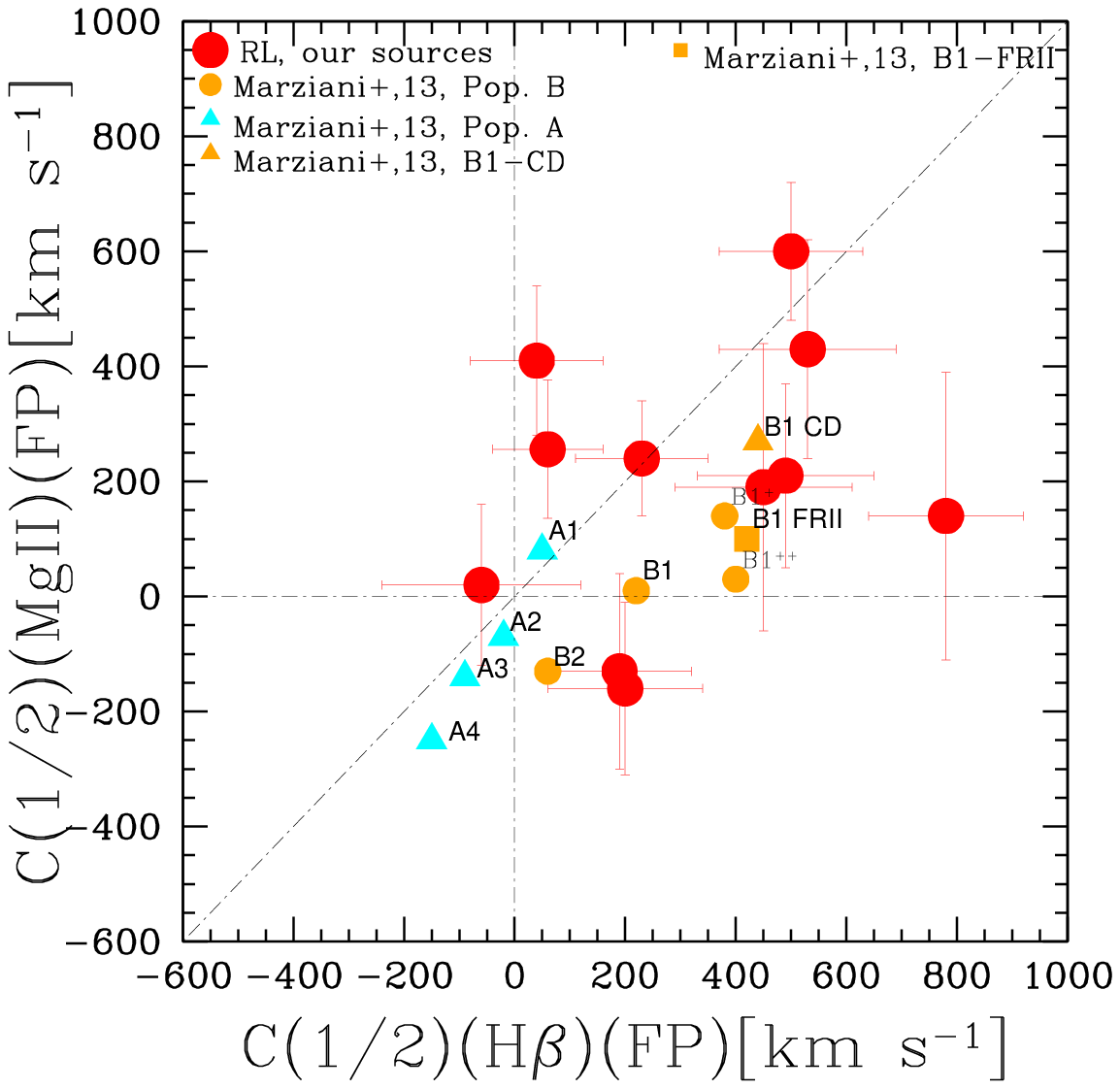}
\includegraphics [width=0.32\textwidth]{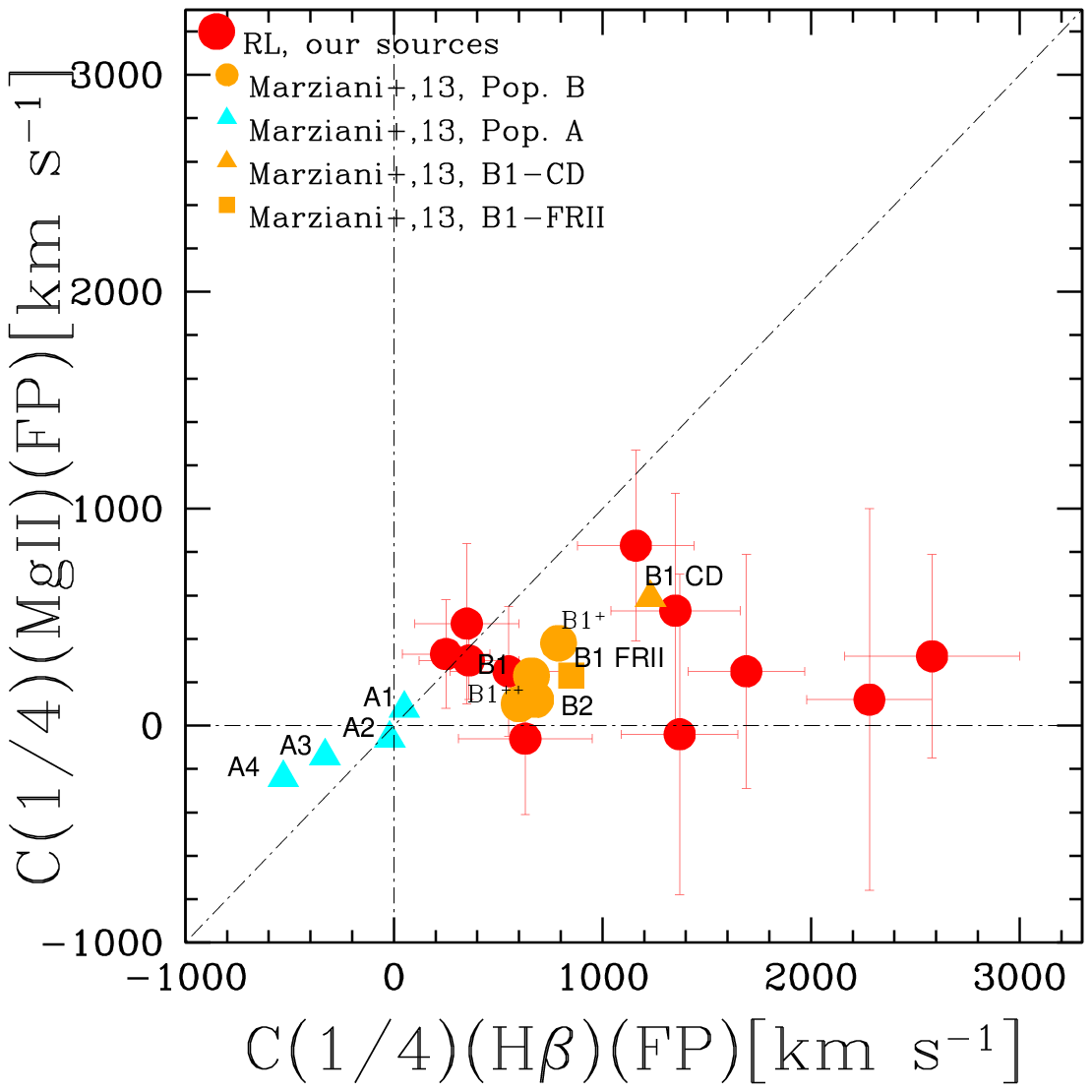}
\includegraphics [width=0.32\textwidth]{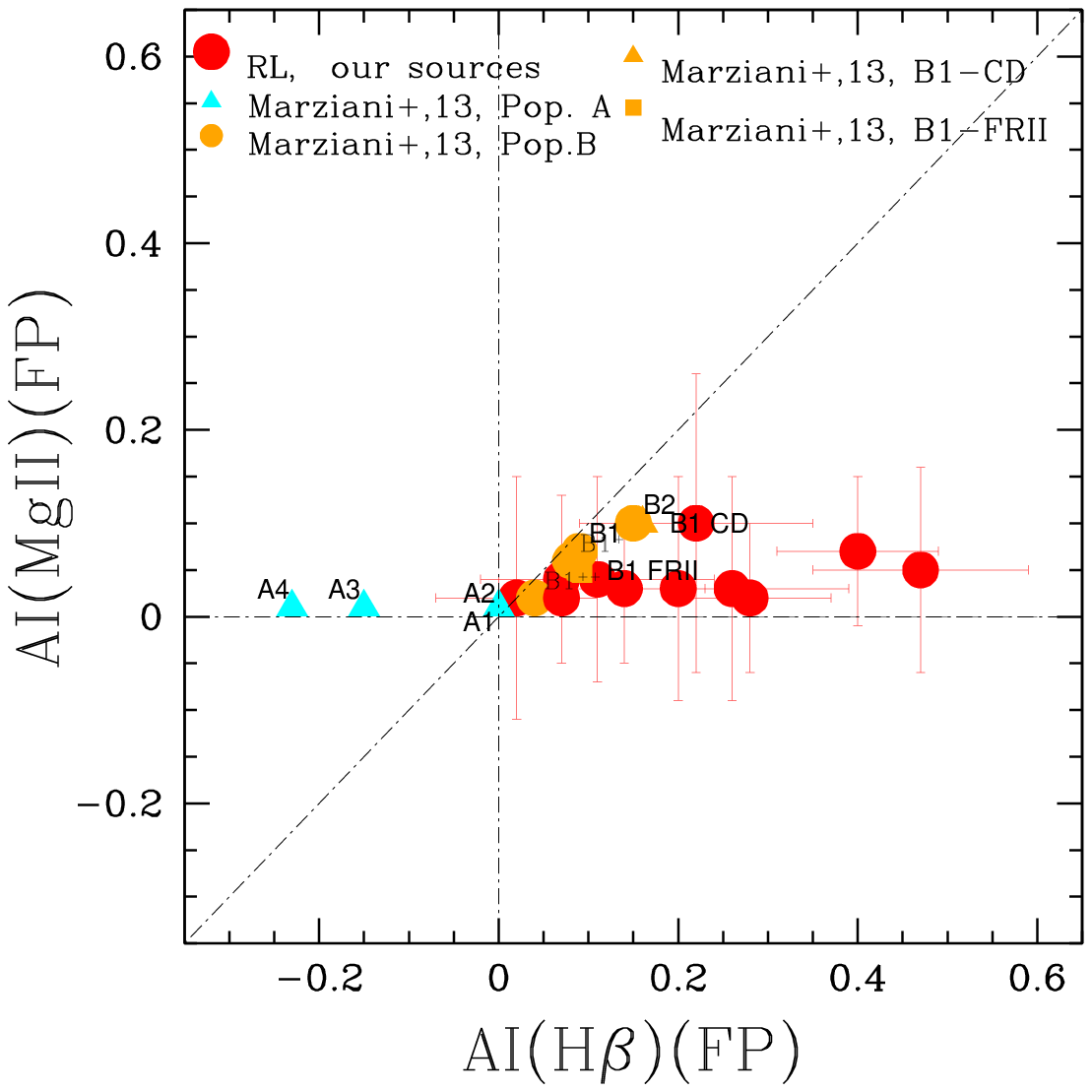}

\caption {Comparison between \hb{} and \mgiionly{} c($\frac{1}{2}$) (left), c($\frac{1}{4}$) (middle) and asymmetry index (right) for our eRk quasars (red filled circles).
Additional comparison sample was taken from composite spectra of \citet{2013A&A...555A..89M}, where symbols are as in Fig. \ref{fig:chcqfwmg}. The diagonal dot-dashed line in each plot represents the one-to-one line. }

\label{fig:fwhbmg}
\end{figure*} 

A second difference is observed in the FP measurements deviation from the rest-frame spectra at specified fractional intensities, mainly at c($\frac{1}{2}$) and c($\frac{1}{4}$),  as we mentioned above (see also Figs. \ref{fig:chcqfwhmhb} and \ref{fig:chcqfwmg}). The results of the centroid measurements on \hbFP\ and \mgiionlyFP\  indicate a shift towards the red for both lines. The shift has larger amplitude in \hb\ than \mgiionly, as shown in Fig. \ref{fig:fwhbmg} (left and middle panels). This amplitude difference becomes more evident when considering the line base: the value of c($\frac{1}{4}$) is larger than the value of c($\frac{1}{2}$), suggesting that \hb\ is more strongly affected than \mgiionly\ by contribution of the VBC  \citep{2000ApJ...545L..15S, 2002ApJ...566L..71S}. 

The same trend was shown when considering the AI (see Table \ref{tab:meashb} and \ref{tab:measmg}, Col.  11, respectively).  Fig. \ref{fig:fwhbmg} (right) indicates that both \hb\ and \mgiionly\ profiles show redward asymmetry, although the AI is significantly larger for \hb\ than for \mgiionly, i.e \mgiionly\ is more symmetric than \hb. Higher KI found for \mgiionly\ than \hb\ are explained by the broadening of the \mgiionly\ line due to its doublet components separated by 300\,\kms\ and by the broader \hb\ profile at the line base.

The redward asymmetry in \hb\ for Pop. B sources have been previously observed  \citep[e.g.,][]{2002ApJ...566L..71S,2009A&A...495...83M,2020MNRAS.492.3580W}, and can be linked to the existence of a distinct kinematic emitting region, the VBLR. The difference in the profile shape between \hb\ and \mgiionly\ and the {\tt specfit} line profile analysis provides empirical evidence that the stronger \hb\ VBC gives rise to a stronger redward asymmetry than in \mgiionly. 

\vspace {-\topsep }
\subsubsection{ Equivalent widths and Intensities}
\label{sec:ewopuv}

\begin{figure}
\includegraphics[width=0.5\linewidth]{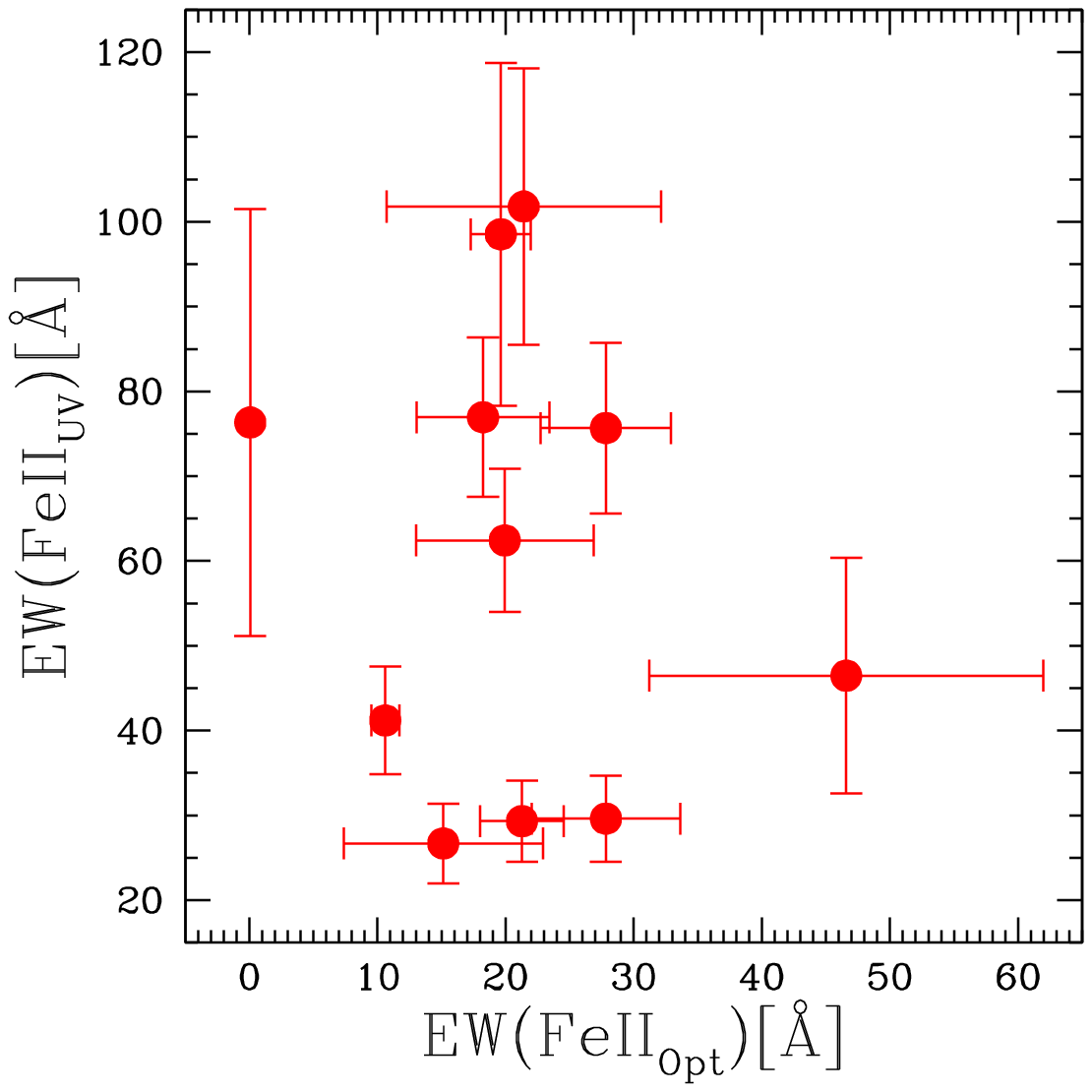}
\includegraphics[width=0.5\linewidth]{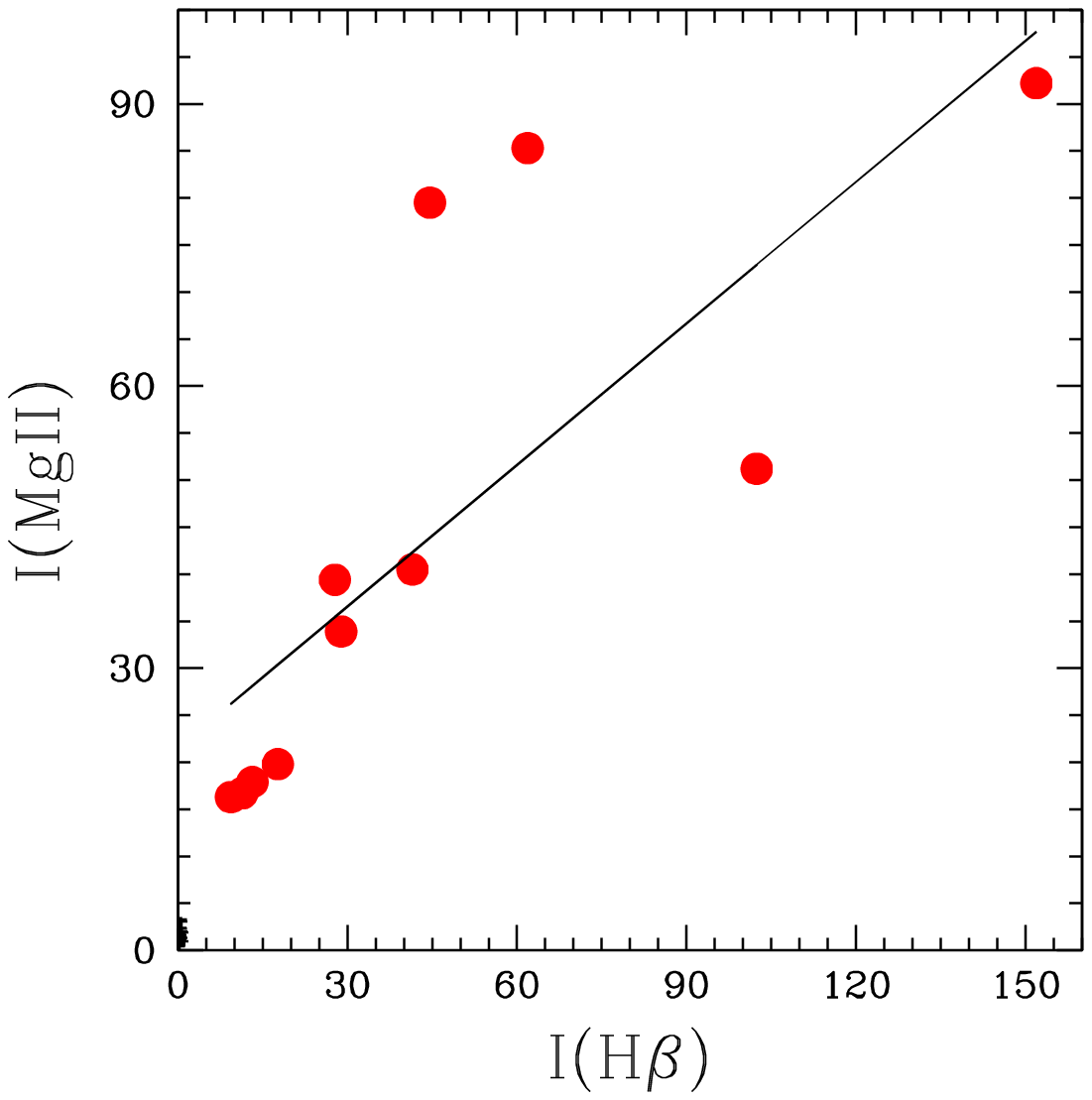}
\vspace{-0.2cm}
\caption{Left: relation between EW of \feiiopt\, and \feiiuv. The uncertainties in the EW  are taken to be proportional to the uncertainty in the flux of the \feii\ emissions in the two regions (see Sect. \ref{sec:errest}).  Right: correlation between the intensities of \hb\, and \mgiionly.  The intensities are highly correlated with $r \approx 0.77$  and a p-value $\approx$\,0.01.}
\label{fig:ewf}
\end{figure}

 The EW reflects the emission line strength relative to the total continuum and was obtained in each object from the best spectral fit.
 Fig. \ref{fig:ewf} (left) shows the EW of \,\feiiuv\ versus the EW of \,\feiiopt. A correlation is not found between the two quantities. This is in agreement with the previous result by \citet{2015ApJS..221...35K} who do not find any correlation between the EWs of these lines. The discrepancy could be due to a difference in the emitting regions: the \feiiopt\ emission is usually thought to arise in the outer BLR before the  inner radius of the  torus  \citep[e.g.,][]{2009AJ....137.3548P, 2010ApJS..189...15K,  2012ApJS..202...10S, 2013ApJ...769..128B,  2015ApJS..221...35K}. The \feiiuv\  might be preferentially  emitted in clouds closer to the continuum source and strongly affected by the X-ray emitting corona believed to be present in most AGN \citep{2019ApJ...875..133P}. Excitation mechanisms are also expected to be different: \feiiuv\ is produced by recombination following photoionization and enhance by florescence phenomena with continuum and Ly$\alpha$, while collisional excitation contributes to optical emission   \citep[see][for a Grotrian diagram showing the channels leading to \feiiuv\ and \feiiopt\ production by \lya\ fluorescence; c.f. \citealt{sigutpradhan03}]{marinelloetal20}. A loose correlation between the intensities of the \feiiopt\ and \feiiuv\ and \rfeUV\ and \rfeopt\ might be  expected in large samples of quasars because optical and UV  emissions are both   dependent on chemical abundances that are widely and systematically different along the quasar main sequence \citep{panda2019quasar,sniegowskaetal21,marzianietal23}.

 The intensities of \hb\, and \mgiionly\, are highly correlated (Fig. \ref{fig:ewf}, right): Pearson's $r \approx0.77$, with a probability $P \approx0.01$\ of a stochastic correlation. The \mgiionly/\hb\ intensity ratio varies in the range 0.5--1.8, with an average of 
 $\approx1.2$, a value consistent with the ones derived for composites RL spectra in Population B \citep[see Table 3 of][]{2013A&A...555A..89M}.

\section{Discussion}
\label{sec:discu}

As pointed out in the previous sections, quasars broad line spectra show a wide range of line profiles, line shifts as well as line intensities. In order to explain this spectroscopic diversity, much emphasis was placed on the connection between profile parameters \citep[e.g.,][]{2015ApJS..217....3M, 2022MNRAS.516.1624R, 2022A&A...659A.130K}, \mbh\ \citep[e.g.,][]{2002ApJ...565...78B, 2016CoKon.105...85H,2022ApJS..262...14B}, luminosity \citep[e.g.,][]{2009A&A...495...83M, 2011ApJ...738...68P, 2017A&A...603A..49R} and accretion rate \citep[e.g.,][]{2001NewA....6..321M, 2001ApJ...558..553M}. In this section, we  present a discussion on  \mbh\ and  Eddington ratio (\lE) (Sect. \ref{sec:mbhld}), and their effects on line profile shapes (Sect. \ref{sec:prof}). Finally, we consider the role of  radio loudness (Sect. \ref{sec:rlrq2}). 

\vspace{-0.5cm}

\subsection{Black hole mass and Eddington ratio }
\label{sec:mbhld}
The estimation of \mbh\ and Eddington ratio (\lE) is crucial for understanding the AGN phenomenon, its evolution across cosmic time, and the properties of the host galaxies \citep[e.g.,][]{2003ApJ...589L..21M}.  In addition, \mbh\ is a fundamental parameter that relates to the evolutionary stages and the accretion processes occurring within them, as the power output is directly proportional to \mbh\ \citep[e.g.,][]{2006apj...650...42L, 2017FrASS...4....1F}. 
\lE\ is a parameter that expresses the relative balance between gravitational and radiation forces  \citep[e.g.,][]{2010MNRAS.409.1033M, 2010ApJ...724..318N}, a major factor influencing both the dynamics and the physical conditions of the line emitting gas \citep{2018FrASS...5....6M}.

\begin{figure}
\centering
\includegraphics[width=0.85\columnwidth]{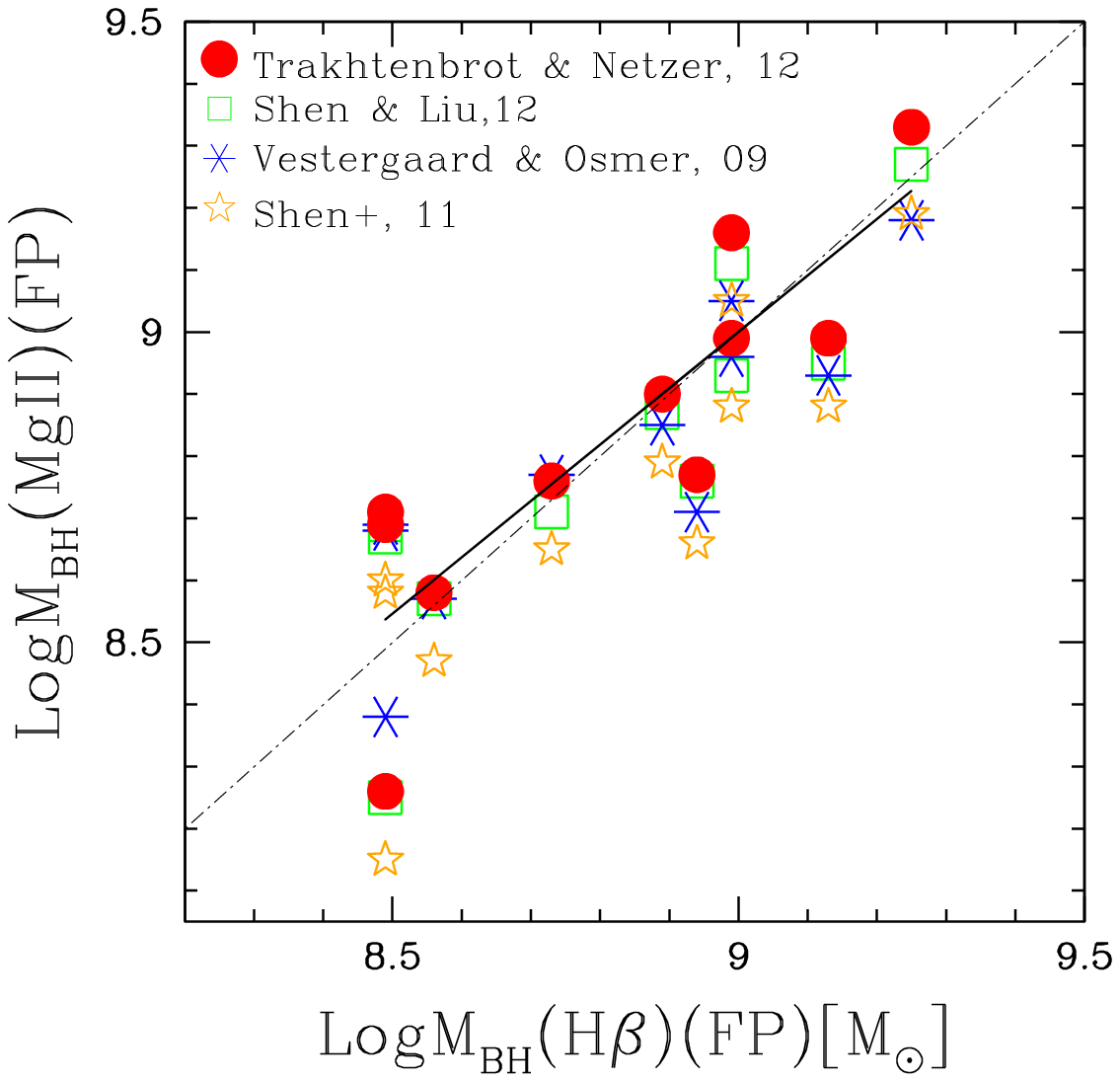}
\caption{\mbh\ comparison for estimates from \hb\ and \mgiionly, FP FWHM measurement using different scaling laws. The abscissa is \mbh\ estimated by using the formula from \citet{2006ApJ...641..689V}. In  ordinate we plot different \mbh\ estimates for \mgiionly, from  \citet{2009ApJ...699..800V}, \citet{2011ApJS..194...45S}, \citet{2012MNRAS.427.3081T}, and \citet{ 2012ApJ...753..125S}. The dot-dashed line represents the 1:1 line and the solid one represents the correlation between the masses estimated by using \citet{2006ApJ...641..689V} for \hb{} and \citet{2012MNRAS.427.3081T} for \mgiionly{}. The Pearson correlation coefficient between the two estimates is $r = 0.91$.}
\label{fig:bhall}
\end{figure}

To estimate \mbh\ for type I AGN, we used an empirically calibrated formalism (scaling laws) that is based on single-epoch spectra  \citep[e.g.,][]{2006ApJ...641..689V, 2015ApJ...809..123H} and that has been applied to  large and diverse samples of AGN \citep[e.g.,][]{2012NewAR..56...49M, 2013BASI...41...61S}.  
At low redshift (z $\lesssim$ 0.8), the lines of choice for estimation can be \hb\ and \mgiionly\ \citep[e.g,][]{2000ApJ...533..631K,2006ApJ...641..689V, 2006NewAR..50..782M}. 
 Table \ref{tab:masses} reports the estimations of the accretion parameters for our eRk quasars by using both the BC and FP FWHM measurements.  Cols. 2 and 5 report the bolometric luminosity ($L_\mathrm{bol}$) estimated from the optical (5100\AA) and UV (3000\AA) contunuum luminosity respectively, as $\lambda L_{\lambda}$, by using the luminosity dependent relation from \citet{2019MNRAS.488.5185N}  
 to calculate the bolometric correction factor ($K_{bol}$) for the luminosity in question ($L_{5100}$ or $L_{3000}$). 
  The bolometric luminosity  of our eRk quasars estimated  
 with log$L_\mathrm{bol}$ between 45.15--46.57 and 44.98--46.53 [ergs\,s$^{-1}$] from \hb{} and \mgiionly, respectively.  
 Cols. 3--4 and 6--7 list the \mbh{} and \lE{} by using the FP FWHM measures and the relations from \citet{2006ApJ...641..689V} for \hb{} and  \citet{2012MNRAS.427.3081T} for the \mgiionly\ line, respectively. The same estimation was also done by using FWHM of the BC alone rather than the FP and reported in Cols. 8--11.
\mbh{} values computed from the FP FWHM range from log\mbh\,[$M_{\odot}$] $\approx$ 8.49 to 9.25 (\hb) and from log\mbh\ $\approx$ 8.26 to 9.33 (\mgiionly). Fig. \ref{fig:bhall} indicates that the mass estimation  using \hb\ and \citet{2006ApJ...641..689V} relation is in a very good agreement with the mass estimation from \mgiionly, when using four different \mgiionly\ scaling laws, namely the ones from \citet{2009ApJ...699..800V}, \citet{2011ApJS..194...45S}, \citet{2012MNRAS.427.3081T} and  \citet{2012ApJ...753..125S}. This result holds also if only the BC is used (see Table \ref{tab:masses}, Cols. 8 and 10).  We also checked the mass estimation by using the \mgiionly\ with  \citet{2002MNRAS.337..109M} relation and found a lower estimation by about 0.25 dex compared to the other scaling relations  (not shown in Fig. \ref{fig:bhall} to avoid confusion).

The log\,(\lE) values  for our eRk quasars range from -1.45 to {-0.55} and from -1.40 to -0.78 (see Table \ref{tab:masses},  Cols. 4 and 7) when using the \hb{} and \mgiionly{} lines, respectively. Previous studies suggested that around some critical value of \lE{}$\approx\,$0.2$\pm$0.1, there could be an accretion mode change i.e., a change in the structure of the accretion disk  \citep[e.g.,][and references therein]{2014SchpJ...9.2408A, 2018FrASS...5....6M, 2019A&A...630A..94G}, or at least in the BLR dynamics \citep{2006A&A...456...75C}. { }For a given log\,\mbh\,$\approx$\,8.5,  Pop. A sources show log\,(\lE)\,=\,-0.7\,--\,0, and Pop. B show log(\lE)\,= -2 -- -0.7 \citep{2011BaltA..20..427S}. All our quasars fit into the Pop. B domain of low accretion rates.  The two sources  (3C 179 and PKS 2208-137) with FWHM\,<\,4000 \kms{} have $\log$(\lE) $\approx$ -0.64 and -0.75, respectively. These  values are at the \lE\d\ boundary between Pop. A and B, signifying a possible \mbh\ underestimate due to  low S/N (i.e., loss of line wings in the case of 3C 179) or perhaps to a pole-on orientation of the emitting regions  \citep[e.g.,][]{1986ApJ...302...56W, 2003ApJ...597L..17S, 2003MNRAS.340.1298R, 2008MNRAS.387..856Z}. 

\begin{table*}
\caption{Physical Parameters measurement from the FP (BC + VBC) and BC of \hb\ and \mgiionly\ lines.}
\centering 

\scalebox{0.96}{  
\begin{tabular}{@{\extracolsep{1.2pt}}lcccccccccc} 
\hline\hline 
\multirow{1}{*}{Object}&\multicolumn{6}{c}{Measurements from full broad profile (BC + VBC)}& \multicolumn{4}{c}{Measurement from broad component only (BC)}\\ \cline{2-7}\cline{8-11}

&\multicolumn{3}{c}{Accretion parameters (\hb)}&\multicolumn{3}{c}{Accretion parameters (\mgiionly)}& \multicolumn{2}{c}{Accretion parameters (\hb)}&\multicolumn{2}{c}{Accretion parameters (\mgiionly)}\\ \cline{2-4} \cline{5-7}\cline{8-9}\cline{10-11}

 &\multicolumn{1}{c}{logL$_\mathrm{bol,5100}$} &\multicolumn{1}{c}{log$M_{BH}^{a}$}&\multicolumn{1}{c}{log\lE$^{a}$} & \multicolumn{1}{c}{logL$_\mathrm{bol,3000}$}&\multicolumn{1}{c}{log$M_{BH}^{b}$}&\multicolumn{1}{c}{log\lE$^{b}$} &\multicolumn{1}{c}{log$M_{BH}^{a}$} &\multicolumn{1}{c}{log\lE$^{a}$} &\multicolumn{1}{c}{log$M_{BH}^{b}$}&\multicolumn{1}{c}{log\lE$^{b}$}\\ 
 
(1)&(2)&(3)&(4)&(5)&(6)&(7)&(8)&(9)&(10)&(11) \\
\hline 
PHL923 &45.78   &8.56  &
-0.90  & 45.80    & 8.58 &-0.89   & 8.39 &-0.73  &  8.43  & -0.74\\ 
B2 0110+29 &45.15  &8.49  &-1.45  & 44.98    & 8.26 & -1.40   & 8.21 &-1.18  &  8.15  &-1.29 \\ 
3C37  & 45.73   &8.73  &-1.11 &45.66    & 8.76 & -1.21   & 8.60 & -0.98  &  8.64  &-1.10 \\ 
PKS 0230-051  & 46.11   &8.89  &-0.90  &46.10    & 8.90 & -0.92   & 8.66 &-0.66  &  8.75  &-0.77 \\ 
3C94  &46.57   &9.25  &-0.74 &46.53    & 9.33 &  -0.92  & 9.35 &-0.70  & 9.33  &-0.73\\ 
PKS 0420-01  &46.49   &8.94  &-0.57  &46.17    & 8.77 &  -0.71  & 8.83 &-0.45  &  8.62  &-0.56 \\ 
3C179   &45.95   &8.49  &-0.64 & 45.86    & 8.71 &-0.96   & 8.22 &-0.38 &  8.54  &-0.79 \\ 
3C380    &46.33  &9.13  &-0.92  & 46.18    & 8.99 & -0.93  & 8.82 & -0.60 &  8.80  &-0.73\\ 
S5 1856+73   &46.11   &8.99  &-1.00  & 45.93    & 8.99 &-1.18   & 8.92 & -0.93  &  8.84  &-1.03 \\ 
PKS 2208-137 &45.86   &8.49  &-0.75  &45.83   & 8.69 &-0.97   & 8.43 &-0.69  &  8.56  &-0.84 \\ 
PKS 2344+09  &46.55   &8.99  &-0.55  &46.49   &9.16 & -0.78   &8.87 & -0.44  &  9.01  &-0.64\\ 
 \hline
\end{tabular}
}
{\raggedright 
{$^{(a)}$Estimated from \citet{2006ApJ...641..689V}.$^{(b)}$Estimated from \citet{2012MNRAS.427.3081T}. Cols. 2 and 5 in ergs\,s$^{-1}$. \mbh\, cols. in units of $M_{\odot}$. \par}}
\label{tab:masses}
\end{table*}

\subsection{Correlations between profile and physical parameters}
\label{sec:prof}

\begin{figure*}
\includegraphics [scale=0.29]{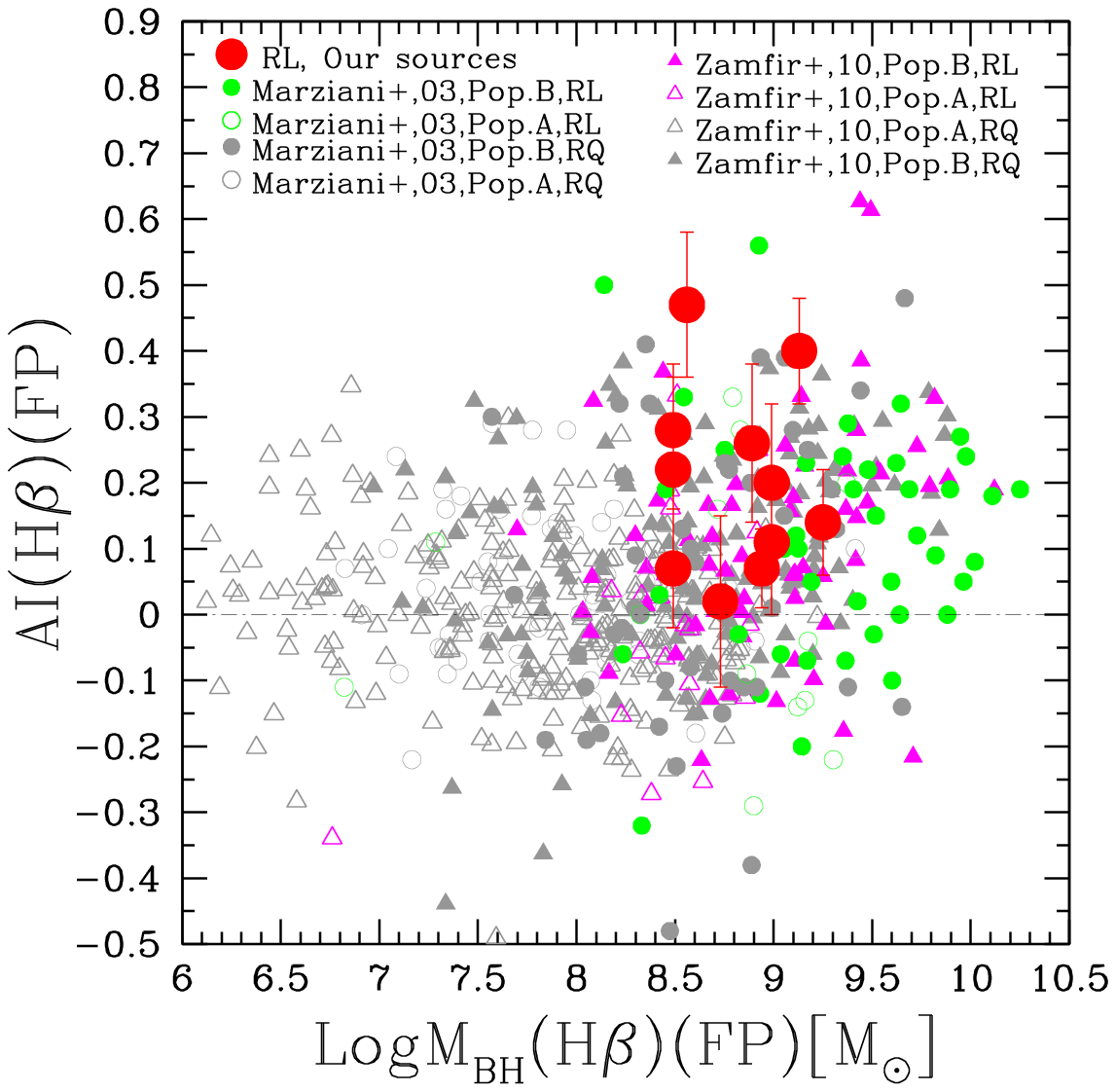} 
\includegraphics [scale=0.29]{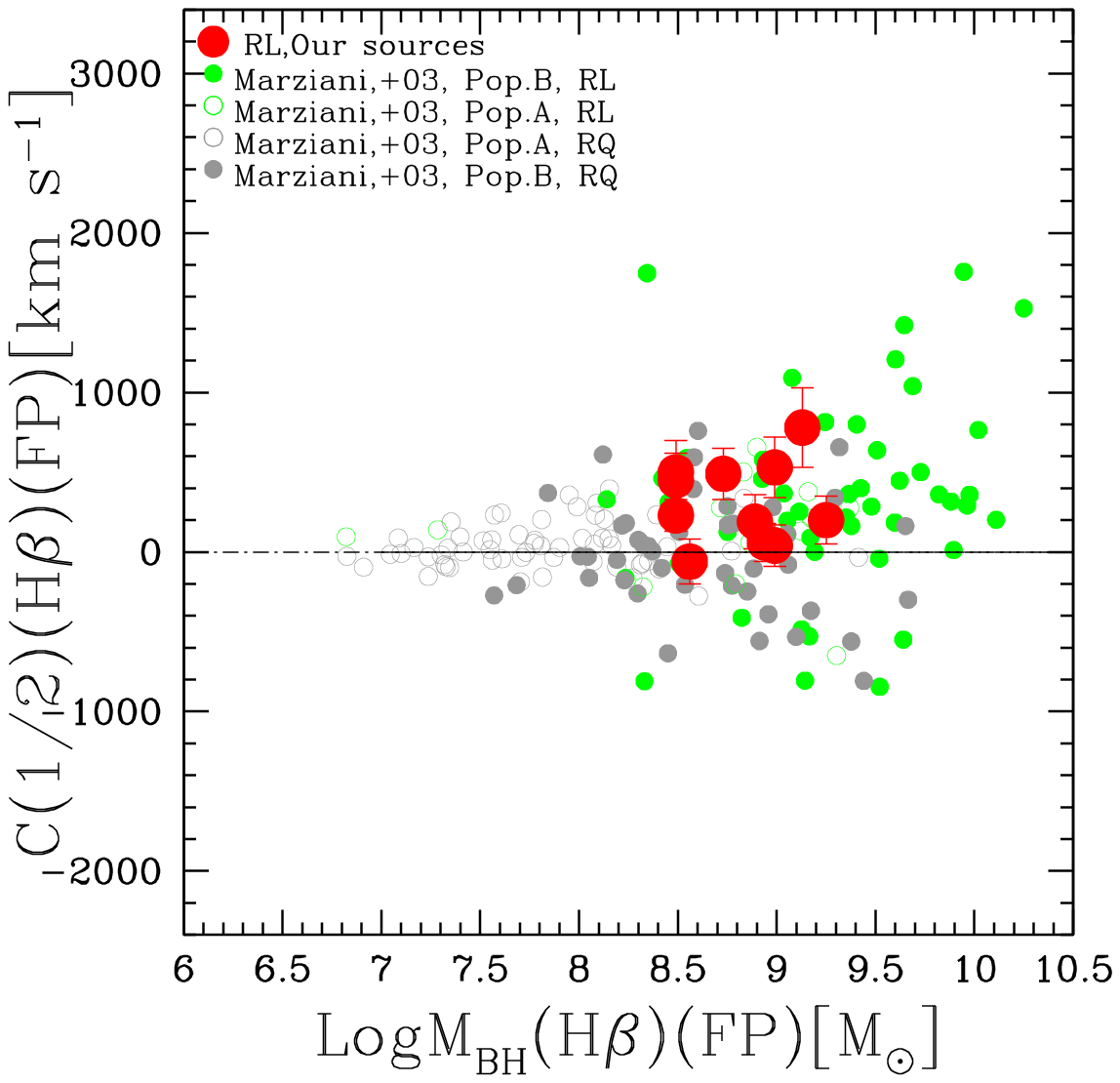} 
\includegraphics [scale=0.29]{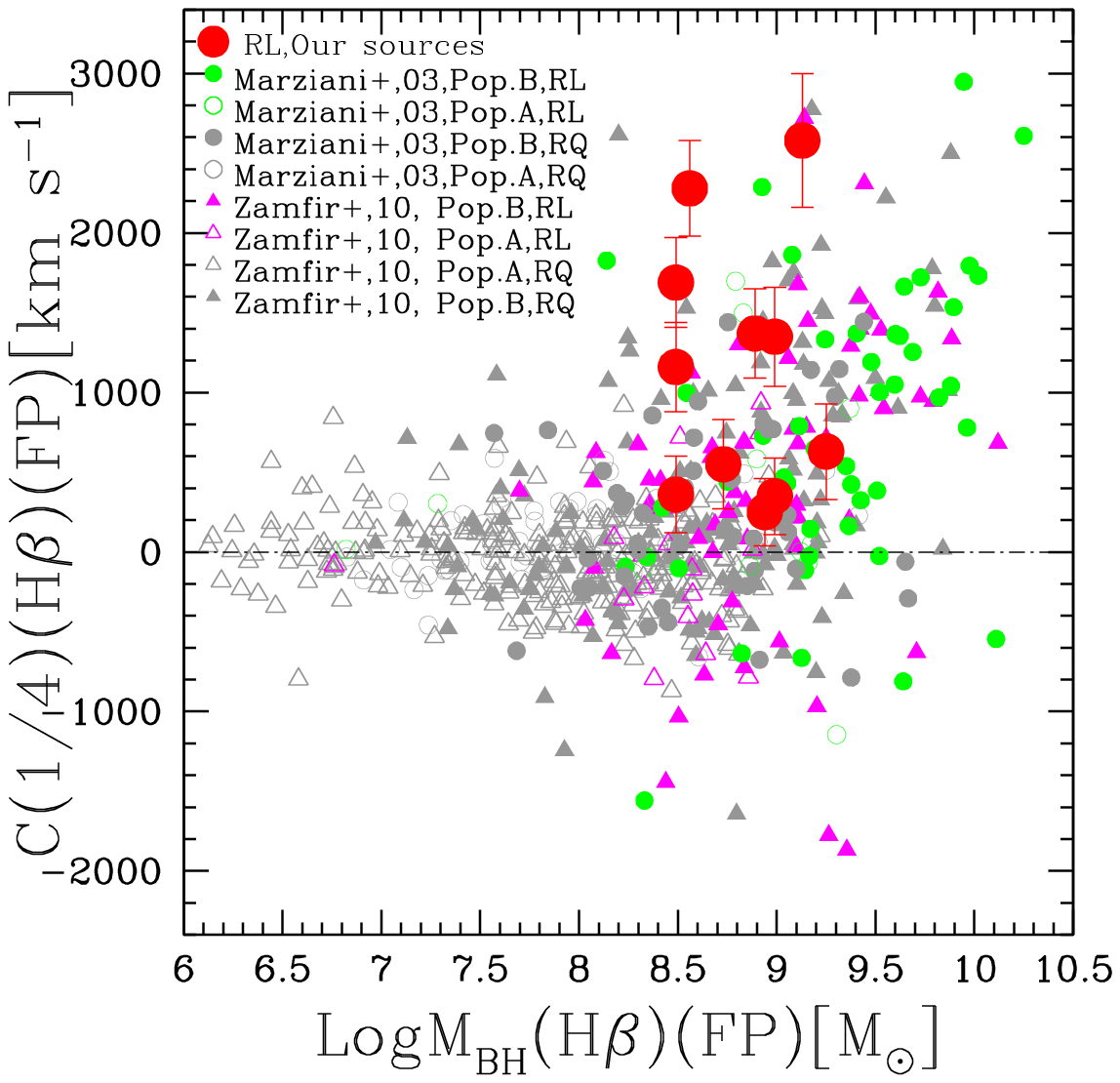} 

\caption{ {\mbh\ effect on line profile parameters, AI (left), on c($\frac{1}{2}$)  and c($\frac{1}{4}$) (middle and right plots, respectively). Comparison samples were taken from \citet{2003ApJS..145..199M} for c($\frac{1}{2} $) and \citet{2003ApJS..145..199M} and \citet{2010MNRAS.403.1759Z} for AI and c($\frac{1}{4} $). The large red solid circles represent the results from our RL spectra. 
The horizontal dot-dashed lines  trace the symmetry line for zero AI, c($\frac{1}{2} $) and c($\frac{1}{4}$). 
}}
\label{fig:bhhbch}
\end{figure*}
\begin{figure}
\centering
\includegraphics [width=0.85\columnwidth]{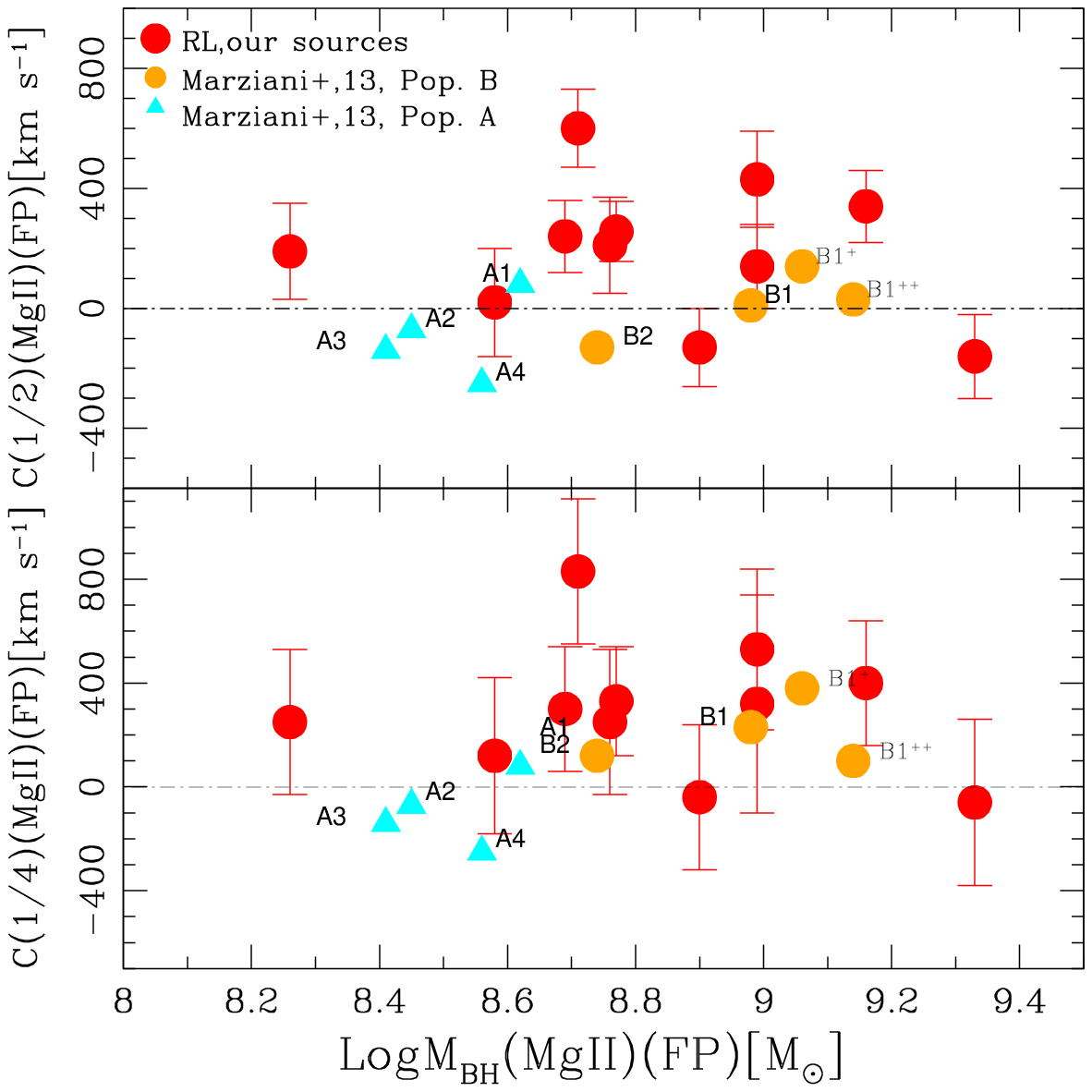}
\caption{\mbh\ of \mgiionlyFP\ and centroid shift comparison at c($\frac{1}{2}$) (upper) and at c($\frac{1}{4}$) (lower) of \mgiionly. The large red solid circles represent the results from our RL spectra. A comparison sample was taken from the composite spectra of \citet{2013A&A...555A..89M} and the meaning of symbols is as for  Fig.\ref{fig:chcqfwmg}.}
\label{fig:bhmg}
\end{figure} 

\subsubsection{ \mbh\ correlations} 
\label{sec:mbhcorrel}

 Fig. \ref{fig:bhhbch} and Fig. \ref{fig:bhmg}  show  profile parameters, as AI and centroids, as a function of \mbh, for \hb\ and \mgiionly{} respectively, including our eRk quasars and the comparison samples. The immediate result is that there is no  strong, clear correlation with \mbh\ -- although an intriguing trend appears for the joint comparison sample. This is  also supported by previous statistical tests for low-$z$ Pop. B sources \citep{2010MNRAS.403.1759Z}.  

Restricting the attention to \hb,  for the joint comparison sample and irrespective of whether the sources are Pop. A or B,  below log\mbh\ $\sim$ 8.5[\msol] RL sources show a weak median redshift for c($\frac{1}{4}$) $\approx$ 280$\pm$270\,\kms, while RQs show no significant shift nor asymmetry (median c($\frac{1}{4}$) $\approx$ -20 $\pm$180\,\kms, $\mu_{1/2}$AI\,$\approx$\,0.01\,$\pm$\,0.08). Above log\,\mbh\,$\sim$\,8.5\,[\msol], there is an increase in scatter in the values of the AI and centroids, and a predominance of shifts to the red appears, especially for RL sources ($\mu_{1/2}$c($\frac{1}{4})\approx$\,660\,$\pm$\,650\,\kms, $\mu_{1/2}$AI $\approx$ 0.1$\pm$0.1), while RQs 
remain more symmetric (median c($\frac{1}{4})\approx $150\,$\pm$\,490\,\kms, AI $\approx$ 0.04$\pm$0.10), albeit with a slight net shift to the red. 
Restricting now the attention to only Pop. B sources, \hb{} yields similar results: 
 RL show a trend towards the red with a median c($\frac{1}{4})\approx$ 300$\pm$270\,\kms{} for log\,\mbh\,$\lesssim$ 8.5\,[\msol], while the RQ counterpart a median centroid consistent with no shift (c($\frac{1}{4})\approx$ 30$\pm$280\,\kms). For the higher \mbh{} range, both RQ and RL Pop. B sources show a net shift to the red that reaches a median c($\frac{1}{4})\approx$ 680\,\kms for the RL subsample.
 Consistently, for the whole comparison sample, the ratio of the number of sources with negative and positive c$(\frac{1}{4})$ is $\approx$\,1.2 for RQ for $\log$\mbh\,< 8.5 and 0.61 for $\log$\mbh\,>\,8.5. The same ratio for RL sources is 0.45 and 0.31 for the two mass ranges, showing a net predominance of redshift. If the attention is restricted to Pop. B, we see a net predominance of redshifts for both
RQ and RL in the higher mass range (ratio negative-to-positive c$(\frac{1}{4})$ is 
 0.38 and 0.30, respectively). In the lower \mbh{} range, the prevalence
of redshifts is higher for RL (ratio $\approx$ 0.54) than for RQ (ratio $\approx$ 0.95)
for which the distribution is fairly symmetric, as expected also by the almost zero median shift amplitude of c$(\frac{1}{4})$.

 The intrinsic shift to red might be associated with gravitational redshift \citep[e.g.][]{2015Ap&SS.360...41B,2020ApJ...903...44P} or with infall plus obscuration \citep{2017NatAs...1..775W}. However, the origin of the redward asymmetry is still unclear and the subject of ongoing investigations \citep[e.g.,][]{2022ApJS..262...14B}. Infall of gas toward the center  is an alternative to gravitational redshift that may also produce significant shifts to the red expected to grow in amplitude toward the line base \citep{1977MNRAS.181P..89N,1990MNRAS.244..357P,2001ApJ...549..205F,2017NatAs...1..775W}. 

Related to \mgiionly{}, Figure \ref{fig:bhmg} shows the centroid shift at the two different fractional intensities of the \mgiionly\ line as a function of \mbh\ computed from FWHM \mgiionlyFP. The values of  c($\frac{1}{2}$)  and c($\frac{1}{4}$) are comparable, with a  slight systematic difference of about 100 \kms. Considering the comparison composite samples from \citet{2013A&A...555A..89M},   Pop. B \mgiionly\ centroid shifts close to the line base exceed the ones at the center by a modest amount of 200 -  300 \kms, at variance with \hb:  the \hb\  line base is significantly more redshifted than the center. For instance,   the eRk sample sources show an average c($\frac{1}{4}$)  - c($\frac{1}{2}$) $\approx$ 800 \kms.  As mentioned above, the \mgiionly\ profile retains a higher degree of symmetry than \hb\ because of a less prominent VBC. Whatever the cause of the reward asymmetry might be, the \mgiionly\ profile is apparently less affected than \hb.

\subsubsection{ \lE\ correlations} 
\label{sec:lleddcorrel}
\begin{figure}
\centering
\includegraphics [scale=0.4]{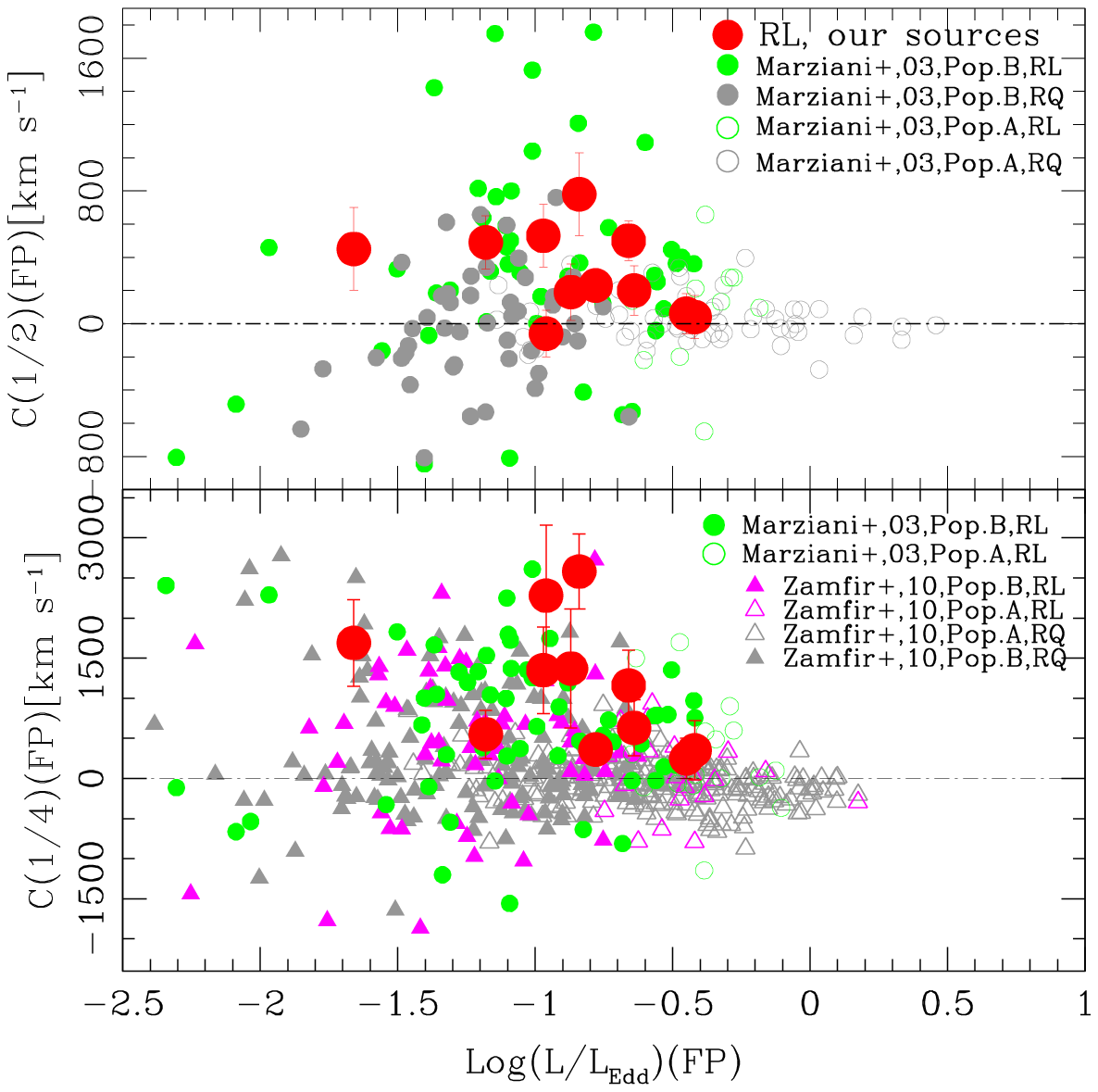}
\vspace{-0.2cm}
\caption{{Dependence of \hb\ c($\frac{1}{2} $) (upper) and  c($\frac{1}{4} $) (lower) on log\lE\ calculated from the FP. Comparison samples were taken from \citet{2003ApJS..145..199M} for c($\frac{1}{2}$) and \citet{2003ApJS..145..199M} and \citet{2010MNRAS.403.1759Z} for c($\frac{1}{4} $)..
}}  
\label{fig:eddchcq}
\end{figure}

Several authors suggested  that  outflows are apparently more related to \lE\ than to luminosity or \mbh\ \citep[e.g., ][]{2014SchpJ...9.2408A, 2017A&A...608A.122S}.  How c($\frac{1}{2}$),  c($\frac{1}{4}$) and  \lE{} are related,  in our eRk quasars and the comparison samples, is shown in Fig. \ref{fig:eddchcq}.  
 
Pop. B RL and RQ quasars belonging to the comparison samples are redshifted, and they generally possess low \lE. 
If the shift to the red is gravitational in origin,  the line might be emitted from a region closer to the central SMBH which results in larger shifts if the \lE{} is lower. As there is a general consensus that outflows produce blueshifts \citep{2012AstRv...7d..33M},   
 the lower shifts at higher \lE{} may be due to the increased relevance of radiation forces with respect to gravitation, which may push outward the emitting gas if there is a sort of radiation pressure/gravitation balance \citep{1993ApJ...412L..17M,2009ApJ...698L.103M,2010ApJ...724..318N,2015MNRAS.446.1848K}. Therefore, a difference between RL and RQ might be due to a combination of inflows and outflows that may result in a net slightly redshifted profile for RQ where   winds are stronger  \citep[e.g.,][]{2004ApJ...617..171B,2007ApJ...666..757S,2011AJ....141..167R},  and more redshift profiles for RLs, due to less prominent outflows.  
 
\begin{figure}
\centering
\includegraphics[width=0.9\columnwidth]{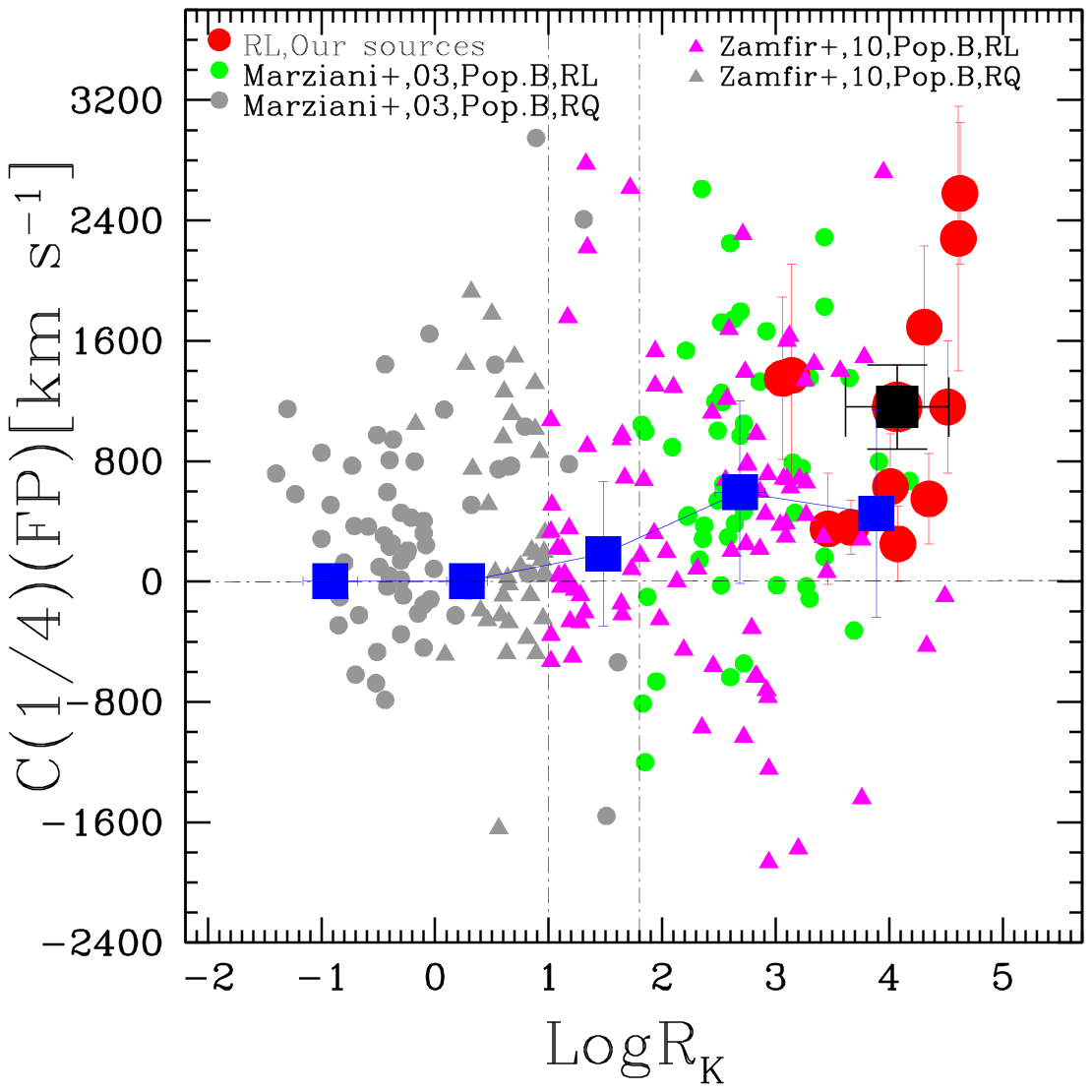}
\vspace{-0.2cm}
\caption{{Relation between c($\frac{1}{4}$) of \hb{} and radioloudness parameter. Sources from \citet{2003ApJS..145..199M} and \citet{2010MNRAS.403.1759Z} were taken as comparison sample. The connected blue solid squares represent the median value of the comparison sample distribution in equally-spaced bins where the vertical bars were computed by using the semi-interquartile range as an estimate of the sample dispersion. The horizontal bins denote, for each bin, the mid-point of the bin interval. The black solid square is the median value for our eRk quasars. 
The horizontal dot-dashed line traces the symmetric line in c($\frac{1}{4}$). The vertical lines at 1 and 1.8 mark the nominal RQ-radio intermediate and radio intermediate-RL boundaries    \citep{2008MNRAS.387..856Z}.} }
\label{fig:logrk}
\end{figure} 

\subsection{Similarities and differences between RL and RQ quasars}
\label{sec:rlrq2}
 For better visualization of the trends using the main comparison samples as well as the eRk sources, we analyzed the distribution of the velocity centroids at 
1/4 based on their population as well as radio type. 
In addition, we showed  the median measurements of the comparison samples in equally-spaced bins by using the semi-interquartile range as an estimate of the sample dispersion. 
Our  eRk quasars are strong radio emitters with log\,\rk{} > 3 and very powerful with log\, $P_{\nu}$\,>\,33.3\,[ergs\,s$^{-1}$\,Hz$^{-1}$], and include some of the highest log\,\rk{} values ever observed.  Objects with $\log\,$\rk $\ge3$ are most likely highly beamed, but they are still expected to be  very powerful radio sources as suggested previously by \citet{2007ApJ...658..232C}.  

 A first result of the inter-comparison between RL and RQ sources is their distribution  in the optical plane of the MS (see Fig. \ref{fig:optplane}). In that figure, all of  the eRk sources show a restricted domain occupation (mostly B1)  compared to the RQ majority (taken from \citet{2010MNRAS.403.1759Z}.   RQ sources  are found in both Pop. A and Pop. B  and  are distributed along the MS in all bins. The RLs, with CD and FR II morphology, show a distribution centered in the region of low \rfeopt\ and broader FWHM. A restricted domain space occupation of Pop. B RL sources is also shown in the previous studies of \citet{2000ARA&A..38..521S, 2003ApJ...597L..17S}. Similar considerations apply to the sources that are used to make the composite in \citet[][see their Table 1 for the number of sources in different bins]{2013A&A...555A..89M},

Figure \ref{fig:logrk} relates the 
velocity centroid shift, c($\frac{1}{4}$) of \hb\ to the radio loudness parameter, \rk, for our eRk, and the \citet{2003ApJS..145..199M} and \citet{ 2010MNRAS.403.1759Z} Pop. B sub-samples, represented in equally-spaced bins of \rk{} (see also figure in Appendix B of the online supplementary material for the relation with c($\frac{1}{2}$)). The second result is that the Pop. B RL source distribution doesn't show a clear trend with the \rk. However, there is a significant increase in the centroid shift between Pop. B RQ and Pop. B RL: the median c($\frac{1}{4}$) is $\approx$ 233 \kms\ and 144 \kms\ for Pop. B RQ, and becomes $\approx$ 720 \kms\ and 440 \kms\  for the RL sources of \citet{2003ApJS..145..199M} and \citet{2010MNRAS.403.1759Z} samples, respectively. The centroid shift for the eRk sources even reaches 1350 \kms. The significance of this difference needs to be further investigated in samples with matching \lE\ and \mbh\ distributions, as systematic differences between RL and RQ sources are also found in terms of \mbh.   
Considering the  two comparison samples  from  \citet{2003ApJS..145..199M} and \citet{ 2010MNRAS.403.1759Z}  and our eRk, and then subdividing the Pop. B quasars only into RLs and RQs, the RQ sources have $\mu_\frac{1}{2}$log\mbh$ \approx$ 8.18 $\pm $ 0.53 [\msol] and the RL's, $\mu_\frac{1}{2}$log\mbh$ \approx$ 8.98 $\pm$ 0.54 [\msol],  a 0.8 dex difference.  In addition, considering the \lE, the RQs  may be  slightly higher accretors compared to the RLs, 
$\mu_\frac{1}{2}$log\lE$\,  \approx$ -1.081 $\pm $ 0.156 and $\mu_\frac{1}{2}$log\lE$\, \approx$ -1.121 $ \pm$ 0.224, respectively that become $\mu_\frac{1}{2}$log\lE$\,  \approx$ -0.785 $\pm $ 0.343 and $\mu_\frac{1}{2}$log\lE$\, \approx$ -1.009 $ \pm$ 0.348 with a difference of -0.224 dex  if  a constant optical bolometric correction of 10 is applied. 
Systematically lower \lE\ and larger \mbh\ for Pop. B RL sources have been also found in several past works   \citep[e.g.,][]{2002ApJ...565...78B,2003MNRAS.340.1095D, 2004mas..conf..389M}. 

\begin{figure*}
\centering
\includegraphics [scale=0.46]{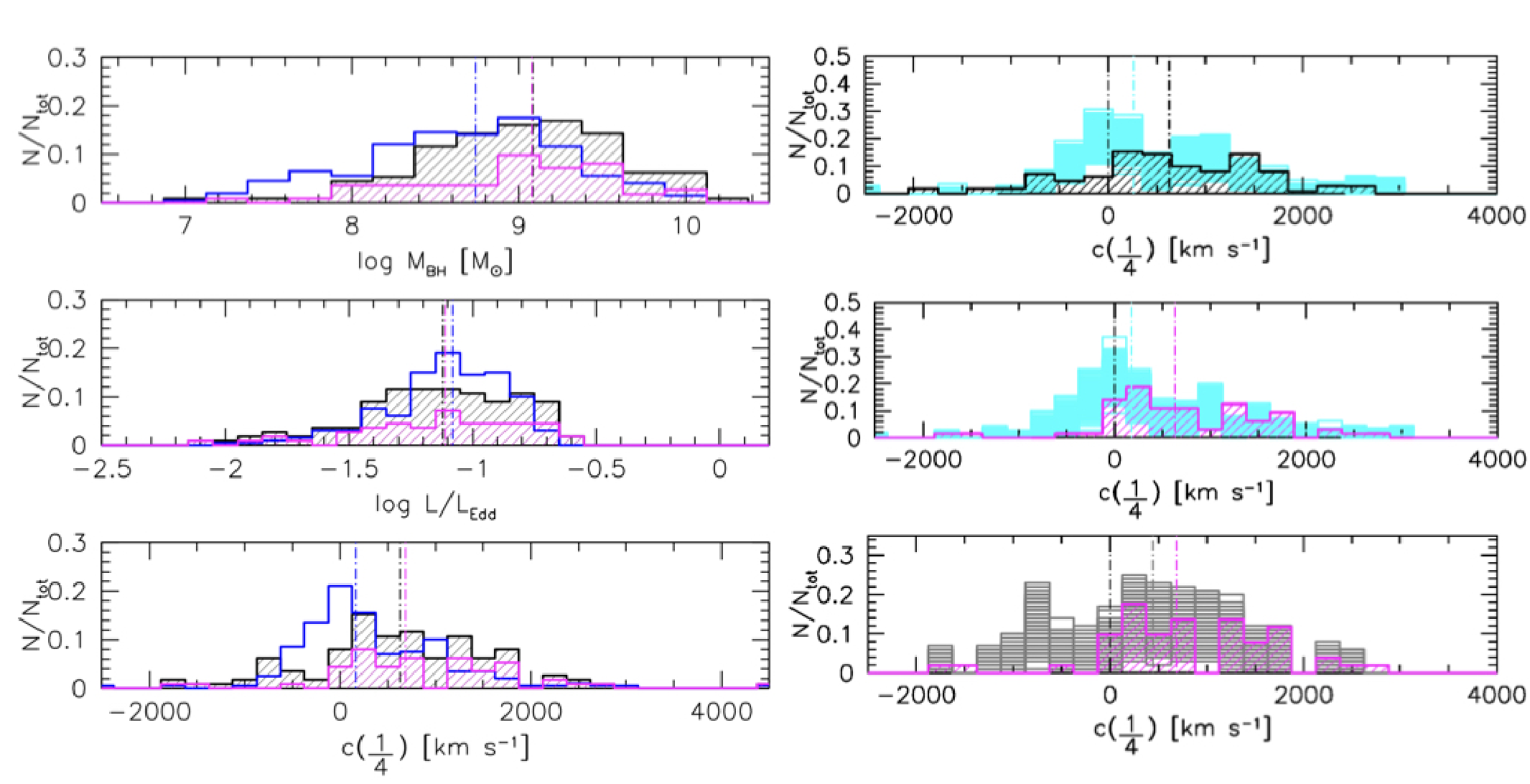}
\vspace{-0.2cm}
\caption{ Results of the bootstrap analysis. Left panels, from top to bottom: distributions of \mbh, \lledd, and \hb\ c($\frac{1}{4}$) for the RQ (blue), RL (black shaded), and extreme RL ($\log$ \rk $\ge 3$, magenta). Right panels: comparison between  \hb\ c($\frac{1}{4}$) resampled distributions matching \mbh\ and \lE\ and original distributions. Top: RL vs RQ bootstrap replications; middle: extreme RL vs RQ bootstrap; bottom: extreme RL vs. rest of RL i.e., objects with $1.8 \le \log$\rk$<3$. 
Vertical dot dashed lines indicate medians.   
\label{fig:boo}}
\end{figure*}

 Therefore, it is not necessarily appropriate to correlate radio loudness to a single variable such as c($\frac{1}{4}$), since other parameters such as \lE\ and \mbh\ are expected to  play a role \citep{2002ApJ...579..530W,2006MNRAS.372.1366K}. We need to take this into account if we compare the radio parameters and centroid shifts: 
the distribution of shifts appears to be slightly dependent on \mbh, i.e. largest c($\frac{1}{4}$) values occur for the highest \mbh\ (Fig. \ref{fig:bhhbch}) and lowest \lE\  (Fig. \ref{fig:eddchcq}). 
To verify whether there is a genuine effect of radio loudness on the c($\frac{1}{4}$),  
 we considered the samples of \citet{2010MNRAS.403.1759Z},  and of \citet{2003ApJS..145..199M}, added the 11 RLs of the present work and separated RQ and RL within Pop. B only, where most RL reside.  This gave us 169 RQ and 145 RL Pop. B sources, of which 53 are eRk with $\log$\rk $\ge$ 3. The c($\frac{1}{4}$) distributions remain different ($P \sim 3 \cdot 10^{-5}$), suggesting that the shift amplitudes are higher in RL than in RQ. 
  However, the \mbh{} distributions of RQ and RL are also markedly different  ($P \sim 1 \cdot 10^{-4}$\ that they are drawn from the same parent populations), with medians differing by $\delta$log\mbh\ $\approx$ 0.35, while the two \lE\  distributions are similar (Fig. \ref{fig:boo}). 

Bootstrap replications of the RQ Pop. B samples were computed considering only distributions of \mbh\ and \lE{} that were consistent 
with the ones of RL Pop. B  (Fig. \ref{fig:boo}\ upper right panel). Of the 1000 bootstrap distributions for the RQ Pop. B c($\frac{1}{4}$)  $\lesssim$ 85\%  were different from the RL one by a confidence level more than 2$\sigma$. Therefore, within the limitations  of our sample, we are unable to detect a highly significant effect of radio loudness on the c($\frac{1}{4}$) shift amplitude. The majority of the RQ Pop. B bootstrapped distributions of  c($\frac{1}{4}$) are largely overlapping with the RL one. The analysis thus confirms the statement of  \citet{2003MNRAS.345.1133M} that the RQ and RL show similar velocity shift amplitudes, with the most extreme values (the ones attracting more attention) occurring for RLs. 

 Extreme RL sources, i.e. the 53 Pop. B sources with $\log$\rk $\ge$ 3, 
 have a larger c($\frac{1}{4}$) shift with respect to the full population of RL, with  distributions that are different at an even higher significance ($P \sim 10^{-6}$). We repeat the bootstrap analysis matching the RQ \mbh\ and \lE\ distributions to the ones of the extreme RLs, and in this case, the  difference at a  $2\sigma$ confidence level is highly significant:   less than 1 out of 1000 RQ c($\frac{1}{4}$) resampled distributions are statistically indistinguishable from the extreme RL one. The extreme RL sources also show c($\frac{1}{4}$) in excess with respect to the RL sources with  1.8 $\le$ log\rk\ < 3, and the  $2\sigma$ excess is still  confirmed  at a high confidence level. Therefore,  the significance of the difference between RQ and RL sources  is strongly dependent on sample biases, as it is mainly driven by the fraction of sources with high Kellermann ratio.  

Equally important is not to forget  the similarity between RQ and RL Pop. B sources. The same line profile phenomenology is observed in both classes, provided that a restriction to Pop. B is done for RQ quasars: RLs  appear more extreme but with a large overlap in the line profile parameter distribution with RQs. This indicates that there could be only a quantitative effect on the BLR associated with the jet, without inducing any strong structural or dynamical change.  Investigations that are beyond the scope of the present paper are needed to ascertain the physical origins of the excess c($\frac{1}{4}$) shift in the most powerful RL.

\section{Conclusions}
\label{sec:con}

 This work presented new long-slit simultaneous near-UV and optical spectra of 11 relativistically jetted quasars selected on the basis of their extreme radio emission (log\,\rk\,>\,3, the eRk sources) with redshifts 0.35\,$\lesssim$\,z\,$\lesssim$\,1, and studied their spectroscopic properties by using the optical projection of the four-dimensional eigenvector 1 (4DE1) parameter space i.e., the so-called quasar main sequence (MS). Our analysis focused mainly on the spectral fitting of the strongest emission lines recorded on our spectra, \mgiionly,  and  \hb, by using {\tt specfit} routine within the {\tt IRAF} package.
We aimed to quantify broad emission line differences between radio-loud (RL) and radio-quiet (RQ)  quasars, paying special attention to the quasars with extreme radio emission. 

The main findings that we draw from this study are:

\begin{itemize}
\item The eRk quasars  presented in this paper occupy a much more restricted domain in the optical plane ($\sim$ bin B1) compared to the RQ sources of the comparison samples. This confirms the result of \citet{2008MNRAS.387..856Z} who found that powerful jetted sources tend to occupy mainly the Pop. B region of the MS.
    
\item The UV plane formed from FWHM of \mgiionly\, and \feiiuv\, looks similar to the optical plane. As in the optical plane, our quasars occupy a more restricted domain than the full quasar population, with low \rfeUV. There is no correlation between the EW of \feiiuv{} and \feiiopt{} in our eRk sample. 
   
\item The FWHM of \mgiionly\ is systematically narrower than FWHM \hb\ by about 10\%. This holds for both the full profile (BC+VBC) and if only the broad component is considered for our eRk sample, a result consistent with previous studies. 

    \item Both  \hb\ and \mgiionly\  lines show profile shifts and asymmetries towards the red. The centroid shift of the line base c($\frac{1}{4}$), as well as the asymmetry index, are larger in \hb\ than \mgiionly.

    \item \hb\ and \mgiionly\ appear to be  provide consistent virial \mbh\  estimates. 
     The eRk quasars  of this paper lie within the range of log\mbh\ = 8.49 -- 9.25[\msol]  and  8.26 -- 9.33[\msol]  when using \hbFP\ and \mgiionlyFP, respectively. The resulting log\lE\ has a range of [-1.65, -0.42]  and [-1.40,-0.71].   
\item Joining the sources studied in this work with comparison samples, we find that the distribution of shifts appears to be slightly dependent on \mbh\ in which larger c($\frac{1}{4}$) values occur for the highest \mbh\ and lowest \lE\ values. A possible explanation is offered by a combination of outflow  and infall (or gravitational redshift) contributing to blueshifted and redshifted excesses, respectively, with the outflow component being minimal in the RL \hb{} and \mgiionly{} profiles. 

\item There is a trend between the velocity shifts,  stronger for c($\frac{1}{4}$), and \rk: Pop. B RL quasars tend to have larger velocity shifts to the red than RQs. The difference  is found to be only marginally significant if the RQ and RL \mbh\ and \lE\ distributions are matched. However, the difference becomes highly significant if the comparison is carried between eRk  and RQ Pop. B AGN.

\end{itemize}

There is apparently no evidence of outflow in the broad line profiles of  our eRk sources. The only evidence is provided by a slight asymmetry of  \oiiionly.  Further observations of HILs such as \civ\ would be needed to assess the extent of any mildly ionized outflow origination from the accretion disk.

\section*{Acknowledgements}
The authors thank the anonymous referees  for their valuable
suggestions that helped us to significantly improve the paper.
STM acknowledges the support from Jimma University under the Ministry of Science and Higher Education. STM and MP acknowledge financial support from the Space Science and Geospatial Institute (SSGI) under the Ethiopian Ministry of Innovation and Technology (MInT). STM, ADO, and MP acknowledge financial support  through the grant CSIC I-COOP+2020, COOPA20447. STM especially acknowledges the IAA for all the support received during the two stays. ADO, MP, JP, PM, and IM acknowledge financial support from the Spanish Ministerio de Ciencia e Innovaci\'on - Agencia Estatal de Investigaci\'on through projects PID2019-106027GB-C41 and    PID2019–106027GB–C43, and from the Severo Ochoa grants SEV-2017-0709 and CEX2021-001131-S funded by MCIN/AEI/ 10.13039/501100011033. MAMC has been supported by the Spanish Research project PID 2021-122961NB-IOO. 

\section*{Data Availability}
This work used original data observed by the group using the Cassegrain TWIN spectrograph of the 3.5m telescope at the Calar Alto Observatory (CAHA) (Almería, Spain)\footnote{\url{http://www.caha.es/}} and present new optical and near-UV spectra of 11 powerful jetted quasars. In support of this study,  the additional data used in this article were obtained from the public sources cited in the article (or references therein).




\bibliographystyle{mnras}
\bibliography{1_myreferences} 



%
\appendix
\onecolumn


\vspace{0.1in}

\begin{figure}
	\label{appendix:morp}
	\section{Radio information and overlay with optical Pan-STARRS images.}
	\Large The figures show the images we used for the radio morphology determination. We considered seven out of 11 RL sources that have FIRST cutout image (5 sources) as well as two sources (B2 0110+29 and S5 1856+73) in which our classification is based on NVSS and an overlay with the optical Pan-STARRS.
	\vspace{1cm}
	
	\centering
	\includegraphics[scale = 0.54 ]{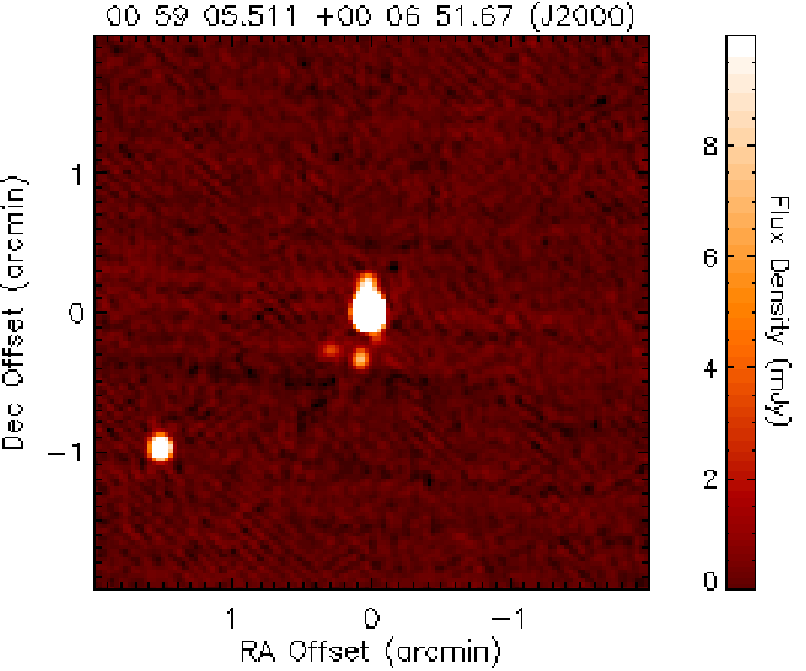}~~~~~~~~~~~~~
	\includegraphics[scale = 0.26, bb =0in -1in 8in 0in]{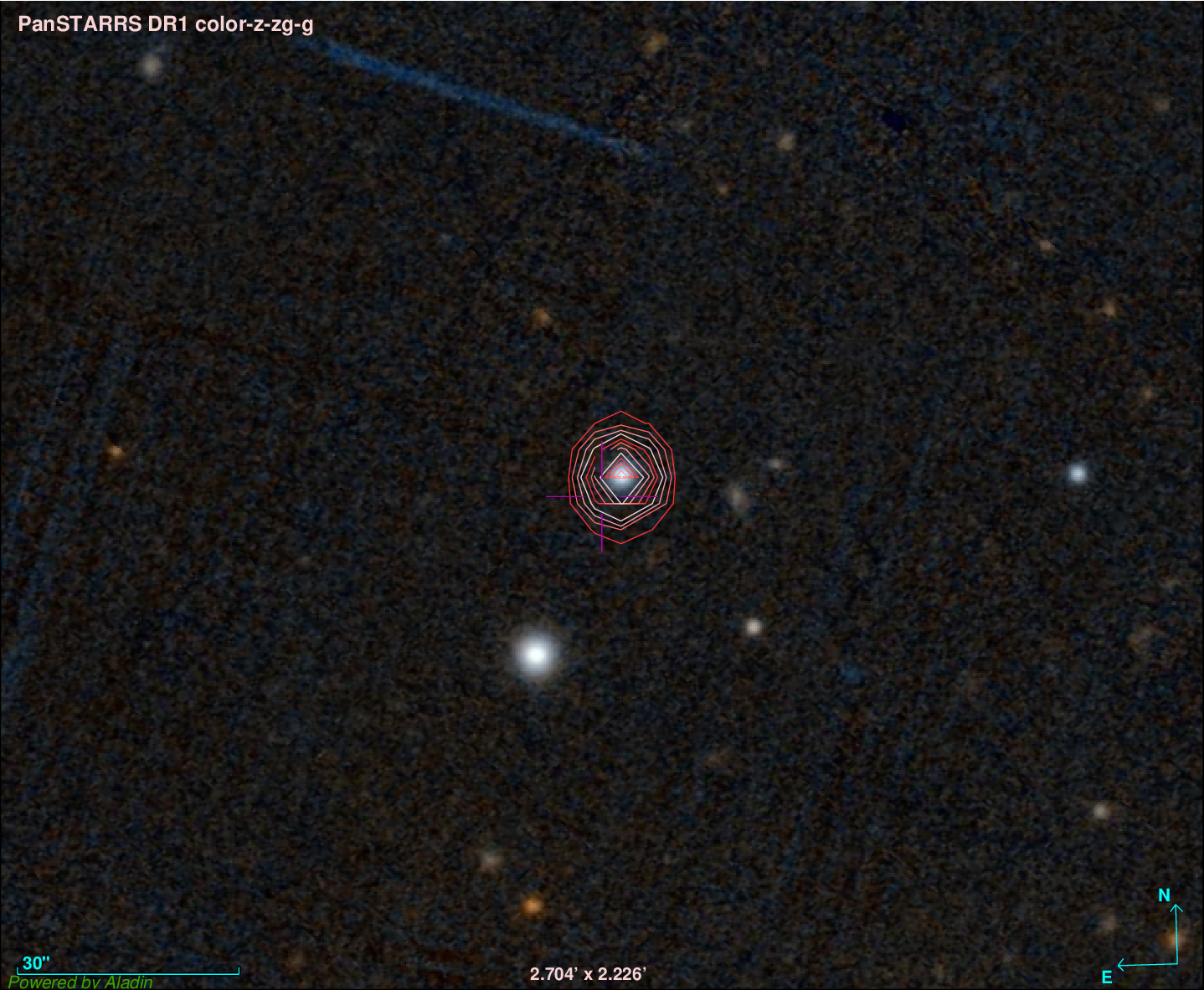}\\

	\label{fig:p1}
	\caption{\large{FIRST cutout image in arcmin ({\bf left}) and an overlay of FIRST cutout image on the Optical Pan-STARRS image ({\bf right}) of the source \textbf{PHL 923, (00 59 05.5148 +00 06 51.621).}}}
\end{figure} 
\begin{figure}
	\centering
	\includegraphics [scale = 0.35]{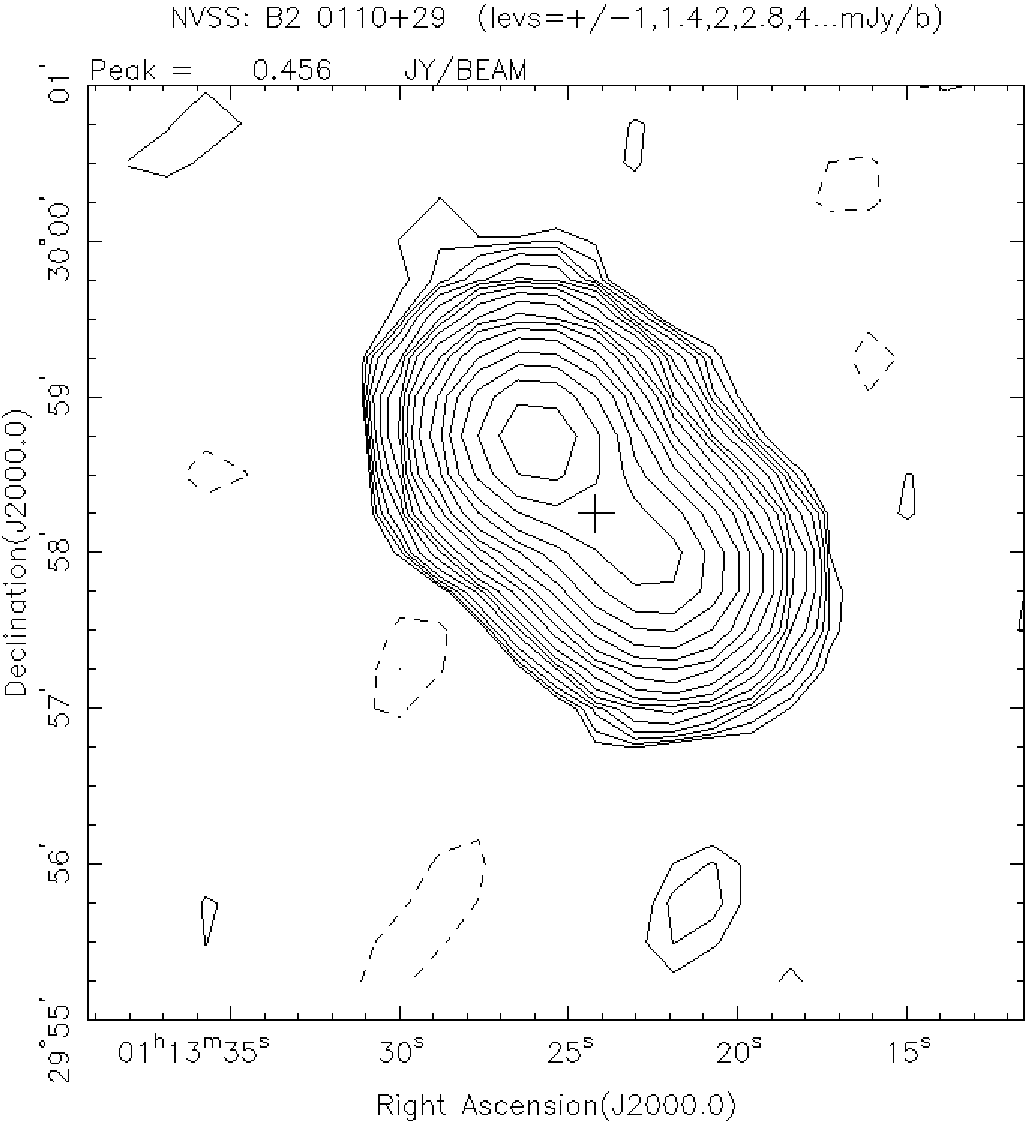}~~~~~~~~~~~~~
	\includegraphics[scale = 0.29, bb = -2in -1in 8in 0in]{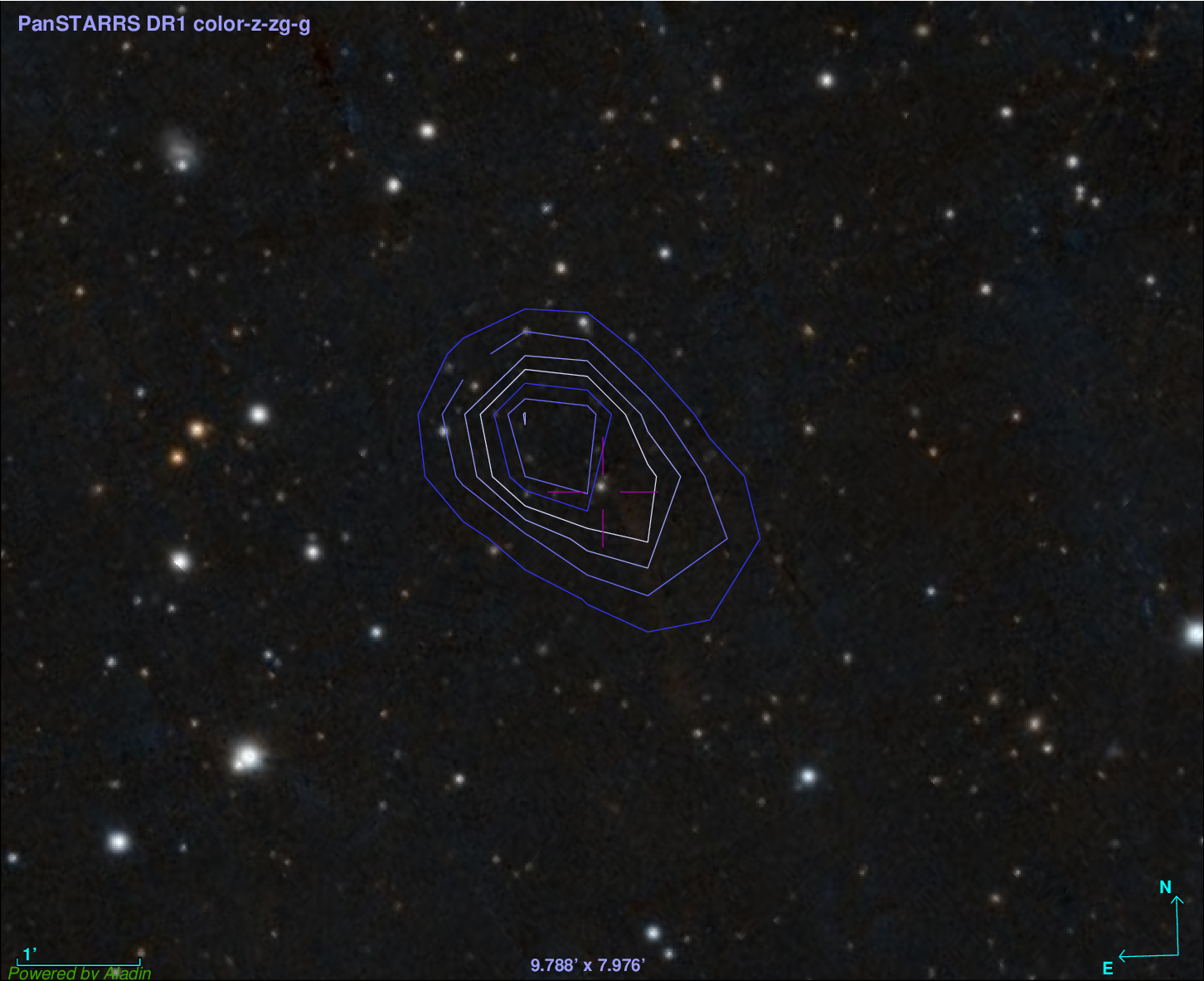}\\
	\caption{\large{NVSS contour map ({\bf left}) and an overlay of NVSS contour map on the Optical Pan-STARRS image ({\bf right}) of \textbf{B2 0110+29, (01 13 24.200 +29 58 15.00)}}}
	\label{fig:Ap2}
\end{figure} 


\begin{figure}
	\centering
	\includegraphics [scale = 0.8 ]{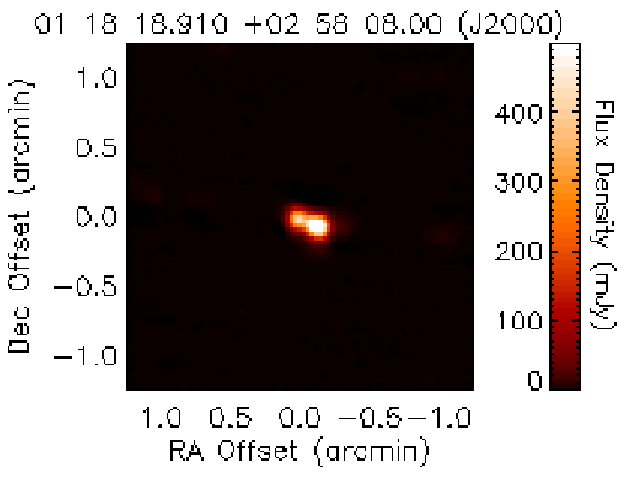}~~~~~~~~~~~~~
	\includegraphics[scale = 0.25, bb = -1.in -1.8in 8in 0in]{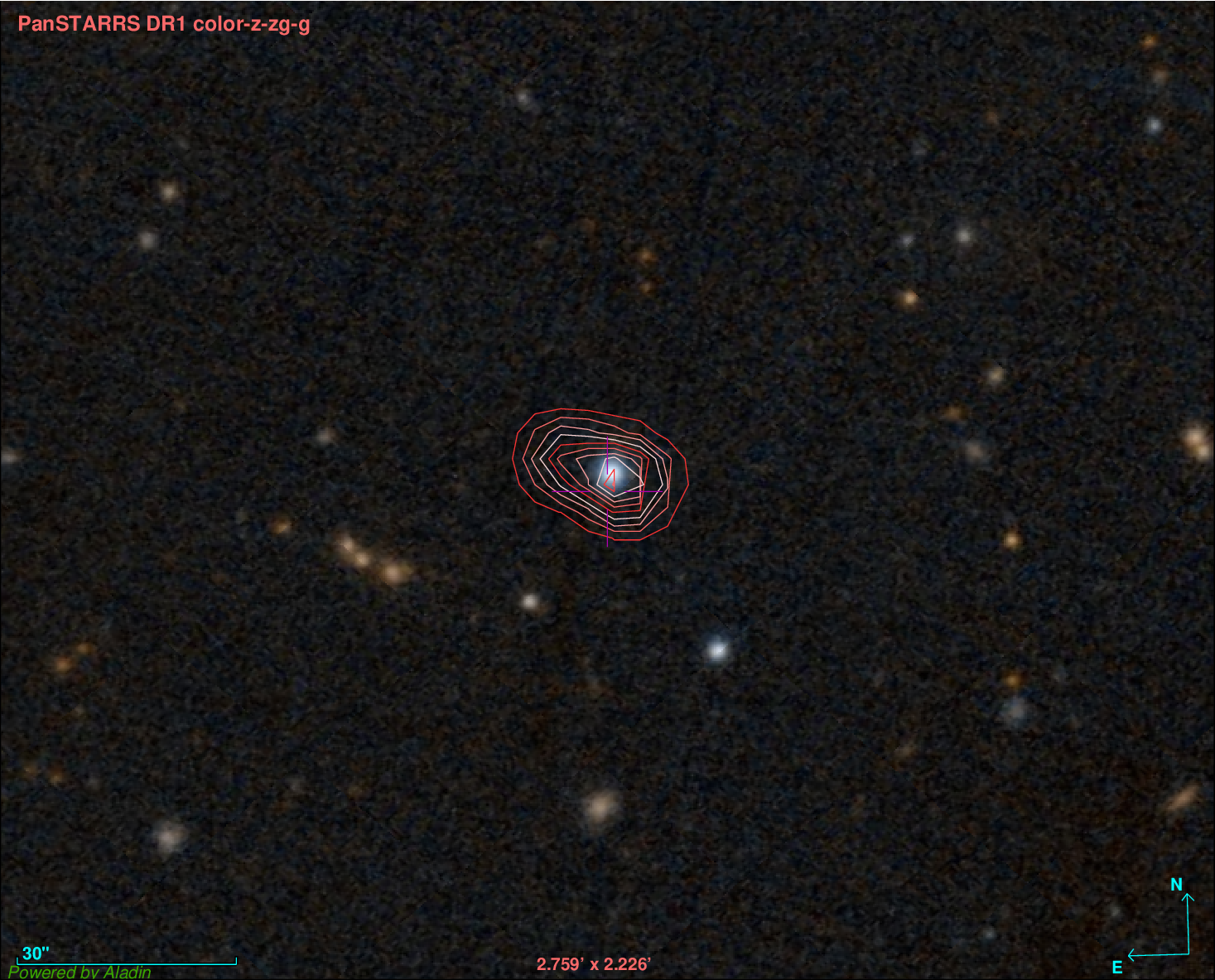}\\

	\caption{\large{FIRST cutout image obtained with maximum scaling to be 1000\,mJy ({\bf left}) and an overlay of FIRST cutout image on the Optical Pan-STARRS image ({\bf right}) for the source \textbf{3C 37, (01 18 18.489 +02 58 05.97).}}}
	\label{fig:Ap3}
\end{figure}

\begin{figure}
	\centering
	\includegraphics [scale = 0.74 ]{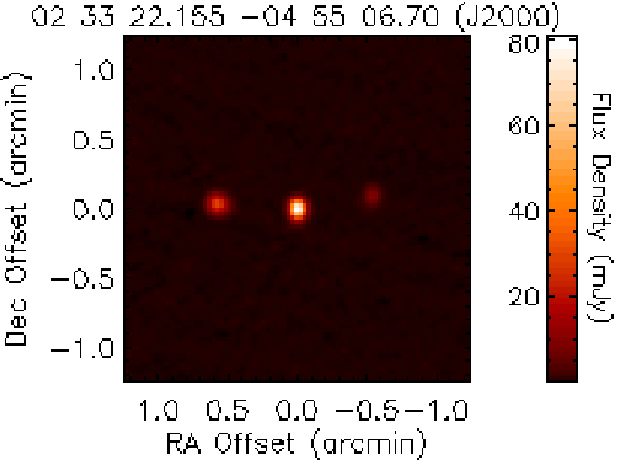}~~~~~~~~~~~~~
	\includegraphics[scale = 0.24, bb = -2in -1.2in 8in 0in]{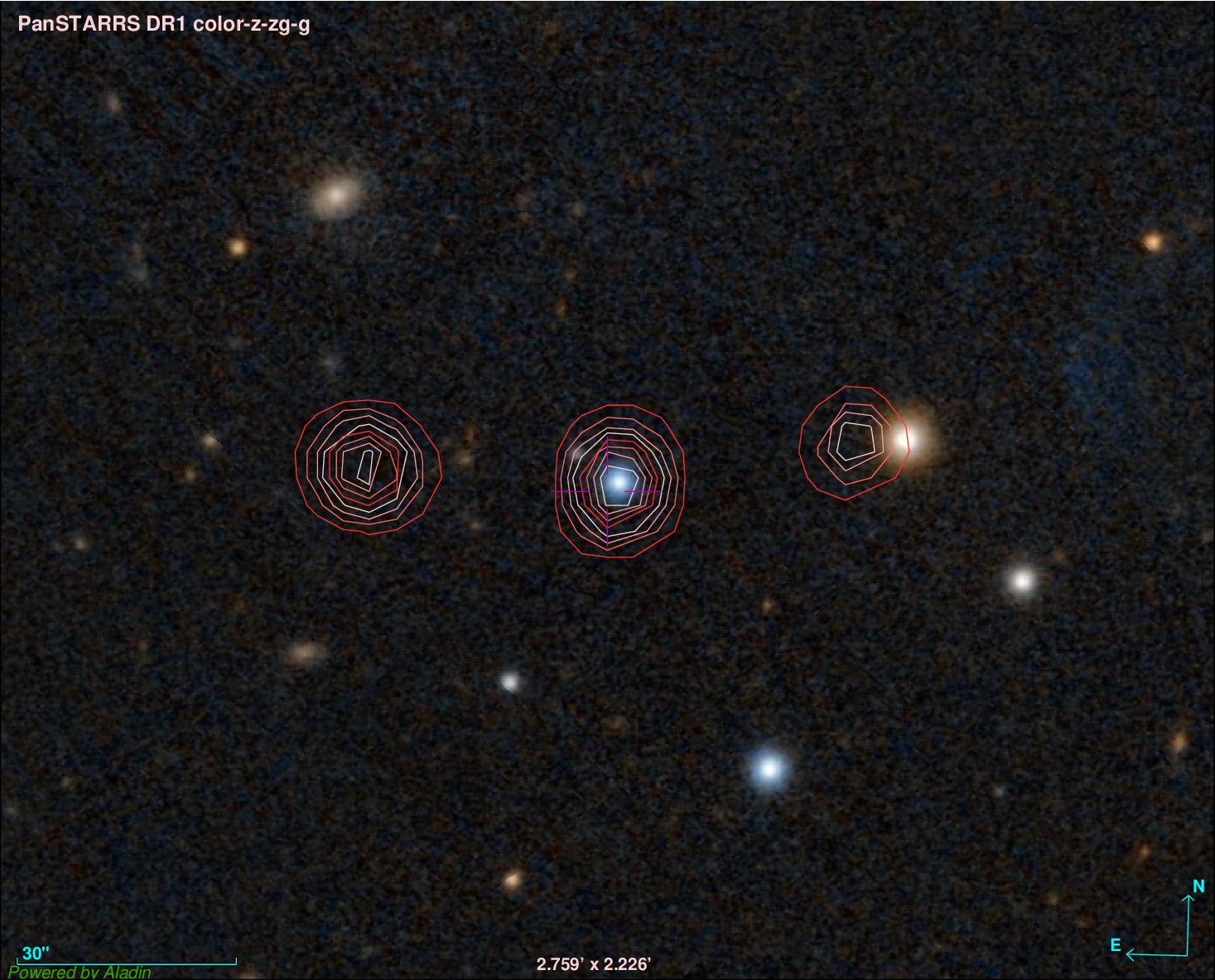}\\
	\caption{\large{FIRST cutout image obtained with maximum scaling to be 1000\,mJy to show the separate components ({\bf left}) and an overlay of FIRST cutout on the Optical Pan-STARRS image ({\bf right}) for the source \textbf{PKS 0230-051, (02 33 22.18  -04 55 06.8.)}}}
	\label{fig:AP4}
\end{figure} 


\begin{figure}
	\centering
	\includegraphics [scale = 0.73 ]{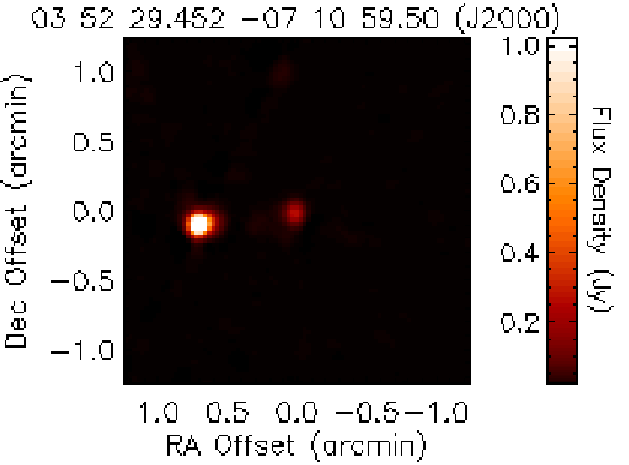}~~~~~~~~~~~~~
	\includegraphics[scale = 0.22, bb = -1.5in -1.5in 10in 0in]{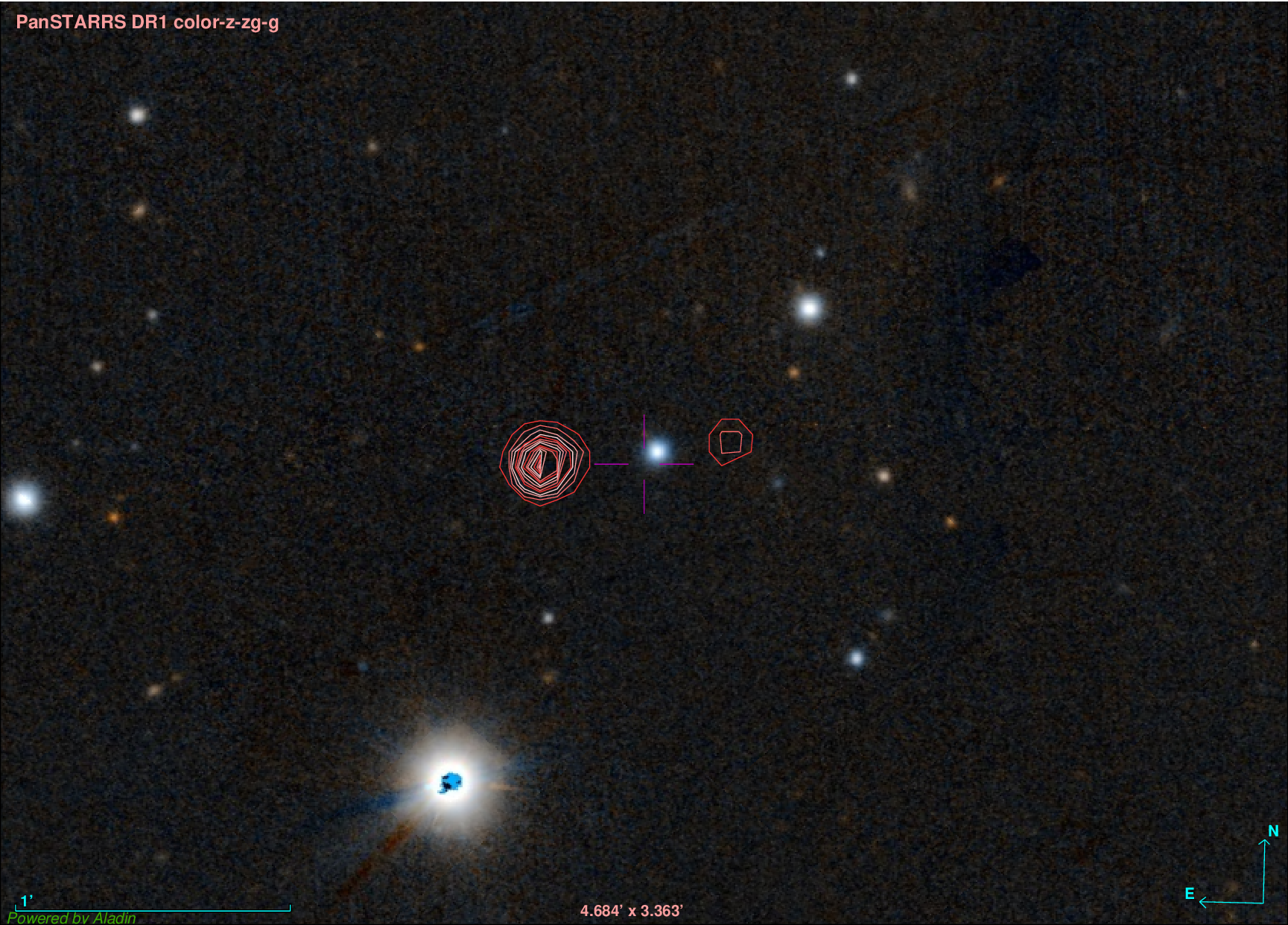}\\

	\caption{\large{FIRST cutout image obtained with maximum scaling to be 1000\,mJy to show the separate components ({\bf left}) and an overlay of FIRST cutout on the Optical Pan-STARRS image ({\bf right}) for the source \textbf{3C 94, (03 52 30.552 -07 11 02.32). }}}
	\label{fig:Ap5}
\end{figure} 


\begin{figure}
	\centering
	\includegraphics [scale = 0.35]{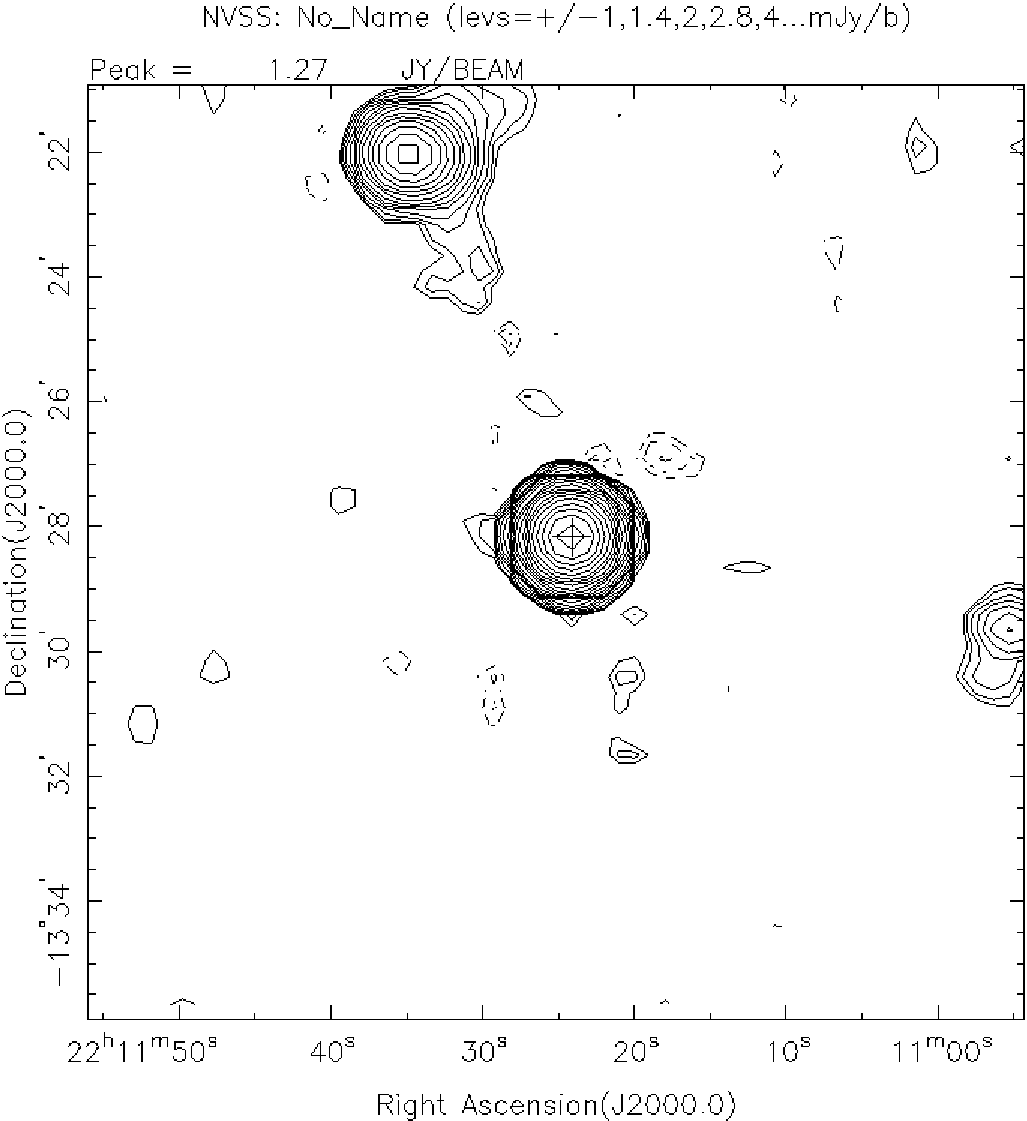}~~~~~~~~~~~~~
	\includegraphics[scale = 0.25, bb = -2.5in -1.99in 10in 0in]{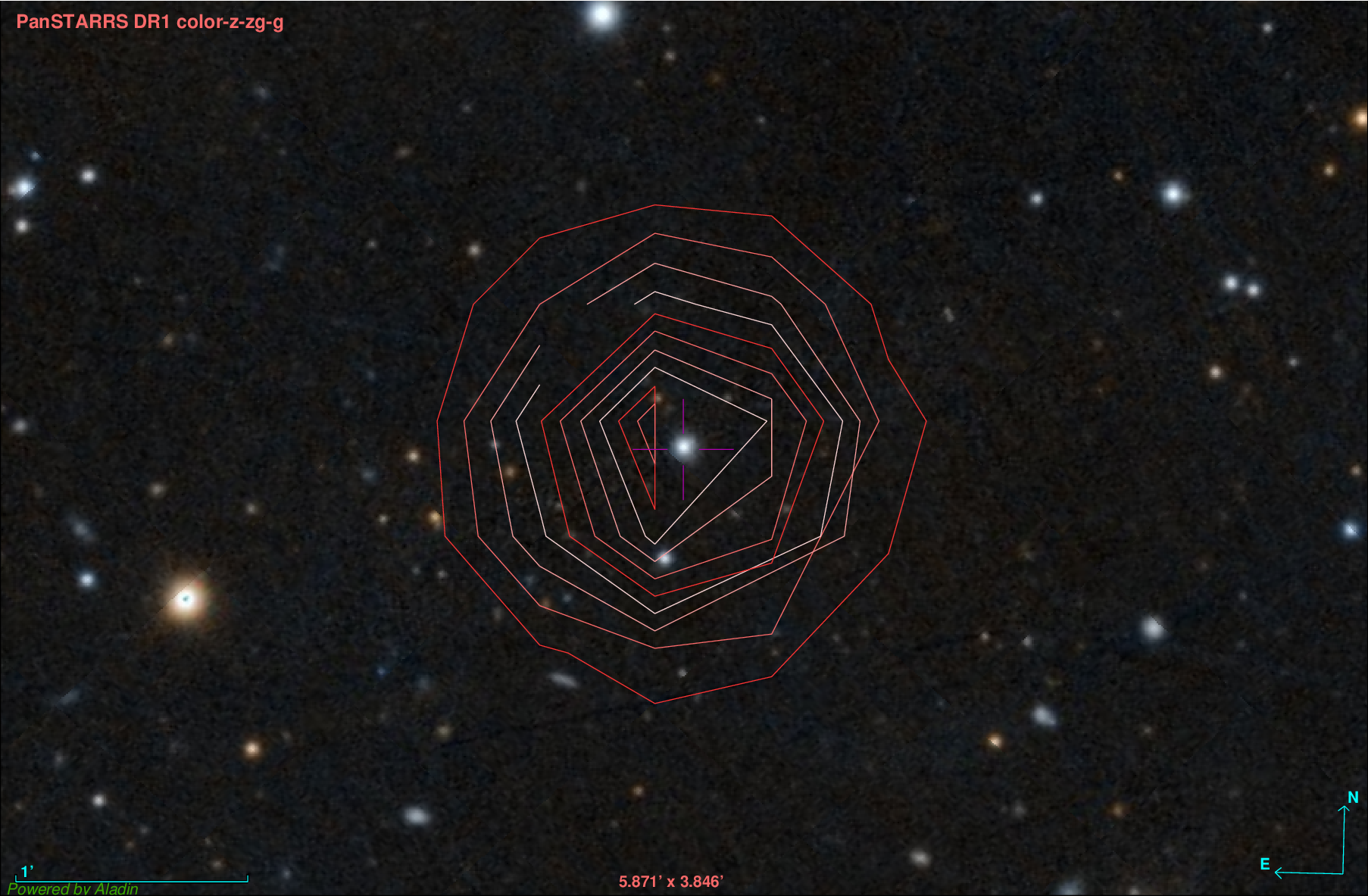}\\
	\caption{\large{NVSS contour map ({\bf left}) and an overlay of NVSS contour map on the Optical Pan-STARRS image ({\bf right}) of the source \textbf{ PKS 2208-137, (22 11 24.0994 -13 28 09.723).}}}
	\label{fig:Ap6}
\end{figure} 

\begin{figure}
	\centering
	\includegraphics [scale = 0.7]{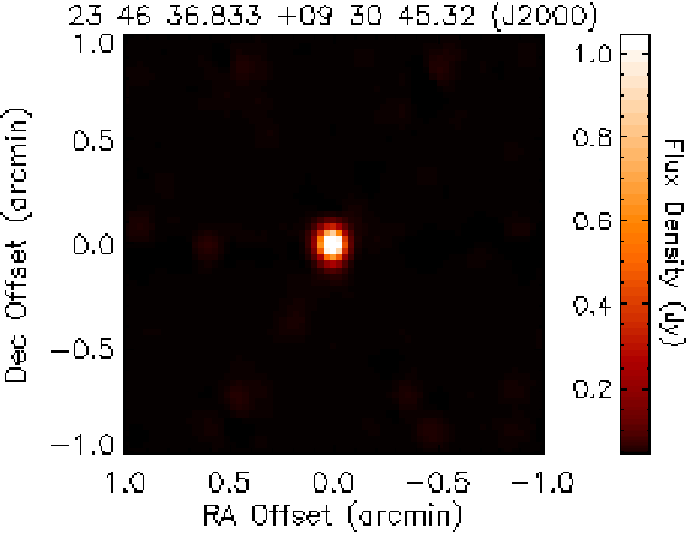}~~~~~~~~~~~~~
	\includegraphics[scale = 0.27, bb = -2.5in -1in 10in 0in]{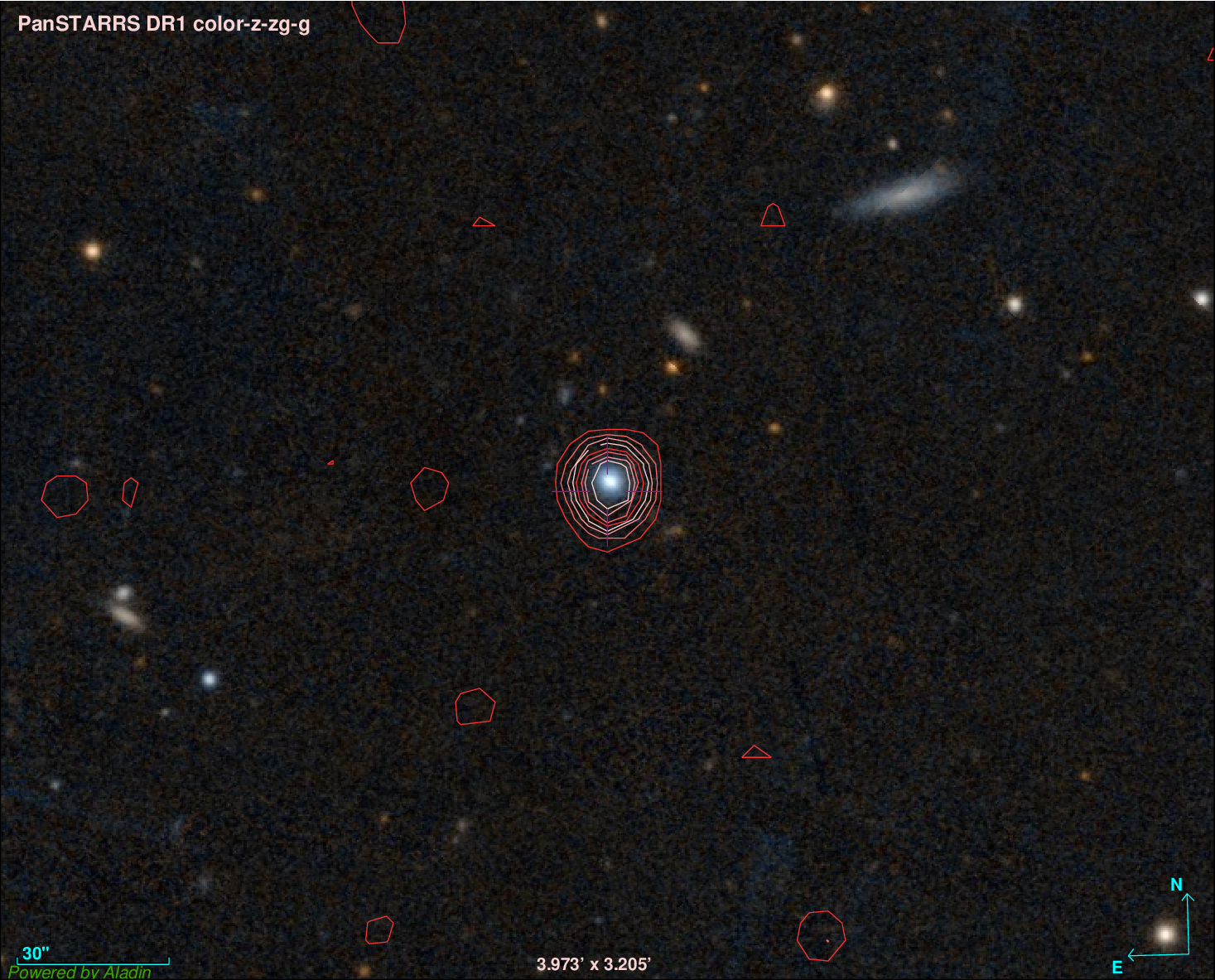}\\
	\caption{\large{FIRST cutout image obtained with maximum scaling to be 1000\, mJy to show the separate components ({\bf left}) and an overlay of FIRST cutout on the Optical Pan-STARRS image ({\bf right}) for the source \textbf{PKS 2344+09, (23 46 36.8385 +09 30 45.515)}}}.
\end{figure} 
\vfill



. 

\begin{figure}
	
	\section{Distribution of median measurement for centroid shift and radio loudness parameter }
	\label{appendix:median}
	\vspace{0.3cm}
	\large The distribution of median measurement in equal bin intervals by using the semi-interquartile range as an estimate of the sample dispersion for the centroid velocity shift at half fractional intensity and the radio loudness parameter of the sources from \citet{marziani2003optical} and our eRK quasars is shown below.
	\begin{center}

		\includegraphics [scale=0.5]{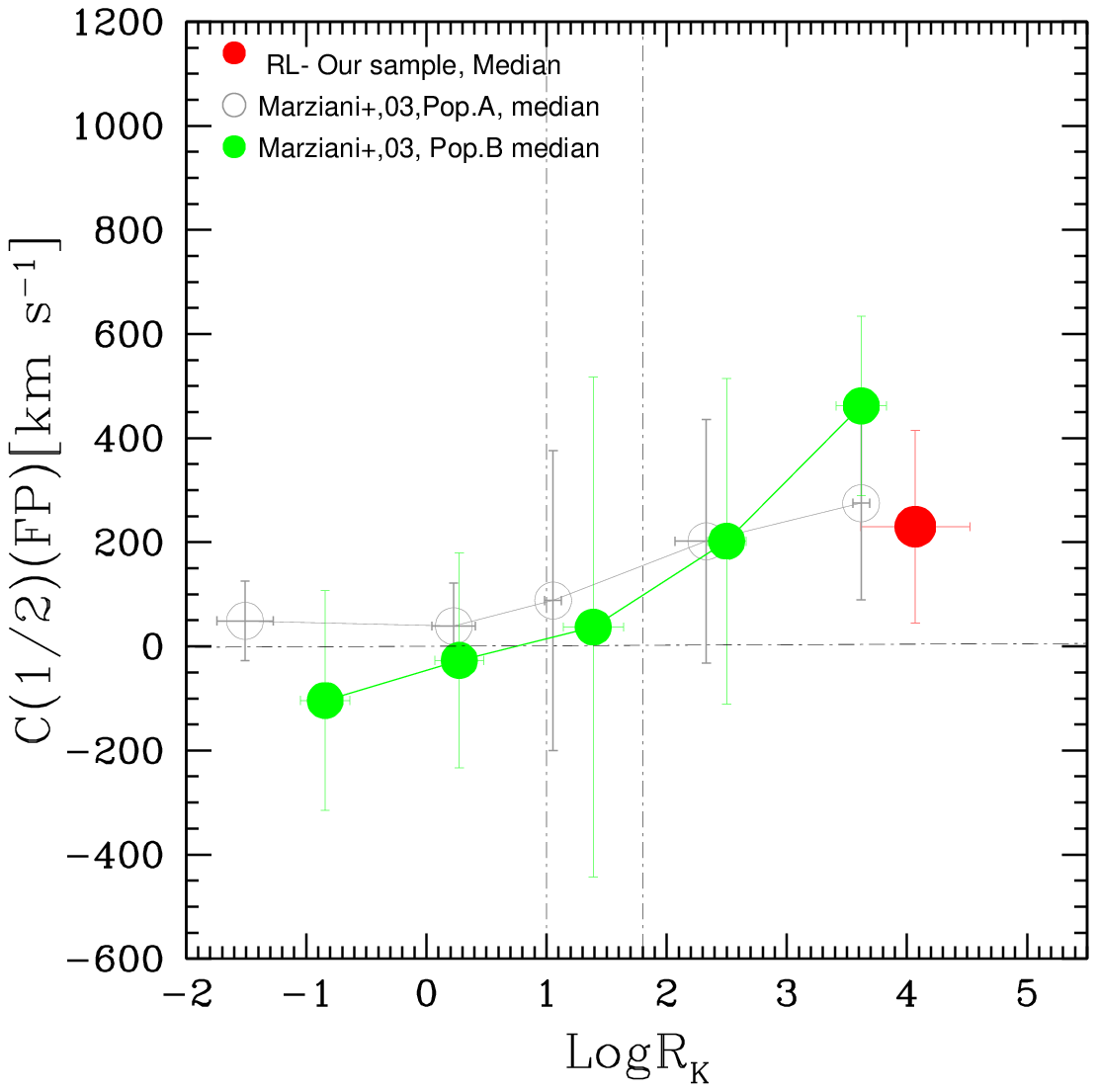}
	\end{center}
	\caption{\large{ Relation between \hb\ c($\frac{1}{2}$) with the logarithm of radio loudness parameter that indicates the median distribution to show how the trend is with the comparison sample in which Pop.A and Pop.B are shown from \citet{marziani2003optical}. The semi-interquartile range was used as an error bar. The solid red circle represents the median result from our RL spectra with \hb\ profile parameter. 
	The horizontal dot dashed line  marks the symmetric line in c(1/2). The vertical lines at 1 and 1.8 mark the nominal RQ-radio intermediate and radio intermediate-RL boundaries    \citep{1989AJ.....98.1195K,2008MNRAS.387..856Z}.
	}}
	\label{fig:logrkmed}
\end{figure}

\bsp	
\label{lastpage}
\end{document}